\documentclass[aps,prd, twocolumn, showkeys, groupedaddress,nofootinbib,amssymb]{revtex4-1}
\usepackage[T1]{fontenc}
\usepackage[latin1]{inputenc}
\usepackage{graphicx}
\usepackage[english]{babel}
\usepackage{graphicx}
\usepackage{bm}
\usepackage{amsmath}
\usepackage{amssymb}
\usepackage{amsfonts}
\usepackage{epsfig}
\usepackage{colordvi}
\usepackage{color}

\def\a {\alpha}
\def\bc{\begin{center}}
\def\ec{\end{center}}
\def\ie{{\it i.e.}}
\def\eg{{\it e.g.}}
\def\etc{{\it etc}}
\def\etal{{\it et al.}}

\def\to{\rightarrow}
\def\lsim{\mathrel{\mathpalette\atversim<}}
\def\gsim{\mathrel{\mathpalette\atversim>}}

\def\a {\alpha}
\def\bc{\begin{center}}
\def\ec{\end{center}}

\def\gsim{\mathrel{\rlap{\lower4pt\hbox{\hskip1pt$\sim$}}

    \raise1pt\hbox{$>$}}}       

\def\gsim{\mathrel{\rlap{\lower4pt\hbox{\hskip1pt$\sim$}}
    \raise1pt\hbox{$>$}}}       

\begin{document}
\def\ben{\begin{eqnarray}}
\def\een{\end{eqnarray}}
\def\be{\begin{equation}}
\def\ee{\end{equation}}
\def\ba{\begin{eqnarray}}
\def\ea{\end{eqnarray}}
\def\bea{\begin{eqnarray*}}
\def\eea{\end{eqnarray*}}
\def\D{\partial}
\def\na{\nabla}
\def\ie{{\it i.e.} }
\def\eg{{\it e.g.} }
\def\etc{{\it etc.} }
\def\etal{{\it et al.}}
\def\nn{\nonumber}
\def\ra{\rightarrow}
\def\lra{\leftrightarrow}
\def\lsim{\buildrel{<}\over{\sim}}
\def\gsim{\buildrel{>}\over{\sim}}
\def\G{\Gamma}


\def\jmp{{\it J. Math. Phys.}\ }
\def\pr{{\it Phys. Rev.}\ }
\def\prl{{\it Phys. Rev. Lett.}\ }
\def\pl{{\it Phys. Lett.}\ }
\def\np{{\it Nucl. Phys.}\ }
\def\modpl{{\it Mod. Phys. Lett.}\ }
\def\ijmp{{\it Int. Journ. Mod. Phys.}\ }
\def\ijtp{{\it Int. Journ. Theor. Phys.}\ }
\def\cmp{{\it Commun. Math. Phys.}\ }
\def\cqg{{\it Class. Quantum Grav.}\ }
\def\ap{{\it Ann. Phys. (N.Y.)}\ }
\def\spj{{\it Sov. Phys. JETP}\ }
\def\spjl{{\it Sov. Phys. JETP Lett.}\ }
\def\prs{{\it Proc. R. Soc.}\ }
\def\grg{{\it Gen. Relativ. Grav.}\ }
\def\nat{{\it Nature}\ }
\def\apj{{\it Ap. J.}\ }
\def\aa{{\it Astron. Astrophys.}\ }
\def\ncim{{\it Nuovo Cim.}\ }
\def\ptp{{\it Prog. Theor. Phys.}\ }
\def\aip{{\it Adv. Phys.}\ }
\def\jpamg{{\it J. Phys. A: Math. Gen.}\ }
\def\mnras{{\it Mon. Not. R. Ast. Soc.}\ }
\def\prep{{\it Phys. Rep.}\ }
\def\ncb{{\it Il Nuovo Cimento ``B''}}
\def\ssr{{\it Space Sci. Rev.}\ }
\def\pasp{{\it Pub. A. S. P.}\ }
\def\epl{{\it Europhys. Lett.}\ }
\def\araa{{\it Ann. Rev. Astr. Ap.}\ }
\def\asr{{\it Adv. Space Res.}\ }
\def\rmp{{\it Rev. Mod. Phys.}\ }
\def\etal{{\it et al.}}
\def\ie{{\it i.e. }}
\def\eg{{\it e.g. }}
\def\a{\`a }\def\o{\`o }\def\ii{\`\i{} }
\def\u{\`u  }\def\e{\`e }\def\ke{ch\'e }

\def\psiul{\overline\psi}
\def\pb{\not\!\partial}
\def\p{\phi}
\def\pv{\phi}
\def\v{V(\phi)}
\def\vp{V'(\phi)}

\def\ad{\dot{a}}
\def\vol{\int d^4x\,\sqrt{-g}}
\def\grav{\frac{1}{16 \pi G}}
\def\half{\frac{1}{2}}
\def\gu{g^{\mu\nu}}
\def\gd{g_{\mu\nu}}

\def\umu{^{\mu}}
\def\unu{^{\nu}}
\def\dmu{_{\mu}}
\def\dnu{_{\nu}}
\def\umunu{^{\mu\nu}}
\def\dmunu{_{\mu\nu}}
\def\ua{^{\alpha}}
\def\ub{^{\beta}}
\def\da{_{\alpha}}
\def\db{_{\beta}}
\def\ug{^{\gamma}}
\def\dg{_{\gamma}}
\def\uamu{^{\alpha\mu}}
\def\uanu{^{\alpha\nu}}
\def\uab{^{\alpha\beta}}
\def\dab{_{\alpha\beta}}
\def\dabgd{_{\alpha\beta\gamma\delta}}
\def\uabgd{^{\alpha\beta\gamma\delta}}
\def\udeab{^{;\alpha\beta}}
\def\ddeab{_{;\alpha\beta}}
\def\ddemunu{_{;\mu\nu}}
\def\udemunu{^{;\mu\nu}}
\def\ddemu{_{;\mu}}  \def\udemu{^{;\mu}}
\def\ddenu{_{;\nu}}  \def\udenu{^{;\nu}}
\def\ddea{_{;\alpha}}  \def\udea{^{;\alpha}}
\def\ddeb{_{;\beta}}  \def\udeb{^{;\beta}}

\def\naba{\nabla_{\alpha}}
\def\nabb{\nabla_{\beta}}
\def\pmu{\partial_{\mu}}
\def\pnu{\partial_{\nu}}
\def\pa{\partial}

\let\lam=\lambda  \let\Lam=\Lambda
\let\eps=\varepsilon
\let\gam=\gamma
\let\alp=\alpha
\let\sig=\sigma

\def\jmp{{\it J. Math. Phys.}\ }
\def\pr{{\it Phys. Rev.}\ }
\def\prl{{\it Phys. Rev. Lett.}\ }
\def\pl{{\it Phys. Lett.}\ }
\def\np{{\it Nucl. Phys.}\ }
\def\modpl{{\it Mod. Phys. Lett.}\ }
\def\ijmp{{\it Int. Journ. Mod. Phys.}\ }
\def\ijtp{{\it Int. Journ. Theor. Phys.}\ }
\def\cmp{{\it Commun. Math. Phys.}\ }
\def\cqg{{\it Class. Quantum Grav.}\ }
\def\ap{{\it Ann. Phys. (N.Y.)}\ }
\def\spj{{\it Sov. Phys. JETP}\ }
\def\spjl{{\it Sov. Phys. JETP Lett.}\ }
\def\prs{{\it Proc. R. Soc.}\ }
\def\grg{{\it Gen. Relativ. Grav.}\ }
\def\nat{{\it Nature}\ }
\def\apj{{\it Ap. J.}\ }
\def\aa{{\it Astron. Astrophys.}\ }
\def\ncim{{\it Il Nuovo Cim.}\ }
\def\ptp{{\it Prog. Theor. Phys.}\ }
\def\aip{{\it Adv. Phys.}\ }
\def\jpamg{{\it J. Phys. A: Math. Gen.}\ }
\def\mnras{{\it Mon. Not. R. Ast. Soc.}\ }
\def\prep{{\it Phys. Rep.}\ }
\def\ncb{{\it Il Nuovo Cimento ``B''}}
\def\ssr{{\it Space Sci. Rev.}\ }
\def\pasp{{\it Pub. A. S. P.}\ }
\def\araa{{\it Ann. Rev. Astr. Ap.}\ }
\def\asr{{\it Adv. Space Res.}\ }
\def\rmp{{\it Rev. Mod. Phys.}\ }
\def\etal{{\it et al.}}
\def\ie{{\it i.e. }}
\def\eg{{\it e.g. }}
\def\jmp{{\it J. Math. Phys.}\ }
\def\pr{{\it Phys. Rev.}\ }
\def\pl{{\it Phys. Lett.}\ }
\def\nat{{\it Nature}\ }
\def\apj{{\it Ap. J.}\ }
\def\mnras{{\it Mon. Not. R. Ast. Soc.}\ }
\def\araa{{\it Ann. Rev. Astr. Ap.}\ }
\def\rmp{{\it Rev. Mod. Phys.}\ }
\def\arns{{\it Ann. Rev. Nucl. Part. Sci.}\ }

\def\ad{\dot{a}}
\def\vol{\int d^4x\,\sqrt{-g}}
\def\grav{\frac{1}{16 \pi G}}
\def\half{\frac{1}{2}}
\def\gu{g^{\mu\nu}}
\def\gd{g_{\mu\nu}}
\def\umu{^{\mu}}
\def\unu{^{\nu}}
\def\dmu{_{\mu}}
\def\dnu{_{\nu}}
\def\umunu{^{\mu\nu}}
\def\dmunu{_{\mu\nu}}
\def\ua{^{\alpha}}
\def\ub{^{\beta}}
\def\da{_{\alpha}}
\def\db{_{\beta}}
\def\ug{^{\gamma}}
\def\dg{_{\gamma}}
\def\uamu{^{\alpha\mu}}
\def\uanu{^{\alpha\nu}}
\def\uab{^{\alpha\beta}}
\def\dab{_{\alpha\beta}}
\def\dabgd{_{\alpha\beta\gamma\delta}}
\def\uabgd{^{\alpha\beta\gamma\delta}}
\def\udeab{^{;\alpha\beta}}
\def\ddeab{_{;\alpha\beta}}
\def\ddemunu{_{;\mu\nu}}
\def\udemunu{^{;\mu\nu}}
\def\ddemu{_{;\mu}}  \def\udemu{^{;\mu}}
\def\ddenu{_{;\nu}}  \def\udenu{^{;\nu}}
\def\ddea{_{;\alpha}}  \def\udea{^{;\alpha}}
\def\ddeb{_{;\beta}}  \def\udeb{^{;\beta}}

\def\naba{\nabla_{\alpha}}
\def\nabb{\nabla_{\beta}}
\def\pmu{\partial_{\mu}}
\def\pnu{\partial_{\nu}}
\def\pa{\partial}

\let\lam=\lambda
\let\Lam=\Lambda
\let\eps=\varepsilon
\let\gam=\gamma
\let\alp=\alpha
\let\sig=\sigma
\let\lb=\label
\renewcommand{\epsilon}{\varepsilon}
\let\no=\nonumber
\let\noin=\noindent

\def\CS{\cal{CS}}
\def\beqt{\begin{tabular}}
\def\eet{\end{tabular}}
\def\beqf{\begin{figure}}
\def\eef{\end{figure}}
\def\beqa{\begin{eqnarray}}
\def\eeqa{\end{eqnarray}}
\def\vol{\int d^4x\,\sqrt{-g}}
\def\grav{\frac{1}{16 \pi G}}
\def\half{\frac{1}{2}}
\def\ra{\rightarrow}
\def\Gu{G^{\mu\nu}}
\def\Gd{G_{\mu\nu}}
\def\gu{g^{\mu\nu}}
\def\gd{g_{\mu\nu}}
\def\hu{h^{\mu\nu}}
\def\hd{h_{\mu\nu}}
\def\umu{^{\mu}}
\def\unu{^{\nu}}
\def\dmu{_{\mu}}
\def\dnu{_{\nu}}
\def\umunu{^{\mu\nu}}
\def\dmunu{_{\mu\nu}}
\def\ua{^{\alpha}}
\def\ub{^{\beta}}
\def\da{_{\alpha}}
\def\db{_{\beta}}
\def\ug{^{\gamma}}
\def\dg{_{\gamma}}
\def\uamu{^{\alpha\mu}}
\def\uanu{^{\alpha\nu}}
\def\uab{^{\alpha\beta}}
\def\dab{_{\alpha\beta}}
\def\dabgd{_{\alpha\beta\gamma\delta}}
\def\uabgd{^{\alpha\beta\gamma\delta}}
\def\udeab{^{;\alpha\beta}}
\def\ddeab{_{;\alpha\beta}}
\def\ddemunu{_{;\mu\nu}}
\def\udemunu{^{;\mu\nu}}
\def\ddemu{_{;\mu}}
\def\udemu{^{;\mu}}
\def\ddenu{_{;\nu}}
\def\udenu{^{;\nu}}
\def\ddea{_{;\alpha}}
\def\udea{^{;\alpha}}
\def\ddeb{_{;\beta}}
\def\udeb{^{;\beta}}
\def\naba{\nabla_{\alpha}}
\def\nabb{\nabla_{\beta}}
\def\pmu{\partial_{\mu}}
\def\pnu{\partial_{\nu}}
\def\pa{\partial}
\def\etal{{\it et al.}}
\def\ie{{\it i.e. }}
\def\eg{{\it e.g. }}
\def\psiul{\overline\psi}
\def\pb{\not\!\partial}
\def\fp{F'(\phi)}
\def\fpp{F''(\phi)}
\def\p{\phi}
\def\pv{\phi}
\def\v{V(\phi)}
\def\vp{V'(\phi)}
\def\l{\cal L}
\def\g{\cal G}
\def\fo{\cal F}
\def\s{\cal S}
\def\n{\cal N}

\let\lam=\lambda
\let\Lam=\Lambda
\let\eps=\varepsilon
\let\gam=\gamma
\let\alp=\alpha
\let\sig=\sigma
\let\lb=\label
\renewcommand{\epsilon}{\varepsilon}
\let\no=\nonumber
\let\noin=\noindent

\def\disp{\displaystyle}

\def\vol{d^4x\,\sqrt{-g}}
\def\volbr{d^4x\,\sqrt{-\bar{g}}}
\def\grav{\frac{1}{16 \pi G}}
\def\half{\frac{1}{2}}
\def\gu{g^{\mu\nu}}
\def\gd{g_{\mu\nu}}
\def\gbru{{\bar{g}}^{\mu\nu}}
\def\gbrd{{\bar{g}}_{\mu\nu}}

\def\naba{\nabla_{\alpha}}
\def\nabb{\nabla_{\beta}}
\def\pmu{\partial_{\mu}}
\def\pnu{\partial_{\nu}}
\def\pa{\partial}

\def\al{{\alpha}}
\def\gam{{\gamma}}
\def\del{{\delta}}
\def\epsi{{\varepsilon}}
\def\lam{{\lambda}}
\def\sig{{\sigma}}
\def\ome{{\omega}}
\def\GAM{{\Gamma}}
\def\LAM{{\Lambda}}
\def\OME{{\Omega}}

\def\ua{^{\alpha}}
\def\ub{^{\beta}}
\def\ug{^{\gamma}}
\def\da{_{\alpha}}
\def\db{_{\beta}}
\def\dg{_{\gamma}}
\def\umu{^{\mu}}
\def\unu{^{\nu}}
\def\dmu{_{\mu}}
\def\dnu{_{\nu}}
\def\umunu{^{\mu\nu}}
\def\dmunu{_{\mu\nu}}
\def\daub{_{\alpha}^{~\beta}}
\def\dbua{_{\beta}^{~\alpha}}
\def\dmuunu{_{\mu}^{~\nu}}
\def\dnuumu{_{\nu}^{~\mu}}
\def\uamu{^{\alpha\mu}}
\def\uanu{^{\alpha\nu}}
\def\uab{^{\alpha\beta}}
\def\dab{_{\alpha\beta}}
\def\dabgd{_{\alpha\beta\gamma\delta}}
\def\uabgd{^{\alpha\beta\gamma\delta}}
\def\uvmu{_{,}^{\mu}}
\def\uvnu{_{,}^{\nu}}
\def\uva{_{,}^{\alpha}}
\def\uvb{_{,}^{\beta}}
\def\dvmu{_{,\mu}}
\def\dvnu{_{,\nu}}
\def\dva{_{,\alpha}}
\def\dvb{_{,\beta}}
\def\udemu{_{;}^{\mu}}
\def\udenu{_{;}^{\nu}}
\def\udea{_{;}^{\alpha}}
\def\udeb{_{;}^{\beta}}
\def\ddemu{_{;\mu}}
\def\ddenu{_{;\nu}}
\def\ddea{_{;\alpha}}
\def\ddeb{_{;\beta}}
\def\ddemunu{_{;\mu\nu}}
\def\udemunu{^{;\mu\nu}}
\def\udeab{^{;\alpha\beta}}
\def\ddeab{_{;\alpha\beta}}

\def\L{{{\cal L}}}
\def\f{{\phi}}
\def\ka{{\kappa}}
\def\gbr{{\bar{g}}}
\def\ad{{\dot{a}}}
\def\add{{\ddot{a}}}
\def\ap{{a^{'}}}
\def\app{{a^{''}}}
\def\apq{{a^{'2}}}
\def\abr{{\bar{a}}}
\def\abrd{{\dot{\bar{a}}}}
\def\abrdd{{\ddot{\bar{a}}}}
\def\abrp{{{\bar{a}}^{'}}}
\def\abrpp{{{\bar{a}}^{''}}}
\def\abrpq{{{\bar{a}}^{'2}}}
\def\fd{{\dot{\phi}}}
\def\fdd{{\ddot{\phi}}}
\def\fp{{\phi^{'}}}
\def\fpp{{\phi^{''}}}
\def\fpq{{\phi^{'2}}}
\def\fbr{{\bar{\phi}}}
\def\fbrd{{\dot{\bar{\phi}}}}
\def\fbrdd{{\ddot{\bar{\phi}}}}
\def\fbrp{{{\bar{\phi}}^{'}}}
\def\fbrpp{{{\bar{\phi}}^{''}}}
\def\fbrpq{{{\bar{\phi}}^{'2}}}
\def\VDEF{{V_{\phi}}}
\def\VBR{{\bar{V}}}
\def\VBRDEF{{\bar{V}_{\bar{\phi}}}}
\def\FDEF{{F_{\phi}}}
\def\FD{{\dot{F}}}
\def\FDD{{\ddot{F}}}
\def\tbr{{\bar{t}}}
\def\LBR{{\overline{L}}}
\def\EBR{{\overline{E}}}

\def\ie{{\it i.e. }}
\def\eg{{\it e.g. }}
\def\etal{{\it et al.}}

\def\jmp{{\it J. Math. Phys.}\ }
\def\pr{{\it Phys. Rev.}\ }
\def\prl{{\it Phys. Rev. Lett.}\ }
\def\pl{{\it Phys. Lett.}\ }
\def\np{{\it Nucl. Phys.}\ }
\def\modpl{{\it Mod. Phys. Lett.}\ }
\def\ijmp{{\it Int. Journ. Mod. Phys.}\ }
\def\ijtp{{\it Int. Journ. Theor. Phys.}\ }
\def\aph{{\it Ann. Phys. (N.Y.)}\ }
\def\cmp{{\it Commun. Math. Phys.}\ }
\def\cqg{{\it Class. Quantum Grav.}\ }
\def\spj{{\it Sov. Phys. JETP}\ }
\def\spjl{{\it Sov. Phys. JETP Lett.}\ }
\def\prs{{\it Proc. R. Soc.}\ }
\def\grg{{\it Gen. Relativ. Grav.}\ }
\def\nat{{\it Nature}\ }
\def\apj{{\it Ap. J.}\ }
\def\aa{{\it Astron. Astrophys.}\ }
\def\ncim{{\it Nuovo Cim.}\ }
\def\ptp{{\it Prog. Theor. Phys.}\ }
\def\adp{{\it Adv. Phys.}\ }
\def\jpamg{{\it J. Phys. A: Math. Gen.}\ }
\def\mnras{{\it Mon. Not. R. Ast. Soc.}\ }
\def\prep{{\it Phys. Rep.}\ }
\def\ncb{{\it Il Nuovo Cimento ``B''}\ }
\def\ssr{{\it Space Sci. Rev.}\ }
\def\pasp{{\it Pub. A. S. P.}\ }
\def\araa{{\it Ann. Rev. Astr. Ap.}\ }
\def\asr{{\it Adv. Space Res.}\ }
\def\rmp{{\it Rev. Mod. Phys.}\ }
\def\hpa{{\it Helv. Phys. Acta}\ }
\def\phs{{\it Physica Scripta}\ }

\def\rjmp{ J. Math. Phys. }
\def\rpr{ Phys. Rev. }
\def\rprl{ Phys. Rev. Lett. }
\def\rpl{ Phys. Lett. }
\def\rnp{ Nucl. Phys. }
\def\rmodpl{ Mod. Phys. Lett. }
\def\rijmp{ Int. Journ. Mod. Phys. }
\def\rcmp{ Commun. Math. Phys. }
\def\rcqg{ Class. Quantum Grav. }
\def\raph{ Ann. Phys. (N.Y.) }
\def\rspj{ Sov. Phys. JETP }
\def\rspjl{ Sov. Phys. JETP Lett. }
\def\rprs{ Proc. R. Soc. }
\def\rgrg{ Gen. Relativ. Grav. }
\def\rnat{ Nature }
\def\rapj{ Ap. J. }
\def\raaa{ Astron. Astrophys. }
\def\rncim{ Nuovo Cim. }
\def\rptp{ Prog. Theor. Phys. }
\def\raip{ Adv. Phys. }
\def\rjpamg{ J. Phys. A: Math. Gen. }
\def\rmnras{ Mon. Not. R. Ast. Soc. }
\def\rprep{ Phys. Rep. }
\def\rncb{ Il Nuovo Cimento ``B'' }
\def\rssr{ Space Sci. Rev. }
\def\rpasp{ Pub. A. S. P. }
\def\raraa{ Ann. Rev. Astr. Ap. }
\def\rasr{ Adv. Space Res. }
\def\rhpa{ Helv. Phys. Acta }
\def\rphs{ Physica Scripta }

\def\bib#1{$^{\ref{#1}}$}

\let\lam=\lambda  \let\Lam=\Lambda
\let\eps=\varepsilon
\let\gam=\gamma
\let\alp=\alpha
\let\sig=\sigma

\let\lb=\label
\renewcommand{\epsilon}{\varepsilon}
\let\no=\nonumber

\def\jmp{{\it J. Math. Phys.}\ }
\def\pr{{\it Phys. Rev.}\ }
\def\prl{{\it Phys. Rev. Lett.}\ }
\def\pl{{\it Phys. Lett.}\ }
\def\np{{\it Nucl. Phys.}\ }
\def\modpl{{\it Mod. Phys. Lett.}\ }
\def\ijmp{{\it Int. Journ. Mod. Phys.}\ }
\def\ijtp{{\it Int. Journ. Theor. Phys.}\ }
\def\cmp{{\it Commun. Math. Phys.}\ }
\def\cqg{{\it Class. Quantum Grav.}\ }
\def\spj{{\it Sov. Phys. JETP}\ }
\def\spjl{{\it Sov. Phys. JETP Lett.}\ }
\def\prs{{\it Proc. R. Soc.}\ }
\def\grg{{\it Gen. Relativ. Grav.}\ }
\def\nat{{\it Nature}\ }
\def\apj{{\it Ap. J.}\ }
\def\aa{{\it Astron. Astrophys.}\ }
\def\ncim{{\it Nuovo Cim.}\ }
\def\ptp{{\it Prog. Theor. Phys.}\ }
\def\aip{{\it Adv. Phys.}\ }
\def\jpamg{{\it J. Phys. A: Math. Gen.}\ }
\def\mnras{{\it Mon. Not. R. Ast. Soc.}\ }
\def\prep{{\it Phys. Rep.}\ }
\def\ncb{{\it Il Nuovo Cimento ``B''}}
\def\ssr{{\it Space Sci. Rev.}\ }
\def\pasp{{\it Pub. A. S. P.}\ }
\def\araa{{\it Ann. Rev. Astr. Ap.}\ }
\def\asr{{\it Adv. Space Res.}\ }
\def\rmp{{\it Rev. Mod. Phys.}\ }
\def\etal{{\it et al.}}
\def\ie{{\it i.e. }}
\def\eg{{\it e.g. }}

\def\psiul{\overline\psi}
\def\pb{\not\!\partial}
\def\fp{F'(\phi)}
\def\fpp{F''(\phi)}

\def\p{\phi}
\def\pv{\phi}
\def\v{V(\phi)}
\def\vp{V'(\phi)}
\def\l{\cal L}

\def\beqa{\begin{eqnarray}}
\def\eeqa{\end{eqnarray}}
\def\beq{\begin{equation}}
\def\eeq{\end{equation}}
\def\ad{\dot{a}}
\def\vol{\int d^4x\,\sqrt{-g}}
\def\grav{\frac{1}{16 \pi G}}
\def\half{\frac{1}{2}}
\def\gu{g^{\mu\nu}}
\def\gd{g_{\mu\nu}}

\def\umu{^{\mu}}
\def\unu{^{\nu}}
\def\dmu{_{\mu}}
\def\dnu{_{\nu}}
\def\umunu{^{\mu\nu}}
\def\dmunu{_{\mu\nu}}
\def\ua{^{\alpha}}
\def\ub{^{\beta}}
\def\da{_{\alpha}}
\def\db{_{\beta}}
\def\ug{^{\gamma}}
\def\dg{_{\gamma}}
\def\uamu{^{\alpha\mu}}
\def\uanu{^{\alpha\nu}}
\def\uab{^{\alpha\beta}}
\def\dab{_{\alpha\beta}}
\def\dabgd{_{\alpha\beta\gamma\delta}}
\def\uabgd{^{\alpha\beta\gamma\delta}}
\def\udeab{^{;\alpha\beta}}
\def\ddeab{_{;\alpha\beta}}
\def\ddemunu{_{;\mu\nu}}
\def\udemunu{^{;\mu\nu}}
\def\ddemu{_{;\mu}}  \def\udemu{^{;\mu}}
\def\ddenu{_{;\nu}}  \def\udenu{^{;\nu}}
\def\ddea{_{;\alpha}}  \def\udea{^{;\alpha}}
\def\ddeb{_{;\beta}}  \def\udeb{^{;\beta}}

\def\naba{\nabla_{\alpha}}
\def\nabb{\nabla_{\beta}}
\def\pmu{\partial_{\mu}}
\def\pnu{\partial_{\nu}}
\def\pa{\partial}

\let\lam=\lambda  \let\Lam=\Lambda
\let\eps=\varepsilon
\let\gam=\gamma
\let\alp=\alpha
\let\sig=\sigma

\def\mpl{{\it Mod. Phys. Lett.}\ }
\def\jmp{{\it J. Math. Phys.}\ }
\def\pr{{\it Phys. Rev.}\ }
\def\prl{{\it Phys. Rev. Lett.}\ }
\def\pl{{\it Phys. Lett.}\ }
\def\np{{\it Nucl. Phys.}\ }
\def\modpl{{\it Mod. Phys. Lett.}\ }
\def\ijmp{{\it Int. Journ. Mod. Phys.}\ }
\def\ijtp{{\it Int. Journ. Theor. Phys.}\ }
\def\cmp{{\it Commun. Math. Phys.}\ }
\def\cqg{{\it Class. Quantum Grav.}\ }
\def\ap{{\it Ann. Phys. (N.Y.)}\ }
\def\spj{{\it Sov. Phys. JETP}\ }
\def\spjl{{\it Sov. Phys. JETP Lett.}\ }
\def\prs{{\it Proc. R. Soc.}\ }
\def\grg{{\it Gen. Relativ. Grav.}\ }
\def\nat{{\it Nature}\ }
\def\apj{{\it Ap. J.}\ }
\def\aa{{\it Astron. Astrophys.}\ }
\def\ncim{{\it Nuovo Cim.}\ }
\def\ptp{{\it Prog. Theor. Phys.}\ }
\def\aip{{\it Adv. Phys.}\ }
\def\jpamg{{\it J. Phys. A: Math. Gen.}\ }
\def\mnras{{\it Mon. Not. R. Ast. Soc.}\ }
\def\prep{{\it Phys. Rep.}\ }
\def\ncb{{\it Il Nuovo Cimento ``B''}}
\def\ssr{{\it Space Sci. Rev.}\ }
\def\pasp{{\it Pub. A. S. P.}\ }
\def\epl{{\it Europhys. Lett.}\ }
\def\araa{{\it Ann. Rev. Astr. Ap.}\ }
\def\asr{{\it Adv. Space Res.}\ }
\def\rmp{{\it Rev. Mod. Phys.}\ }
\def\etal{{\it et al.}}
\def\ie{{\it i.e. }}
\def\eg{{\it e.g. }}
\def\a{\`a }\def\o{\`o }\def\ii{\`\i{} }
\def\u{\`u  }\def\e{\`e }\def\ke{ch\'e }
\def\psiul{\overline\psi}
\def\pb{\not\!\partial}
\def\f{F(\phi)}
\def\fp{F'(\phi)}
\def\fpp{F''(\phi)}
\def\p{\phi}
\def\v{V(\phi)}
\def\vp{V'(\phi)}
\def\P{\bf\Phi}
\def\l{\cal L}

\title{Deriving the mass of particles  from Extended Theories of Gravity in LHC era}

\author{Salvatore Capozziello\footnote{corresponding author: {\it capozziello@na.infn.it}}}
\affiliation{Dipartimento di Scienze Fisiche, Università di Napoli {}``Federico II'' and INFN Sez. di Napoli, Compl. Univ. di Monte S. Angelo, Edificio G, Via Cinthia, I-80126, Napoli, Italy,}

\author{  Giuseppe Basini}
\affiliation{ CERN,  CH-1211,  Geneva 23, Switzerland and Laboratori Nazionali di Frascati, INFN, Via E. Fermi,  C.P. 13, I-0044 Frascati, Italy,}

\author{Mariafelicia De Laurentis }
\affiliation{Dip.  di Scienze Fisiche, Università di Napoli {}``Federico II'' and INFN Sez. di Napoli, Compl. Univ. di Monte S. Angelo, Edificio G, Via Cinthia, I-80126, Napoli,   Italy.}

\date{\today}

\begin{abstract}
We derive a geometrical approach to produce the mass of particles  that could be suitably tested  at LHC. Starting from a 5D unification scheme, we show that all the  known interactions could be suitably deduced as an induced symmetry breaking of the non-unitary $GL(4)$-group of diffeomorphisms. The deformations inducing such a breaking  act as vector bosons that, depending on the gravitational mass states, can assume the role of interaction bosons like gluons, electroweak bosons or photon. The further gravitational degrees of freedom, emerging from the reduction mechanism in 4D, eliminate the hierarchy problem since generate a cut-off comparable with electroweak one at TeV scales.  In this "economic" scheme, gravity should induce  the other interactions in a non-perturbative way.

\end{abstract}

\maketitle

\section{Introduction}
\label{uno}
 The  {\it Standard Model of Particles}  can be considered a
 successful relativistic quantum field theory both from
particle physics and group theory points of view. Technically, it is a
non-Abelian gauge
theory (a Yang-Mills theory) associated with the tensor product of the internal symmetry groups $%
SU(3)\times SU(2)\times U(1)$, where the $SU(3)$ color symmetry for
 quantum chromodynamics, is treated as exact, whereas the $%
SU(2)\times U(1)$ symmetry, responsible for generating the
electro-weak gauge fields, is considered spontaneously broken.

So far, as we
know, there are four fundamental forces in Nature; namely,
electromagnetic, weak, strong and gravitational
forces. The Standard Model well represents  the first three, but not the
gravitational interaction. On the other hand,  General Relativity (GR) is a geometric theory of the
gravitational field  which is described by the metric tensor $g_{\mu \nu
}$ defined on  pseudo-Riemannian space-times.  The Einstein field equations are nonlinear and have to be satisfied by
the metric tensor. This nonlinearity is
indeed a source of difficulty in quantization of General
Relativity. Since the  Standard Model is a gauge theory where all the fields mediating the
interactions are represented by gauge potentials, the question is
  why the fields mediating the gravitational
interaction are different from those of  the other fundamental forces.
It is reasonable to expect that there may be a gauge theory in
which the gravitational fields stand on the same footing as those
of other fields \cite{oqr}.  As it is well-known, this expectation has prompted a
re-examination of GR from the point of view of gauge theories.

While the gauge groups involved in the Standard Model are all
internal symmetry groups, the gauge groups in GR
must be associated with external space-time symmetries. Therefore,
the gauge theory of gravity cannot be dealt under the standard of the  usual Yang-Mills
theories. It must be one in which gauge objects are not only  gauge
potentials but also tetrads that relate the symmetry group to the
external space-time. For this reason we have to consider a more
complex nonlinear gauge theory where all the interactions should be
dealt under the same standard \cite{unification}. In GR, Einstein took the space-time metric components as the basic
set of variables representing gravity, whereas Ashtekar and collaborators employed the
tetrad fields and the connection forms as the fundamental
variables \cite{rovelli}. We also will consider the tetrads and the connection forms
as the fundamental fields but with the difference that this
approach gives rise to a covariant symplectic formalism capable of
achieving the result of dealing with physical fields under the
same standard \cite{basini,symplectism}.

In order to frame historically  our approach, let us sketch a quick summary of the various attempts where  the Standard Model and GR have been considered under the same comprehensive picture.
In 1956, Utiyama suggested that gravitation may be viewed as a
gauge theory \cite{Utiyama} in analogy to the Yang-Mills
\cite{YangMills} theory (1954). He identified the gauge potential
due to the Lorentz group with the symmetric connection of the Riemann
geometry, and reproduced the  Einstein GR as a gauge
theory of the Lorentz group $SO(3$, $1)$ with the help of tetrad
fields introduced in an \textit{ad hoc} manner. Although the
tetrads were necessary components of the theory (to relate the
Lorentz group) adopted as an internal gauge group to the external
space-time, they were not introduced as gauge fields. In 1961,
Kibble \cite{Kibble}
constructed a gauge theory based on the Poincar\'{e} group $P(3$, $1)=T(3$, $1)\rtimes SO(3$, $1)$ (the symbol $\rtimes $ represents the
semi-direct product) which resulted in the Einstein-Cartan theory
characterized by curvature and torsion.
The translation group $T(3$, $1)$ is considered responsible for
generating the tetrads as gauge fields. Cartan \cite{Cartan}
generalized the Riemann geometry in order  to include torsion in addition to
curvature. The torsion (tensor) arises from an asymmetric
connection.  Sciama \cite{Sciama}, and others (Fikelstein
\cite{Finkelstein}, Hehl \cite{Hehl1, Hehl2}) pointed out that
intrinsic spin may be the source of torsion of the underlying
space-time manifold.

Since the form and the role of  tetrad fields are very different
from those of gauge potentials, it has been thought that even
Kibble's attempt is not satisfactory as a full gauge theory. There
have been a number of gauge theories of gravitation based on a
variety of Lie groups \cite{Hehl1, Hehl2, Mansouri1, Mansouri2,
Chang, Grignani, MAG}. It was argued that a gauge theory of
gravitation corresponding to GR can be constructed
with the translation group alone, in the so-called teleparallel
scheme \cite{teleparallelism}.

Inomata \textit{et al.} \cite{Inomata} proposed that Kibble's
gauge theory could be obtained, in a way closer to the
Yang-Mills approach, by considering the de Sitter group $SO(4$,
$1)$, which is reducible to the Poincar\'{e} group by a
group-contraction. Unlike the Poincar\'{e} group, the de Sitter
group is homogeneous and the associated gauge fields are all of
gauge potential type and by the Wigner-In\"{o}nu group contraction
procedure, one of the five vector potentials reduces to the
tetrad.

It is standard to use the fiber-bundle formulation by which gauge
theories can be constructed on the basis of any Lie group.
Works by Hehl \textit{et al.} \cite{MAG}, on the so-called Metric
Affine Gravity (MAG), adopted
as a gauge group the affine group $A(4$, $\mathbf{%
\mathbb{R}
})=T(4)\rtimes GL(4$, $\mathbf{%
\mathbb{R}
})$, which can be linearly  realized. The tetrad has been identified by
the nonlinearly realized translational part of the affine
connection, on the tangent bundle. In MAG theory, the Lagrangian
is quadratic in both curvature and torsion, in contrast to the
Einstein-Hilbert Lagrangian of  GR which is linear
in the scalar curvature. The theory has the Einstein limit on one
hand and leads to the Newtonian inverse distance potential plus
the linear confinement potential (in the weak field approximation)
on the other hand. In summary, as we have seen, there are many attempts
to formulate gravitation as a gauge theory but currently no theory
has been uniquely accepted as the gauge theory of gravity.

The nonlinear approach to group realizations was originally
introduced by  Coleman,  Wess and  Zumino \cite{CCWZ1, CCWZ2} in
the context of internal symmetry groups (1969). It was later
extended to the case of space-time symmetries by Isham, Salam, and
Strathdee \cite{Isham}
considering the nonlinear action of $GL(4$, $\mathbf{%
\mathbb{R}
})$, modulus the Lorentz subgroup. In 1974, Borisov, Ivanov and
Ogievetsky \cite{BorisovOgievetskii, IvanovOgievetskii},
considered the simultaneous nonlinear realization (NLR) of the
affine and conformal groups. They stated that GR
can be viewed as a consequence of spontaneous breakdown of the
affine symmetry, in  the same way that chiral dynamics, in
quantum chromodynamics, is a result of spontaneous breakdown of
chiral symmetry. In their model, gravitons are considered as
Goldstone bosons associated with the affine symmetry breaking. As we will see below, this approach can be pursued in general.

In 1978, Chang and
Mansouri \cite{ChangMansouri} used the NLR scheme adopting $GL(4$, $\mathbf{%
\mathbb{R}
})$ as the principal group. In 1980, Stelle and West \cite{StelleWest}
investigated the NLR induced by the spontaneous breakdown of $SO(3$, $2)$.
In 1982 Ivanov and Niederle considered nonlinear gauge theories of the
Poincar\'{e}, de Sitter, conformal and special conformal groups \cite{Ivanov1, Ivanov2}.
 In 1983, Ivanenko and Sardanashvily \cite{IvanenkoSardanashvily}
 considered gravity to be a spontaneously broken $%
GL(4 $, $\mathbf{%
\mathbb{R}
})$ gauge theory. The tetrads fields arise, in their formulation,
as a result of the reduction of the structure group of the tangent
bundle from the general linear to Lorentz group. In 1987, Lord and
Goswami \cite{Lord1, Lord2} developed the NLR in the fiber bundle
formalism based on the bundle structure $G\left( G/H\text{,
}H\right) $ as suggested by Ne'eman and Regge \cite{NeemanRegge}.
In this approach, the quotient space $G/H$ is identified with
physical space-time. Most recently, in a series of papers,
Lopez-Pinto,  Julve,  Tiemblo,  Tresguerres and  Mielke discussed
nonlinear gauge theories of gravity on the basis of the
Poincar\'{e}, affine and conformal groups \cite{Julve,
Lopez-Pinto, TresguerresMielke, Tresguerres, TiembloTresguerres1,
TiembloTresguerres2}.

Now, following the prescriptions of GR, the
physical space-time is assumed to be a four-dimensional
differential manifold. In Special Relativity (SR), this manifold is the
Minkwoski flat-space-time $M_{4}$ while, in GR, the
underlying space-time is assumed to be curved in order to describe
the effects of gravitation.

As we said, Utiyama \cite{Utiyama}  proposed that GR can be seen as a gauge theory based on the local
Lorentz group  in the same way that the Yang-Mills gauge theory
 \cite{YangMills} is developed on the basis of the internal
iso-spin gauge group. In this formulation, the Riemannian
connection is the gravitational counterpart of the Yang-Mills
gauge fields. While $SU(2)$, in the Yang-Mills theory, is an
internal symmetry group, the Lorentz symmetry represents the local
nature of space-time rather than internal degrees of freedom. The
Einstein Equivalence Principle, asserted for GR,
requires that the local space-time structure can be identified with
the Minkowski space-time possessing Lorentz symmetry.

In order to relate local Lorentz symmetry to the external
space-time, we need to solder the local space to the external
space. The soldering tools can be the tetrad fields. Utiyama
regarded the tetrads as objects given \textit{a priori} while they
can be dynamically generated \cite{unification} and the space-time
has  necessarily to be endowed with torsion in order to
accommodate spinor fields \cite{classification}. In other words, the gravitational
interaction of spinning particles requires the modification of the
Riemann space-time of GR to be a (non-Riemannian)
curved space-time with torsion. Although Sciama used the tetrad
formalism for his gauge-like handling of gravitation, his theory
fell shortcomings in treating tetrad fields as gauge fields.

Following the Kibble approach \cite{Kibble}, it can be
demonstrated how gravitation can be formulated starting from a
pure gauge viewpoint. In particular,  gravity can be seen as  a gauge theory
which can be obtained starting from some local invariance, e.g.
the local Poincar\'{e} symmetry, leading to a suitable unification
scheme \cite{unification}. This  dynamical structure
can be based on a
nonlinear realization of the local conformal-affine  group of
symmetry transformations  \cite{poincare}.

Here, we start from a General  Invariance Principle, as requested in the so called Open Quantum Relativity (OQR)  \cite{oqr,conservation} and consider first the
Global Poincar\'{e} Invariance and then the Local Poincar\'{e}
Invariance. This approach leads to construct a given theory of
gravity as a gauge theory. Such a viewpoint, if considered in
detail, can avoid many shortcomings and could be useful to
formulate self-consistent schemes for quantum gravity and then the
unification of all interactions \cite{unification}.

In particular, the idea of an unification theory, capable of describing all the
fundamental interactions of physics under the same standard, has
been one of the main issues of  modern physics, starting from
the early efforts of Einstein, Weyl, Kaluza and Klein
\cite{kaluza} until the most recent  approaches
\cite{ross}. Nevertheless, the large number of ideas, up to now
proposed,  which we classify as {\it unified theories}, results
unsuccessful due to several reasons: the technical difficulties
connected with the lack of a unitary mathematical description of
all the interactions; the huge number of  parameters introduced
to "build up"  the unified theory and the fact that most of them
cannot be observed neither at laboratory nor at astrophysical (or
cosmological) conditions \cite{morselli};  the very wide (and several times
questionable since not-testable) number of extra-dimensions
requested by several approaches. Due to this situation, it seems
that unification is a useful (and aesthetic) paradigm, but far to
be achieved, if the trend is continuing  to try to unify
interactions (\ie to make something simple) by adding and adding
ingredients: new particles and new parameters (e.g. dark matter forest).
 A  different  approach could be to consider the very essential
physical quantities and try to achieve unification without  any {\it
ad hoc} new ingredients. This approach
can be pursued starting from  straightforward considerations which
lead to reconsider modern physics under a sort of economic issue: let us try to unifying forces approaching new schemes but without adding new parameters
\footnote{Following Occam's Razor prescription: {\it Entia non sunt multiplicanda praeter necessitatem.}}.
 A prominent role
deserves the conservation laws and the fact that each of them
brings out the existence of a symmetry \cite{unification}.

As a general remark, the Noether Theorem  states that, for every
conservation law of Nature, a symmetry {\it must} exist. This
leads to a fundamental result also from a mathematical point of
view since the presence of symmetries technically reduces
dynamics (\ie gives rise to first integrals of motion) and, in
several cases, allows to get the general solution. With these
considerations in mind, we can try to change our point of view
and investigate what will be the consequences of the absolute
validity of conservation laws without introducing any arbitrary symmetry breaking.

In order to see what happens as soon as we ask for the absolute
validity of conservation laws, we could take into account the
Bianchi identities. Such geometrical  identities work in every
covariant field theory (\eg Electromagnetism or GR) and can be read as  equations of motion also in a fiber bundle approach \cite{bundle}.
 We want to
show that, the absolute validity of conservation laws,
intrinsically contains symmetric dynamics; moreover, reducing
dynamics from 5D to 4D, it gives rise to the physical quantities
characterizing particles as the mass.

The {\it minimal} ingredient which we require to achieve  these
results is the fact that a 5-dimensional, singularity free space,
where conservation laws are always and absolutely {\it
conserved}, has to be defined. Specifically, in such
a space, Bianchi identities are asked to be always valid and,
moreover, the process of reduction to 4D-space {\it generates}  the mass spectra of particles. In this sense, a dynamical unification scheme
will be achieved where a fifth dimension has the  physical
meaning of inducing the mass of particles by deformations of space-time. In other words, we will show that deformations can be parameterized as "effective" scalar fields in a $GL(4)$-group of diffeomorphisms.
In this sense, we do not need any spontaneous symmetry breaking but just a self-consistent way to classify deformations as "gauge bosons".
The layout of the paper is the following.
In Sec.II, we discuss in detail the conformal-affine structure of gravitational field showing that the nonlinear realization of a group provides a way to determine the transformation properties of fields defined on a given quotient space $G/H$. In other words, we show that gravitational field can be realized in  many equivalent ways  and we will use this feature to show that gravitational massive states are possible.
Sec. III is devoted to the group structure. We show that  the 4D-group of diffeomorphisms can be embedded in that in 5D. Furthermore, it is straightforward to show that $GL(4)$ contains all the generators of the Standard Model plus generators of the gravitational field.  The space-time deformations as elements of $GL(4)$-group are discussed in Sec.IV. The main result of this section is that deformations can be dealt as effective geometric scalar fields. In Sec.V, the 5D space-time structure and the reduction to 4D-dynamics is discussed. Such a reduction mechanism gives rise to  effective theories of gravity
(Extended Theories of Gravity \cite{odishap, odirev,capfra,book,odino}) where higher-order terms in curvature invariants or nonminimal couplings are naturally achieved.  In Sec.VI, we discuss that these effective theories can be conformally related and the only singular theory (with null Hessian determinant) is GR. The straightforward consequence of such a result is that  gravitational massive modes can be always generated. Sec.VII is devoted to the discussion of the mass generation while, in  Sec.VIII, we derive massless and massive gravitational modes related to  Extended Theories of Gravity. An interesting byproduct is the fact that 6 polarization states emerge and this result is perfectly in agreement with the fundamental  Riemann theorem stating that in a $N$-dimensional space, $N(N-1)/2$ gravitational degrees of freedom are allowed. Sec.IX is devoted to the specific issue that massive gravitons could have observable  effects between GeV-TeV scales and induce a symmetry breaking through a sort of regularization mechanism. Conclusions are drawn in Sec.X.

\section{ The Conformal-Affine  Structure of Gravitational Field}
\label{due}

\subsection{Generalities on fiber bundle formalism}
\label{due.uno}
In this section, we shall take into account the fiber bundle formalism of gravitational field showing that it naturally exhibit a conformal-affine structure.
This feature, in some sense, allow to compare all the theories of gravity, based on diffeomorphism invariance, under the same standard.

Before considering in details the conformal-affine structure of
gravitational theories, let us briefly review the standard bundle
approach to gauge theories. First, let us show that a usual gauge potential
$\Omega $ is the pullback of 1-form connection  $\omega $ by the
local sections of the bundle. Then, the transformation laws of
the $\omega $ and $\Omega $ under the action of the structure
group $G$ are deduced.

Modern formulations of gauge field theories are geometrically
expressible  in the language of principal fiber bundles. A fiber bundle is a structure $%
\left\langle \mathbb{P}\text{, }M\text{, }\pi \text{; }\mathbb{F}%
\right\rangle $ where $\mathbb{P}$ (the total bundle space) and $M$ (the
base space) are smooth manifolds, $\mathbb{F}$ is the fiber space and the
surjection $\pi $\ (a canonical projection) is a smooth map of $\mathbb{P}$
onto $M$,%
\begin{equation}
\pi :\mathbb{P}\rightarrow M\text{.}
\end{equation}%
The inverse image $\pi ^{-1}$ is diffeomorphic to $\mathbb{F}$%
\begin{equation}
\pi ^{-1}\left( x\right) \equiv \mathbb{F}_{x}\approx \mathbb{F}\text{,}
\end{equation}%
and it is called the fiber at $x\in M$. The partitioning $\bigcup\nolimits_{x}%
\pi ^{-1}\left( x\right) =\mathbb{P}$ is referred to as the
{\it fibration}. Note that a smooth map is one whose coordinatization is
$C^{\infty }$ differentiable; a smooth manifold is a space that
can be covered with coordinate patches in such a manner that, a
change from one patch to any overlapping patch is smooth
\cite{Schwarz}. Fiber bundles that admit a  decomposition as a direct product,
locally looking like $\mathbb{%
P\approx }M\times \mathbb{F}$, are called {\it trivial}. Given a set of
open coverings $\left\{ \mathcal{U}_{i}\right\} $ of $M$ with
$x\in \left\{
\mathcal{U}_{i}\right\} \subset M$ satisfying $\bigcup\nolimits_{\alpha }%
\mathcal{U}_{\alpha }=M$, the diffeomorphism map is given by%
\begin{equation}
\chi _{i}:\mathcal{U}_{i}\times _{M}G\rightarrow \pi ^{-1}(\mathcal{U}%
_{i})\in \mathbb{P}\text{,}  \label{diff}
\end{equation}%
($\times _{M}$ represents the fiber product of elements defined over the space $%
M $) such that $\pi \left( \chi _{i}\left( x\text{, }g\right)
\right) =x$ and $\chi _{i}\left( x\text{, }g\right) =\chi
_{i}\left( x\text{, }\left( id\right) _{G}\right) g=\chi
_{i}\left( x\right) g\ \forall x\in \left\{
\mathcal{U}_{i}\right\} $ and $g\in G$. Here, $\left( id\right)
_{G}$ represents the identity element of the group $G$. In order
to obtain the global bundle structure, the local charts $\chi
_{i}$ must be glued together continuously. Consider two patches
$\mathcal{U}_{n}$ and $\mathcal{U}_{m}$ with a non-empty
intersection $\mathcal{U}_{n}\cap \mathcal{U}_{m}\neq \emptyset $.
Let $\rho _{nm}$ be the restriction of $\chi _{n}^{-1}$ to $\pi
^{-1}(\mathcal{U}_{n}\cap \mathcal{U}_{m})$ defined by $\rho _{nm}:\pi ^{-1}(%
\mathcal{U}_{n}\cap \mathcal{U}_{m})\rightarrow (\mathcal{U}_{n}\cap
\mathcal{U}_{m})\times _{M}G_{n}$. Similarly let $\rho _{mn}:\pi ^{-1}(%
\mathcal{U}_{m}\cap \mathcal{U}_{n})\rightarrow (\mathcal{U}_{m}\cap
\mathcal{U}_{n})\times _{M}G_{m}$ be the restriction of $\chi _{m}^{-1}$ to $%
\pi ^{-1}(\mathcal{U}_{n}\cap \mathcal{U}_{m})$. The composite
diffeomorphism $\Lambda _{nm}\in G$%
\begin{equation}
\Lambda _{mn}:(\mathcal{U}_{n}\cap \mathcal{U}_{m})\times G_{n}\rightarrow (%
\mathcal{U}_{m}\cap \mathcal{U}_{n})\times _{M}G_{m}\text{,}
\end{equation}%
defined as%
\begin{equation}
\Lambda _{ij}\left( x\right) \equiv \rho _{ji}\circ \rho _{ij}^{-1}=\chi _{i%
\text{, }x}\circ \chi _{j\text{, }x}^{-1}:\mathbb{F}\rightarrow \mathbb{F}\,,
\end{equation}%
constitutes the transition function between bundle charts $\rho _{nm}$ and $%
\rho _{mn}$ ($\circ $ represents the group composition operation) where the
diffeomorphism $\chi _{i\text{, }x}:\mathbb{F}\rightarrow \mathbb{F}_{x}$ is
written as $\chi _{i\text{, }x}(g):=\chi _{i}\left( x\text{, }g\right) $ and
satisfies $\chi _{j}\left( x\text{, }g\right) =\chi _{i}\left( x\text{, }%
\Lambda _{ij}\left( x\right) g\right) $. The transition functions
$\left\{ \Lambda _{ij}\right\} $ can be interpreted as passive
gauge transformations. They satisfy some consistency conditions,
{\it i.e.} the identity $\Lambda _{ii}\left( x\right) $, the inverse
$\Lambda
_{ij}\left( x\right) =\Lambda _{ji}^{-1}\left( x\right) $ and the cocycle $%
\Lambda _{ij}\left( x\right) \Lambda _{jk}\left( x\right) =\Lambda
_{ik}\left( x\right) $. For trivial bundles, the
transition function reduces to%
\begin{equation}
\Lambda _{ij}\left( x\right) =g_{i}^{-1}g_{j}\text{,}  \label{transition}
\end{equation}%
where $g_{i}:\mathbb{F}\rightarrow \mathbb{F}$ is defined by $g_{i}:=\chi _{i%
\text{, }x}^{-1}\circ \widetilde{\chi }_{i\text{, }x}$, provided
the local
trivializations $\left\{ \chi _{i}\right\} $ and $\left\{ \widetilde{\chi }%
_{i}\right\} $ it gives rise to the same fiber bundle.

A section is defined as a smooth map%
\begin{equation}
s:M\rightarrow \mathbb{P}\text{,}
\end{equation}%
such that $s(x)\in \pi ^{-1}\left( x\right) =\mathbb{F}_{x}$ $\forall x\in M$
and satisfies%
\begin{equation}
\pi \circ s=\left( id\right) _{M}\text{,}
\end{equation}%
where $\left( id\right) _{M}$ is the identity\ element of $M$. It assigns to
each point $x\in M$ a point in the fiber over $x$. Trivial bundles admit
global sections.

A bundle is a principal fiber bundle $\left\langle \mathbb{P}\text{, }%
\mathbb{P}/G\text{, }G\text{, }\pi \right\rangle $ provided that
the Lie group $G$ acts freely ({\it i.e.} if $pg=p$ then $g=\left(
id\right) _{G}$) on $\mathbb{P}$ to the right $R_{g}p=pg$, $p\in
\mathbb{P}$, preserves fibers on $\mathbb{P}$
($R_{g}:\mathbb{P}\rightarrow \mathbb{P}$), and finally is
transitive on fibers. Furthermore, there must exist local
trivializations compatible with the $G$
action. Hence, $\pi ^{-1}(\mathcal{U}_{i})$ is homeomorphic to $\mathcal{U}%
_{i}\times _{M}G$ and the fibers of $\mathbb{P}$ are diffeomorphic to $G$.
The trivialization or inverse diffeomorphism map is given by%
\begin{equation}
\chi _{i}^{-1}:\pi ^{-1}(\mathcal{U}_{i})\rightarrow \mathcal{U}_{i}\times
_{M}G\,  \label{trivial}
\end{equation}%
such that $\chi ^{-1}(p)=\left( \pi (p)\text{, }\phi (p)\right) \in
\mathcal{U}_{i}\times _{M}G$, $p\in \pi ^{-1}(\mathcal{U}_{i})\subset
\mathbb{P}$, where we see from the above definition that $\phi $ is a
local mapping of $\pi ^{-1}(\mathcal{U}_{i})$ into $G$ satisfying $\phi
(L_{g}p)$ $=\phi (p)g$ for any $p\in \pi ^{-1}(\mathcal{U})$ and any $%
g\in G$. Let us observe that the elements of $\mathbb{P}$ which
are projected onto the same $x\in \left\{ \mathcal{U}_{i}\right\}
$ are transformed into one another by the elements of $G$. In
other words, the fibers of $\mathbb{P}$ are the orbits of $G$ and
at the same time, they are the set of elements which are projected
onto the same $x\in \mathcal{U}\subset M$. This observation
motivates calling the action of the group {\it vertical}  and the base
manifold {\it horizontal}. The diffeomorphism map $\chi _{i}$ is called
the local gauge since $\chi _{i}^{-1}$ maps $\pi
^{-1}(\mathcal{U}_{i})$\ onto the direct (Cartesian) product
$\mathcal{U}_{i}\times _{M}G$. The action $L_{g}$ of the structure
group $G$ on $\mathbb{P}$ defines an isomorphism of the Lie
algebra $\mathfrak{g}$ of $G$ onto the Lie algebra of vertical
vector fields on $\mathbb{P}$, tangent to the fiber at each $p\in
\mathbb{P}$ called
fundamental vector fields%
\begin{equation}
\lambda _{g}:T_{p}\left( \mathbb{P}\right) \rightarrow T_{gp}(\mathbb{P}%
)=T_{\pi (p)}\left( \mathbb{P}\right) \text{,}
\end{equation}%
where $T_{p}\left( \mathbb{P}\right) $ is the space of tangents at $p$, {\it i.e.}
$T_{p}\left( \mathbb{P}\right) \in T\left( \mathbb{P}\right) $. The map $%
\lambda $ is a linear isomorphism for every $p\in \mathbb{P}$ and is
invariant with respect to the action of $G$, that is, $\lambda _{g}:\left(
\lambda _{g\ast }T_{p}\left( \mathbb{P}\right) \right) \rightarrow
T_{gp}\left( \mathbb{P}\right) $, where $\lambda _{g\ast }$ is the
differential push forward map induced by $\lambda _{g}$ defined by $\lambda
_{g\ast }:T_{p}\left( \mathbb{P}\right) \rightarrow T_{gp}\left( \mathbb{P}%
\right) $.

Since the principal bundle $\mathbb{P}\left( M\text{, }G\right) $ is a
differentiable manifold, we can define tangent $T\left( \mathbb{P}\right) $
and cotangent $T^{\ast }\left( \mathbb{P}\right) $ bundles. The tangent
space $T_{p}\left( \mathbb{P}\right) $\ defined at each point $p\in \mathbb{P%
}$ may be decomposed into a vertical $V_{p}\left( \mathbb{P}\right) $ and
horizontal $H_{p}\left( \mathbb{P}\right) $ subspace as $T_{p}\left( \mathbb{%
P}\right) :=V_{p}\left( \mathbb{P}\right) \oplus H_{p}\left( \mathbb{P}%
\right) $ (where $\oplus $ represents the direct sum). The space $%
V_{p}\left( \mathbb{P}\right) $ is a subspace of $T_{p}\left( \mathbb{P}%
\right) $ consisting of all tangent vectors to the fiber passing through $%
p\in \mathbb{P}$, and $H_{p}\left( \mathbb{P}\right) $\ is the subspace
complementary to $V_{p}\left( \mathbb{P}\right) $ at $p$. The vertical
subspace $V_{p}\left( \mathbb{P}\right) :=\left\{ X\in T\left( \mathbb{P}%
\right) |\pi \left( X\right) \in \mathcal{U}_{i}\subset M\right\}
$ is uniquely determined by the structure of $\mathbb{P}$, whereas
the horizontal subspace $H_{p}\left( \mathbb{P}\right) $ cannot be
uniquely specified. This result is very important because it makes
possible to fix the Cauchy conditions on the dynamics. Thus we require the following condition: when
$p$ transforms as $p\rightarrow
p^{\prime }=pg$, $H_{p}\left( \mathbb{P}\right) $ transforms as \cite%
{Nakahara},%
\begin{equation}
R_{g\ast }H_{p}\left( \mathbb{P}\right) \rightarrow H_{p^{\prime }}\left(
\mathbb{P}\right) =R_{g}H_{p}\left( \mathbb{P}\right) =H_{pg}\left( \mathbb{P%
}\right) .
\end{equation}%
Let the local coordinates of $\mathbb{P}\left( M\text{, }G\right) $ be $%
p=\left( x\text{, }g\right) $ where $x\in M$ and $g\in G$. Let $\mathbf{G}%
_{A}$ denote the generators of the Lie algebra $\mathfrak{g}$ corresponding
to group $G$ satisfying the commutators $\left[ \mathbf{G}_{A}\text{, }%
\mathbf{G}_{B}\right] =f_{AB}^{\text{ \ \ \ }C}\mathbf{G}_{C}$, where $%
f_{AB}^{\text{ \ \ \ }C}$ are the structure constants of $G$. Let $\Omega $
be a connection form defined by $\Omega ^{A}:=\Omega _{i}^{A}dx^{i}\in
\mathfrak{g}$. Let $\omega $ be a connection 1-form defined by%
\begin{equation}
\omega :=\widetilde{g}^{-1}\pi _{\mathbb{P}M}^{\ast }\Omega \widetilde{g}+%
\widetilde{g}^{-1}d\widetilde{g}\,,
\end{equation}%
($\ast $ represents the differential pullback map) belonging to $\mathfrak{g}%
\otimes T_{p}^{\ast }\left( \mathbb{P}\right) $ where $T_{p}^{\ast
}\left( \mathbb{P}\right) $ is the dual space to $T_{p}\left(
\mathbb{P}\right) $.
In such a case, the differential pullback map, applied to a test function $\phi $ and $p$%
-forms $\alpha $ and $\beta $, satisfies $f^{\ast }\phi =\phi \circ f$, $%
\left( g\circ f\right) ^{\ast }=f^{\ast }g^{\ast }$ and$\ f^{\ast }\left(
\alpha \wedge \beta \right) =f^{\ast }\alpha \wedge f^{\ast }\beta $. If $G$
is represented by a $d$-dimensional $d\times d$ matrix, then $\mathbf{G}_{A}=%
\left[ \mathbf{G}_{\alpha \beta }\right] $,\ $\widetilde{g}=\left[
\widetilde{g}^{\alpha \beta }\right] $, where $\alpha $, $\beta =1$, $2$, $3$%
,$...d$. Thus, $\omega $ assumes the form%
\begin{equation}
\omega _{\alpha }^{\text{ }\beta }=\left( \widetilde{g}^{-1}\right) _{\alpha
\gamma }d\widetilde{g}^{\gamma \beta }+\left( \widetilde{g}^{-1}\right)
_{\rho \gamma }\pi _{\mathbb{P}M}^{\ast }\Omega _{\text{ }\sigma i}^{\rho }%
\mathbf{G}_{\alpha }^{\text{ }\gamma }\widetilde{g}^{\sigma \beta }\otimes
dx^{i}\text{.}
\end{equation}

If $M$ is $n$-dimensional, the tangent space $T_{p}\left( \mathbb{P}\right) $
is $\left( n+d\right) $-dimensional. Since the vertical subspace $%
V_{p}\left( \mathbb{P}\right) $ is tangential to the fiber $G$, it is $d$%
-dimensional. Accordingly, $H_{p}\left( \mathbb{P}\right) $ is $n$%
-dimensional. The basis of $V_{p}\left( \mathbb{P}\right) $ can be taken to
be $\partial _{\alpha \beta }:=\frac{\partial }{\partial g^{\alpha \beta }}$%
. Now, let the basis of $H_{p}\left( \mathbb{P}\right) $ be denoted by%
\begin{equation}
E_{i}:=\partial _{i}+\Gamma _{i}^{\alpha \beta }\partial _{\alpha \beta }%
\text{,}\ i=1\text{, }2\text{, }3,..n\ \text{and}\ \alpha \text{, }\beta =1%
\text{, }2\text{, }3,..d
\end{equation}%
where $\partial _{i}=\frac{\partial }{\partial x^{i}}$. The
connection 1-form $\omega $ projects $T_{p}\left(
\mathbb{P}\right) $ onto $V_{p}\left( \mathbb{P}\right) $. In
order for $X\in T_{p}\left( \mathbb{P}\right) $ to belong to
$H_{p}\left( \mathbb{P}\right) $, it has to be $X\in H_{p}\left(
\mathbb{P}\right) $, $\omega _{p}\left( X\right) =\left\langle
\omega \left( p\right) |X\right\rangle =0$. In other words,
\begin{equation}
H_{p}\left( \mathbb{P}\right) :=\left\{ X\in T_{p}\left( \mathbb{P}\right)
|\omega _{p}\left( X\right) =0\right\} \text{,}
\end{equation}%
from which $\Omega _{i}^{\alpha \beta }$ can be determined. The inner
product appearing in $\omega _{p}\left( X\right) =\left\langle \omega \left(
p\right) |X\right\rangle =0$ is a map $\left\langle \cdot |\cdot
\right\rangle :T_{p}^{\ast }\left( \mathbb{P}\right) \times T_{p}\left(
\mathbb{P}\right) \rightarrow
\mathbb{R}
$ defined by $\left\langle W|V\right\rangle =W_{\mu }V^{\nu }\left\langle
dx^{\mu }|\frac{\partial }{\partial x^{\nu }}\right\rangle =W_{\mu }V^{\nu
}\delta _{\nu }^{\mu }$, where the 1-form $W$ and vector $V$ are given by $%
W=W_{\mu }dx^{\mu }$ and $V=V^{\mu }\frac{\partial }{\partial x^{\nu }}$.
Observe also that, $\left\langle dg^{\alpha \beta }|\partial _{\rho \sigma
}\right\rangle =\delta _{\rho }^{\alpha }\delta _{\sigma }^{\beta }$.

We parameterize an arbitrary group element $\widetilde{g}_{\lambda }$ as $%
\widetilde{g}\left( \lambda \right) =e^{\lambda ^{A}\mathbf{G}%
_{A}}=e^{\lambda \cdot \mathbf{G}}$,\ $A=1$,$..dim\left( \mathfrak{g}\right)
$. The right action $R_{\widetilde{g}\left( \lambda \right) }=R_{\exp \left(
\lambda \cdot G\right) }$ on $p\in \mathbb{P}$, {\it i.e.} $R_{\exp \left( \lambda
\cdot \mathbf{G}\right) }p=p\exp \left( \lambda \cdot \mathbf{G}\right) $,
defines a curve through $p$ in $\mathbb{P}$. Define a vector $G^{\#}\in
T_{p}\left( \mathbb{P}\right) $ by \cite{Nakahara}%
\begin{equation}
G^{\#}f\left( p\right) :=\frac{d}{dt}f\left( p\exp \left( \lambda \cdot
\mathbf{G}\right) \right) |_{\lambda =0}\,,
\end{equation}%
where $f:\mathbb{P}\rightarrow
\mathbb{R}
$ is an arbitrary smooth function. Since the vector $G^{\#}$ is tangent to $%
\mathbb{P}$ at $p$, $G^{\#}\in V_{p}\left( \mathbb{P}\right) $,
the components of the vector $G^{\#}$ are the fundamental vector
fields at $p$ which constitute $V(\mathbb{P})$. We have to stress
that the components of $G^{\#}$ may also be
viewed as a basis element of the Lie algebra $\mathfrak{g}$. Given $%
G^{\#}\in V_{p}\left( \mathbb{P}\right) $, $\mathbf{G}\in \mathfrak{g}$,%
\begin{eqnarray}
\omega _{p}\left( G^{\#}\right) &=&\left\langle \omega \left( p\right)
|G^{\#}\right\rangle =\widetilde{g}^{-1}d\widetilde{g}\left( G^{\#}\right) +%
\widetilde{g}^{-1}\pi _{\mathbb{P}M}^{\ast }\Omega \widetilde{g}\left(
G^{\#}\right)  \notag \\
&=&\widetilde{g}_{p}^{-1}\widetilde{g}_{p}\frac{d}{d\lambda }\left( \exp
\left( \lambda \cdot \mathbf{G}\right) \right) |_{\lambda =0}\text{,}
\end{eqnarray}%
where use was made of $\pi _{\mathbb{P}M\ast }G^{\#}=0$. Hence, $\omega
_{p}\left( G^{\#}\right) =\mathbf{G}$. An arbitrary vector $X\in H_{p}\left(
\mathbb{P}\right) $ may be expanded in a basis spanning $H_{p}\left( \mathbb{%
P}\right) $ as $X:=\beta ^{i}E_{i}$. By direct computation, one can show%
\begin{eqnarray}
\left\langle \omega _{\alpha }^{\text{ }\beta }|X\right\rangle& =&\left(
\widetilde{g}^{-1}\right) _{\alpha \gamma }\beta ^{i}\Gamma _{i}^{\gamma
\beta }+\nonumber\\+&&\left( \widetilde{g}^{-1}\right) _{\alpha \gamma }\pi _{\mathbb{P}%
M}^{\ast }\Omega _{\text{ }\sigma i}^{\rho }\beta ^{i}\mathbf{G}_{\rho
}^{\gamma }\widetilde{g}^{\sigma \beta }=0\text{, }\forall \beta ^{i}
\label{conn-vect}
\end{eqnarray}%
Equation (\ref{conn-vect}) yields%
\begin{equation}
\left( \widetilde{g}^{-1}\right) _{\alpha \gamma }\Gamma _{i}^{\gamma \beta
}+\left( \widetilde{g}^{-1}\right) _{\alpha \gamma }\pi _{\mathbb{P}M}^{\ast
}\Omega _{\text{ }\sigma i}^{\rho }\mathbf{G}_{\rho }^{\gamma }\widetilde{g}%
^{\sigma \beta }=0\text{,}
\end{equation}%
from which we obtain%
\begin{equation}
\Gamma _{i}^{\gamma \beta }=-\pi _{\mathbb{P}M}^{\ast }\Omega _{\text{ }%
\sigma i}^{\rho }\mathbf{G}_{\rho }^{\gamma }\widetilde{g}^{\sigma \beta }%
\text{.}
\end{equation}%
In this manner, the horizontal component is completely determined. An
arbitrary tangent vector $\mathfrak{X}\in T_{p}\left( \mathbb{P}\right) $
defined at $p\in \mathbb{P}$ takes the form%
\begin{equation}
\mathfrak{X}=A^{\alpha \beta }\partial _{\alpha \beta }+B^{i}\left( \partial
_{i}-\pi _{\mathbb{P}M}^{\ast }\Omega _{\text{ }\sigma i}^{\rho }\mathbf{G}%
_{\rho }^{\alpha }\widetilde{g}^{\sigma \beta }\partial _{\alpha \beta
}\right) ,
\end{equation}%
where $A^{\alpha \beta }$ and $B^{i}$ are constants. The vector field $%
\mathfrak{X}$ is comprised of horizontal $\mathfrak{X}^{H}:=B^{i}\left(
\partial _{i}-\pi _{\mathbb{P}M}^{\ast }\Omega _{\text{ }\sigma i}^{\rho }%
\mathbf{G}_{\rho }^{\alpha }\widetilde{g}^{\sigma \beta }\partial _{\alpha
\beta }\right) \in H\left( \mathbb{P}\right) $ and vertical $\mathfrak{X}%
^{V}:=A^{\alpha \beta }\partial _{\alpha \beta }\in V\left( \mathbb{P}%
\right) $ components.

Let $\mathfrak{X}\in T_{p}\left( \mathbb{P}\right) $ and $g\in \mathbf{G}$,\
then%
\begin{eqnarray}
R_{g}^{\ast }\omega \left( \mathfrak{X}\right)& =&\omega \left( R_{g\ast }%
\mathfrak{X}\right) =\nonumber\\&&=\widetilde{g}_{pg}^{-1}\Omega \left( R_{g\ast }%
\mathfrak{X}\right) \widetilde{g}_{pg}+\widetilde{g}_{pg}^{-1}d\widetilde{g}%
_{pg}\left( R_{g\ast }\mathfrak{X}\right) \text{,}\nonumber\\  \label{Rightw}
\end{eqnarray}%
Observing that $\widetilde{g}_{pg}=\widetilde{g}_{p}g$ and $\widetilde{g}%
_{gp}^{-1}=g^{-1}\widetilde{g}_{p}^{-1}$ the first term on the RHS of (\ref%
{Rightw}) reduces to $\widetilde{g}_{pg}^{-1}\Omega \left( R_{g\ast }%
\mathfrak{X}\right) \widetilde{g}_{pg}=g^{-1}\widetilde{g}_{p}^{-1}\Omega
\left( R_{g\ast }\mathfrak{X}\right) \widetilde{g}_{p}g$ while the second
term gives $\widetilde{g}_{pg}^{-1}d\widetilde{g}_{pg}\left( R_{g\ast }%
\mathfrak{X}\right) =g^{-1}\widetilde{g}_{p}^{-1}d\left( R_{g\ast }\mathfrak{%
X}\right) \widetilde{g}_{p}g$. We therefore conclude%
\begin{equation}
R_{g}^{\ast }\omega _{\lambda }=ad_{g^{-1}}\omega _{\lambda }\text{,}
\end{equation}%
where the adjoint map $ad$ is defined by%
\begin{eqnarray}
&& ad_{g}Y:=L_{g\ast }\circ R_{g^{-1}\ast }\circ Y=gYg^{-1}\text{, \ }\nonumber\\%
&& ad_{g^{-1}}Y:=g^{-1}Yg\text{.}
\end{eqnarray}

The potential $\Omega ^{A}$ can be obtained from $\omega $ as $\Omega
^{A}=s^{\ast }\omega $. To demonstrate this, let $Y\in T_{p}\left( M\right) $
and $\widetilde{g}$ be specified by the inverse diffeomorphism or
trivialization map (\ref{trivial}) with $\chi _{\lambda }^{-1}\left(
p\right) =\left( x\text{, }\widetilde{g}_{\lambda }\right) $ for $p\left(
x\right) =s_{\lambda }\left( x\right) \cdot \widetilde{g}_{\lambda }$. We
find $s_{i}^{\ast }\omega \left( Y\right) =\widetilde{g}^{-1}\Omega \left(
\pi _{\ast }s_{i\ast }Y\right) \widetilde{g}+\widetilde{g}^{-1}d\widetilde{g}%
\left( s_{i\ast }Y\right) $, where we  have used $s_{i\ast }Y\in
T_{s_{i}}\left( \mathbb{P}\right) $, $\pi _{\ast }s_{i\ast
}=\left( id\right) _{T_{p}\left( M\right) }$ and
$\widetilde{g}=\left( id\right) _{G}$ at $s_{i}$ implying
$\widetilde{g}^{-1}d\widetilde{g}\left( s_{i\ast
}Y\right) =0$ \cite{Nakahara}. Hence,%
\begin{equation}
s_{i}^{\ast }\omega \left( Y\right) =\Omega \left( Y\right) \text{.}
\end{equation}

To determine the gauge transformation of the connection 1-form
$\omega$, we use the fact that $R_{\widetilde{g}\ast
}X=X\widetilde{g}$ for $X\in T_{p}\left( M\right) $ and the
transition functions $\widetilde{g}_{nm}\in G$ defined between
neighboring bundle charts (\ref{transition}). By direct
computation we get%
\begin{eqnarray}
c_{j\ast }X &=&\frac{d}{dt}c_{j}\left( \lambda \left( t\right) \right)
|_{t=0}=\frac{d}{dt}\left[ c_{i}\left( \lambda \left( t\right) \right) \cdot
\widetilde{g}_{ij}\right] |_{t=0}  \notag \\
&=&R_{\widetilde{g}_{ij}\ast }c_{i}^{\ast }\left( X\right) +\left(
\widetilde{g}_{ji}^{-1}\left( x\right) d\widetilde{g}_{ij}\left(
X\right) \right) ^{\#}\text{,}
\end{eqnarray}%
where $\lambda \left( t\right) $ is a curve in $M$ with boundary values $%
\lambda \left( 0\right) =m$ and $\frac{d}{dt}\lambda \left( t\right)
|_{t=0}=X$. Thus, we obtain the useful result%
\begin{equation}
c_{\ast }X=R_{\widetilde{g}\ast }\left( c_{\ast }X\right) +\left( \widetilde{%
g}^{-1}d\widetilde{g}\left( X\right) \right) ^{\#}\text{.}  \label{trans1}
\end{equation}%
Applying $\omega $ to Eq.(\ref{trans1}), we get%
\begin{equation}
\omega \left( c_{\ast }X\right) =c^{\ast }\omega \left( X\right) =ad_{%
\widetilde{g}^{-1}}c^{\ast }\omega \left( X\right) +\widetilde{g}^{-1}d%
\widetilde{g}\left( X\right) \text{, }\forall X\text{.}
\end{equation}%
Hence, the gauge transformation of the local gauge potential $\Omega $ reads,%
\begin{equation}
\Omega \rightarrow \Omega ^{\prime }=ad_{\widetilde{g}^{-1}}\left( d+\Omega
\right) =\widetilde{g}^{-1}\left( d+\Omega \right) \widetilde{g}\text{.}
\label{Lconn-trans}
\end{equation}%
Since $\Omega =c^{\ast }\omega $ we obtain, from
Eq.(\ref{Lconn-trans}), the
gauge transformation law of $\omega $%
\begin{equation}
\omega \rightarrow \omega ^{\prime }=\widetilde{g}^{-1}\left( d+\omega
\right) \widetilde{g}\text{.}
\end{equation}
We have now all the ingredients to investigate the bundle structure of the gravitation field.

\subsection{The Bundle Structure for Gravitation}
\label{due.due}

Let us recall the definition of gauge transformations in the context of
ordinary fiber bundles. Given a principal fiber bundle $\mathbb{P}(M$, $G$; $%
\pi )$ with base space $M$ and standard $G$-diffeomorphic fiber, gauge
transformations are characterized by bundle isomorphisms  $%
\lambda :\mathbb{P}\rightarrow \mathbb{P}$ exhausting all diffeomorphisms $%
\lambda _{M}$ on $M$ \cite{Giachetta}. This mapping is called an
automorphism of $\mathbb{P}$ provided it is equivariant with
respect to the action of $G$. This amounts to restrict the action
$\lambda $ of $G$ along local fibers leaving the base space
unaffected. Indeed, with regard to gauge theories of internal
symmetry groups, a gauge transformation is a fiber preserving
bundle automorphism, {\it i.e.} diffeomorphisms $\lambda $ with $\lambda
_{M}=\left( id\right) _{M}$. The automorphisms $\lambda $ form a
group called the automorphism group $Aut_{\mathbb{P}}$ of
$\mathbb{P}$. The gauge transformations form a subgroup of
$Aut_{\mathbb{P}}$ called the gauge group $G\left(
Aut_{\mathbb{P}}\right) $ (or $G$ in short) of $\mathbb{P}$.

The map $\lambda $ is required to satisfy two conditions, namely its
commutability with the right action of $G$ $[$the equivariance condition $%
\lambda \left( R_{g}(p)\right) =\lambda \left( pg\right) =\lambda \left(
p\right) g]$%
\begin{equation}
\lambda \circ R_{g}(p)=R_{g}(p)\circ \lambda \text{, \ }p\in \mathbb{P}\text{%
, }g\in G
\end{equation}%
according to which fibers are mapped into fibers, and the verticality
condition%
\begin{equation}
\pi \circ \lambda \left( u\right) =\pi \left( u\right) \text{,}
\end{equation}%
where $u$ and $\lambda \left( u\right) $ belong to the same fiber. The last
condition ensures that no diffeomorphisms $\lambda _{M}:M\rightarrow M$
given by%
\begin{equation}
\lambda _{M}\circ \pi \left( u\right) =\pi \circ \lambda \left( u\right)
\text{,}
\end{equation}%
be allowed on the base space $M$. In a gauge description of
gravitation, one is interested in gauging external transformation
groups. This means that the group action on space-time coordinates
cannot be neglected. The spaces of internal fiber and external
base must be interlocked in the sense that transformations in one
space must induce corresponding transformations in the other. The
usual definition of a gauge transformation, {\it i.e.} as a displacement
along local fibers not affecting the base space, must be
generalized to reflect this interlocking. One possible way of
framing this interlocking is to employ a nonlinear realization of
the gauge group $G$, provided a closed subgroup $H\subset G$
exists. The interlocking requirement is then transformed into the
interplay between groups $G$ and one of its closed subgroups $H$.

Let us denote by $G$ a Lie group with elements $\left\{ g\right\}
$. Let $H$ be a
closed subgroup of $G$ specified by
\begin{equation}
H:=\left\{ h\in G|\Pi \left( R_{h}g\right) =\pi \left( g\right) \text{, }%
\forall g\in G\right\} \text{,}
\end{equation}%
with elements $\left\{ h\right\} $ and known linear representations $\rho
\left( h\right) $. Here $\Pi $ is a projection map and $R_{h}$ is the right group action. Let $M$ be
a differentiable manifold with points $\left\{ x\right\} $ to
which $G$ and $H$ may be referred, {\it i.e.} $g=g(x)$ and $h=h(x)$.
Being that $G$ and $H$ are Lie groups, they are also manifolds.
The right action of $H$ on $G$ induces a
complete partition of $G$ into mutually disjoint orbits $gH$. Since $g=g(x)$%
, all elements of $gH=\left\{ gh_{1}\text{, }gh_{2}\text{, }gh_{3}\text{,}%
\cdot \cdot \cdot \text{, }gh_{n}\right\} $ are defined over the
same $x$. Thus, each orbit $gH$ constitutes an equivalence class
of point $x$, with equivalence relation $g\equiv g^{\prime }$
where\ $g^{\prime }=R_{h}g=gh$.

By projecting each equivalence class onto a single element of the
quotient space $\mathcal{M}:=G/H$, the group $G$ becomes organized
as a fiber bundle \ in\ the sense that
$G=\bigcup\nolimits_{i}\left\{ g_{i}H\right\} $. In
this manner the manifold $G$ is viewed as a fiber bundle $G\left( \mathcal{M}%
\text{, }H\text{; }\Pi \right) $ with $H$-diffeomorphic fibers $\Pi
^{-1}\left( \xi \right) :G\rightarrow \mathcal{M}=gH$ and base space $%
\mathcal{M}$. A composite principal fiber bundle\textit{\ }$\mathbb{P}(M$, $G
$; $\pi )$ is one whose $G$-diffeomorphic fibers possess the fibered
structure $G\left( \mathcal{M}\text{, }H\text{; }\Pi \right) \simeq \mathcal{%
M}\times $ $H$ described above. The bundle $\mathbb{P}$ is then locally
isomorphic to $M\times G\left( \mathcal{M}\text{, }H\right) $. Moreover,
since an element $g\in G$ is locally homeomorphic to $\mathcal{M}\times H$
the elements of $\mathbb{P}$ are - by transitivity - also locally
homeomorphic to $M\times \mathcal{M}\times H\simeq \Sigma \times H$ where
(locally) $\Sigma \simeq M\times \mathcal{M}$. Thus, an alternative view
 of $\mathbb{P}(M$, $G$; $\pi )$ is provided by the $%
\mathbb{P}$-associated $H$-bundle $\mathbb{P}(\Sigma $, $H$; $\widetilde{\pi
})$ \cite{Tresguerres}. The total space $\mathbb{P}$ may be
regarded as $G\left( \mathcal{M}%
\text{, }H\text{; }\Pi \right) $-bundles over the base space $M$
or equivalently as $H$-fibers attached to the manifold $\Sigma
\simeq M\times \mathcal{M}$.

The nonlinear realization (NLR) technique \cite{CCWZ1, CCWZ2} provides a way
to determine the transformation properties of fields defined on the quotient
space $G/H$. The NLR of Diff$\left( 4\text{, }%
\mathbb{R}
\right) $ becomes tractable due to a theorem  by
Ogievetsky. According to this theorem \cite{BorisovOgievetskii},
the algebra
of the infinite dimensional group Diff$\left( 4\text{, }%
\mathbb{R}
\right) $ can be taken as the closure of the finite dimensional algebras of $%
SO(4$, $2)$ and $A(4$, $%
\mathbb{R}
)$. Remind that the Lorentz group generates transformations that preserve
the quadratic form on Minkowski space-time built from the metric tensor,
while the special conformal group generates infinitesimal angle-preserving
transformations on Minkowski space-time.

The affine group is a generalization
of the Poincar\'{e} group where the Lorentz group is replaced by the group
of general linear transformations. As such, the affine group generates
translations, Lorentz transformations, volume preserving shear and volume
changing dilation transformations. As a consequence, the NLR of Diff$\left( 4%
\text{, }%
\mathbb{R}
\right) /SO(3$, $1)$ can be constructed by taking a simultaneous realization
of the conformal group $SO(4$, $2)$ and the affine group $A(4$, $%
\mathbb{R}
):=%
\mathbb{R}
^{4}\rtimes GL(4$, $%
\mathbb{R}
)$ on the coset spaces $A(4$, $%
\mathbb{R}
)/SO(3$, $1)$ and $SO(4$, $2)/SO(3$, $1)$. One possible
interpretation of this theorem is that the conformal-affine group
CA  (defined below) may be the
largest subgroup of Diff$\left( 4\text{, }%
\mathbb{R}
\right) $, whose transformations may be put into the form of a
generalized coordinate transformation. We remark that a NLR can be
made linear by embedding the representation in a sufficiently
higher dimensional space. Alternatively, a linear group
realization becomes nonlinear when subject to constraints. One
type of relevant constraints may be those responsible for
symmetry reduction from Diff$\left( 4\text{, }%
\mathbb{R}
\right) $ to $SO(3$, $1)$ for instance.

We take the group $CA(3$, $1)$ as the basic symmetry group $G$. The CA group
consists of the groups $SO(4$, $2)$ and $A(4$, $%
\mathbb{R}
)$. In particular, CA is proportional to the union $SO(4$, $2)\cup A(4$, $%
\mathbb{R}
)$. We know however that the affine and special conformal groups have
several group generators in common. These common generators reside
in the intersection $SO(4$, $2)\cap
A(4$, $%
\mathbb{R}
)$ of the two groups, within which there are \textit{two copies} of $\Pi
:=D\times P(3$, $1)$, where $D$ is the group of scale transformations
(dilations)\ and $P(3$, $1):=T\left( 3\text{, }1\right) \rtimes SO(3$, $1)$
is the Poincar\'{e} group.

Finally we define the CA group as the union of the affine
and conformal groups minus \textit{one copy} of the overlap $\Pi $, {\it i.e.} $%
CA(3$, $1):=SO(4$, $2)\cup A(4$, $%
\mathbb{R}
)-\Pi $. Being defined in this way we recognize that $CA(3$, $1)$ is a $24$
parameter Lie group representing the action of Lorentz transformations $(6)$,
 translations $(4)$, special conformal transformations $(4)$,
space-time shears $(9)$ and scale transformations $(1)$.
All these transformations can be adopted to define any conformally-affine theory of gravity.

In this
paper, we obtain the NLR of $CA(3$, $1)$ modulo $SO(3$, $1)$ as
4D realizations starting from 5D-manifold  \cite{unification}. This procedure has been recently adopted also in holographic approaches to
Quantum Chromodynamics \cite{cappiello}.

\section{ The group structure in 5D and 4D-spaces}
\label{tre}

In this section, we will discuss the  group structure of a
5D-Riemannian manifold (in particular the Lorentz group) and its reduction to 4D-manifold.   Such an approach gives a  useful tool to deal with  the realization of effective theories of gravity in 4D and the problem of mass generation.  Let us
start with the necessary definition of the Minkowski space-time ${\bf
R}^4$ endowed with the metric

\be (X,X)_4
=(x^0)^2-\overrightarrow{X}\cdot\overrightarrow{X}\,,\ee where
$X=(x^0,\overrightarrow{X})$ is a four-vector, $x^0$ is the  time
coordinate and $\overrightarrow{X}$ is an ordinary vector in
${\bf R}^3$.  The Lorentz transformations are those linear
transformations of Minkowski-type space that leave $(X,X)_4$, the
scalar product of four-vectors, invariant: \be X\longrightarrow
X'=\Lambda X\,,\ee being $(\Lambda X,\Lambda X)=(X,X)$. If $g$
is the Minkowski metric with signature $(+\, -\, -\, -)$,
$\Lambda$ is a Lorentz transformation when $\Lambda^t g\Lambda=g$.
The set of such transformations is the orthogonal group $O(4)$,
namely $O(1,3)$ considering the time-like and space-like
components, group characterized, as well known, by the properties
that det$\Lambda=\pm 1$
 and the number of generators is 6. The coset decomposition of
 such a group is
 \be
 \label{decomposition}
 O(1,3)={\bf I}SO(1,3)\cup\Lambda_P SO(1,3)\cup \Lambda_T SO(1,3)\cup
 \Lambda_{PT}(1,3)\,,\ee
 where ${\bf I}SO(1,3)$ is the {\it proper orthochronus} Lorentz
 group with det$\Lambda=1$, whose elements preserve parity
 (spatial orientation) and the direction of time; $\Lambda_PSO(1,3)$
 is the group of spatial inversion (parity inversion); $\Lambda_TSO(1,3)$
 are the time reversal transformations and $\Lambda_{PT}(1,3)$
 are the total space-time inversion, where we have taken into account
 all the components without arbitrarily discarding any part of them. The covering group
 of $SO(1,3)$ is the simply connected complex group $SL(2,C)$
 whose physical meaning is that particles (or in general fields)
 transform according to its representations.

Now we want to extend this scheme to a 5D-space (which we,
initially, consider a flat manifold), where we do not define {\it a priori}
a signature for the metric and which, after a 4D-reduction
procedure, must be capable of reproducing  all the features of
Lorentz group.  For the sake of generality,
we do not specify the signature   and the number of
dimensions. Below we will assume  $N=5$.

Let ${\cal M}^{(p,q)} = {\bf R}^N$ be a manifold where $p,q\geq 0$
are integers and such that $p+q=N$ with the flat metric
$dS^2=\,^{(p,q)}\eta_{AB}dx^Adx^B$ and $A,B=0,1,2,...,N-1$. A
general signature is
 \be
^{(p,q)}\eta=\left(^{(p,q)}\eta_{AB}\right)=\mbox{diag}\,(\,\underbrace{1,1,..,1}_p\,;\,\underbrace{-1,-1,..,-1}_q\,)\,,
\ee where  $p$  are the time-like directions and $q$ are the
space-like directions.  As particular cases, we have \ba
{\bf E}^N={\cal M}^{(0,N)}\;&=&\;\mbox{Euclidean space,}\nonumber\\
{\cal M}^N={\cal M}^{(1,N-1)}\;&=&\;\mbox{$N$-dimensional
Minkowski space.}\nonumber \ea It is important to stress that the other flat (pseudo)-Riemannian
spaces have more than one equivalent (independent) time-like
directions and hence have no distinction between future and past
time-like directions as they have in  Minkowski space. This fact
means that the space-like pseudo-spheres are connected
hypersurfaces, rather than having two disjoint components as in
Minkowski space. The metric can be written as
\be\label{split7}
dS^2=\left(\sum_{A,B=0}^{p}\delta_{AB}dx^{A}dx^{B}\right)-
\left(\sum_{A,B=p+1}^{N-1}\delta_{AB}dx^{A}dx^{B}\right)\,, \ee
where time-like and space-like components are  clearly separated.
Some considerations are necessary at this point. The metric
(\ref{split7}) is invariant under rotations of the time-like
directions among themselves (except for $E^N$ and ${\cal M}^N$
which are degenerate particular cases, since in the first case
there are no time arrows and in the second case, only one time
arrow exists by definition) and of the space-like directions among
themselves. The remaining independent pseudo-rotations are all
boosts each involving a time-like and a space-like direction. The
physical meaning of such a result is that close time-like paths
are an usual feature in pseudo-Riemannian manifolds, moreover a
definite time arrow distinguishing the past from the future is
only a particular characteristic of Minkowski spaces where Lorentz
transformations work.

Let us now take into account the possible linear transformations
on this ${\cal M}^{(p,q)}$-manifold. A pseudo-orthogonal group
$O(p,q)$ can be defined on this pseudo-Riemannian manifold. This
group consists of all the linear transformations $X^A\rightarrow
\Lambda^A_B X^B$ such that the metric (\ref{split7}) is
invariant, \ie \be dS^2\longrightarrow\eta_{AB}\Lambda^A_C
\Lambda^B_D dx^C dx^D=\eta_{CD}dx^Cdx^D\,,\ee more precisely we can
say that \be  O(p,q)\equiv\left\{(\Lambda^A_B)\in GL(N,{\bf
R})\mid \eta_{AB}\Lambda^A_C \Lambda^B_D=\eta_{CD}\right\}\,,\ee
where $GL(N,{\bf R})$ are non-singular matrices in $N$
dimensions. Note that \be
\mbox{det}\,(\eta_{AB})\left[\mbox{det}\,(\Lambda^A_B)\right]^2=\mbox{det}\,(\eta_{AB})\,\longrightarrow
\,\mbox{det}\,(\Lambda^A_B)=\pm 1\,,\ee where the determinant is
$+1$ for rotations $SO(p,q)$ and $-1$ for inversions, inversions
which do not constitute a sub-group (the product of two inversions
gives a rotation). In the first case, we have \be
SO(p,q)\equiv\left\{(\Lambda^A_B)\in O(p,q)\mid
\mbox{det}\,(\Lambda^A_B)=1\right\}\,, \ee which is a special
pseudo-orthogonal group. An important feature of such a group is
that it consists of two disconnected pieces when both $p$ and $q$
are odd (see \cite{gilmore} for the general demonstration). Special
examples of $SO(p,q)$ are
 \begin{eqnarray}
  && SO(0,N)\equiv SO(N,{\bf R})\nonumber\\
  && \mbox{for}\quad p=0\quad\mbox{special orthogonal group,}\nonumber\\
&& SO(1,N-1)\nonumber\\
&& \mbox{for}\quad p=1\quad\mbox{Lorentz or De
Sitter group.}\nonumber\end{eqnarray}

The group $SO(p,q)$ can be
decomposed as follows

\begin{equation}
 SO(p,q)=\left(
\begin{array}{cccc}
\overbrace{SO(p,{\bf R})}^{p\times p} &\mid & \overbrace{\mbox{boosts}}^{p\times q}\\
&\mid &\\
----&\mid &----\\ \\ \underbrace{\mbox{boosts}}_{q\times p} &\mid & \underbrace{SO(q,{\bf
R})}_{q\times q}
\end{array}
\right)\,,
\end{equation}
where $SO(p,{\bf R})$ are $(p\times p)$ square matrices which
rotate the time-like directions among themselves, $SO(q,{\bf R})$
are $(q\times q)$ square matrices which rotate the space-like
directions among themselves, and the boosts are, in general,
$(p\times q)$ or $(q\times p)$ rectangular matrices which rotate
time-like and space-like directions.

The number of generators of the $SO(p,q)$ group, \ie the number of
independent elements or the {\it dimension} of the group, can be
easily calculated being, in general,
$$\mbox{dim}\, SO(N)=\frac{N(N-1)}{2}\,.$$
An important remark is in order at this point.
It is well-known, since an old result by Riemann \cite{riemann}, that a N-dimensional metric has $ s=N(N - 1)/2$ degrees of freedom, that is, it is locally
equivalent to  give  $s$ independent functions. This feature is related to the choice of local charts but it is also related to the number of degrees of freedom of gravitational field. As we will discuss below, this is a key ingredient of our discussion.
In our case, we have

 \be
  \mbox{dim}\, SO(p,q)= \mbox{dim}\,
SO(p,{\bf R})+\mbox{dim}\, SO(q,{\bf R})+p\cdot q\,. \ee
The result is
 \be
\label{dim}\frac{N(N-1)}{2}=\frac{p(p-1)}{2}+\frac{q(q-1)}{2}+p\cdot
q\,,\ee

 where $p\cdot q$ is the number of independent pairs of
one space-like and one time-like direction. For $N=5$, we have 10
independent elements.

Clearly the rotations $SO(p,{\bf R})\otimes SO(q,{\bf R})$ form a
sub-group of $SO(p,q)$ but the boosts do not; boosts along
different directions combine to give a boost plus a rotation.

Let us now add the $N$ translations $\tilde{X}^A\rightarrow
X^A+a^A$ to the pseudo-orthogonal group $O(p,q)$, consisting of
rotations and inversions. This fact yields the full group of
motions of ${\cal M}^{(p,q)}$, which can be classified in the
most general {\it inhomogeneous pseudo-orthogonal group}
$IO(p,q)$. Not taking into account the inversions, a remarkable
sub-group is $ISO(p,q)$, the inhomogeneous {\it special}
pseudo-orthogonal group, of dimension \be
r=\frac{N(N-1)}{2}+N=\frac{N(N+1)}{2}\,,\ee which gives $r=15$
for $N=5$. More generally, these groups can be realized as matrix
groups in $N+1$ dimensions by adding a trivial row and a
nontrivial column to $O(p,q)$, \ie

\begin{equation}
 IO(p,q)=\left(
\begin{array}{cccc}
 &\mid & a^0\\
O(p,q) &\mid& \cdot \\
&\mid& \cdot \\
 &\mid & a^{N-1}\\
----&\mid &----\\ 0\cdots 0 &\mid & 1
\end{array}
\right)\,.
\end{equation}
This fact is extremely interesting for our purposes since, adding
up a dimension to the manifold in which we define dynamics allows
to remove the singularities. As
special cases, we  have

 \ba
&& \bullet\quad\, IO(0,N)= IO(N,{\bf R})\,=\nonumber\\
 &&\mbox{Euclidean group in $N$ dimensions,}\nonumber\\
&&\bullet\quad\, IO(1,N-1)\,=\nonumber\\
&&\mbox{Poincar\'{e} group in $N$ dimensions},\nonumber\\
&&\mbox{ or
inhomogeneous Lorentz group. }\nonumber
 \ea
The pseudo-spheres at the origin of ${\cal M}^{(p,q)}$ satisfy
the fundamental relation
 \begin{eqnarray}
\eta_{AB}x^Ax^B=\mbox{constant}\,.
\end{eqnarray}
Connected components
of  pseudo-sphere are $(N-1)$-dimensional hypersurfaces on which
acts a $N(N-1)/2$--dimensional group,
under which all points are equivalent. Such spaces are all of
constant curvature and of different signatures (which are
determined by considering coordinate directions at their
intersection with the Cartesian axes;  property which is important
for the following projection process).

This general discussion can be specialized to the 5D-case. First
of all, we assume that a 5D-vector field defines a metric whose
signature is given by

\be
 (X,X)_5=(x^0)^2-\overrightarrow{X}\cdot\overrightarrow{X}+\epsilon (x_4)^2\,,
 \ee

where $\epsilon=\pm 1$, so that, using the
traditional terms, the fifth dimension can be time-like or
space-like. Moreover, as we shall see below, it is the 4D-dynamics
which discriminates, by a bijective correspondence, the signature
giving rise to particle-like solutions $(\epsilon=-1)$ or to
wave-like solutions $(\epsilon=+1)$. When $N=5$, we can obtain
pseudo-spheres of Lorentz signature and thus 4D-space-times of
constant curvature (as Friedmann-Robertson-Walker ones). There
are only two independent different  signatures for $N=5$. They are
$(p,q)=(1,4)$, corresponding to the case $\epsilon=-1$ and
$(p,q)=(2,3)$, corresponding to $\epsilon=+1$. The  5D-manifolds
are ${\cal M}^{(1,4)}={\bf R}^5$ and ${\cal M}^{(2,3)}={\bf R}^5$,
respectively, where ${\bf R}^5$ is the 5D-space. As it is well-known, the former case
is called the De Sitter space, while the latter is the Anti-De
Sitter one. The fact that the standard signature of the universe
is $(+\, -\, -\, -)$ can be derived from an equivalent process
starting
 from ${\cal M}^{(1,4)}$ or ${\cal M}^{(2,3)}$.
 The discrimination is dynamically achieved, as we shall see below, when particle masses,
  after the embedding,  spring out.
   Due to this fact, we are going to deal with the degrees of
   freedom of the
 space-time and of the particles under the same
 standard so that a straightforward decomposition of our 5D-group
 can be

\be {\cal G}_5\supset IO(3,1)\otimes SU(3)\otimes SU(2)\otimes
U(1)\,,\ee
and then if ${\cal G}_5=SL(5)$, this is the minimal
group, with $N^2-1=24$ parameters which is capable of including
all the standard fundamental interactions and the 10 generators
of inhomogeneous Lorentz group. In particular, it can include the
inhomogeneous pseudo-orthogonal group $IO(5)$ (in the two
modalities $IO(1,4)$ or $IO(2,3)$), which is a sub-group of
$SL(5)$,  comprehensive of all space-time rotations, inversions
and translations in 5D.
It is
clear that   4D pseudo-Riemannian manifolds ${\cal M}^{(1,3)}$
 can be
obtained from both ${\cal M}^{(1,4)}$ or ${\cal M}^{(2,3)}$. This reduction procedure, as we will show below,  is a
dynamical process depending on the splitting of the 5D-field
equations.

An important consideration is necessary at this point.
After the reduction to  4D-physics, a given theory of gravity is described by the conformal-affine group of
diffeomorphisms $GL(4)$ which is characterized by $4\times 4=16$ generators.  A straightforward splitting of such a group is

\begin{equation}
 \underbrace{GL(4)}_{\underbrace{4\times 4}}\supset \underbrace{SU(3)}_{\underbrace{3^2-1}} \otimes \underbrace{SU(2)}_{\underbrace{2^2-1}} \otimes \underbrace{U(1)}_{\underbrace{1}}\otimes\underbrace{ GL(2)}_{\underbrace{2\times 2}}
 \end{equation}
 where the number of generators is indicated for any subgroup.  The physical meaning of $SU(3)$, $SU(2)$ and $U(1)$ is clear while $GL(2)$ represents a group of diffeomorphisms with 4 generators.
 The further generators of gravitational field can be recovered in the framework of Extended Theories of Gravity \cite{book,odino} as we will see below.  The experimental consequences of these further gravitational modes could be extremely interesting for the physics at LHC.

\section{Space-time deformations and the $GL(4)$-group}
\label{quattro}
Another ingredient for our considerations is represented by the space-time deformations that are elements of the $GL(4)$-group.
Let us take into account a
metric $\mathbf{g}$ on a space-time manifold $\mathcal{M}$. Such a
metric is assumed to be an exact solution of the gravitational field
equations.  From the discussion in Sec. III,  we can decompose it by a co-tetrad field $\omega^{A}(x)
$
\begin{equation}\label{tetrad}
\mathbf{g}=\eta_{AB}\omega^{A} \omega^{B}.
\end{equation}
Let us define now a new tetrad field
$\widetilde{\omega}=\Phi^{A}_{\phantom{A}C}(x)\,\omega^{C}$, with
$\Phi^{A}_{\phantom{A}C}(x)$ a matrix of scalar fields. Finally we
introduce a space-time  $\widetilde{\mathcal{M}}$ with the metric
$\widetilde{g}$  defined in the following way
\begin{equation}\label{deformed}
\mathbf{\widetilde{g}}=\eta_{AB}\Phi^{A}_{\phantom{A}C}\Phi^{B}_{\phantom{B}D}\,\omega^{C}
\omega^{D} = \gamma_{CD}(x)\omega^{C}
\omega^{D},
\end{equation}
where also $\gamma_{CD}(x)$ is a matrix of fields which are
scalars with respect to the coordinate transformations.

If $\Phi^{A}_{\phantom{A}C}(x)$ is  a Lorentz matrix in any point of  $\mathcal{M}$, then
\begin{equation}\label{i.3}
    \widetilde{g}\equiv g
\end{equation}
otherwise  we say that $\widetilde{g}$ is a  deformation of $g$
and $\widetilde{\mathcal{M}}$ is a deformed $\mathcal{M}$.  If all
the functions of $\Phi^{A}_{\phantom{A}C}(x)$ are continuous, then
there is a {\it one - to - one} correspondence between the points
of $\mathcal{M}$ and the points of $\widetilde{\mathcal{M}}$.
If  $\xi$ is a Killing vector for $g$ on
$\mathcal{M}$, the  corresponding vector $\widetilde{\xi}$ on
$\widetilde{\mathcal{M}}$ could not  necessarily be a Killing
vector.

A particular subset of these deformation matrices is given by
\begin{equation}\label{conformal}
{\Phi}{^{A}_{C}}(x)=\Omega(x)\, \delta^{A}_{\phantom{A}C}.
\end{equation}
which define conformal transformations of the metric,
\begin{equation}\label{confmetric}
  \widetilde{g}=\Omega^{2}(x) g\,.
\end{equation}
In this sense, the  deformations defined by Eq. (\ref{deformed})
can be regarded as a generalization of the conformal
transformations which we will discuss below.

We  call the matrices $\Phi^{A}_{\phantom{A}C}(x)$ {\it first
deformation matrices}, while we can refer to
\begin{equation}\label{seconddeformation}
 \gamma_{CD}(x)=\eta_{AB}\Phi^{A}_{\phantom{A}C}(x)\Phi^{B}_{\phantom{B}D}(x).
\end{equation}
as the {\it second deformation matrices}, which, as seen above,
are also matrices of scalar fields. They generalize  the Minkowski
matrix \(\eta_{AB}\) with constant elements in the definition of
the metric. A further restriction on the matrices
$\Phi^{A}_{\phantom{A}C}$ comes from the above mentioned theorem proved by Riemann
by which an $N$-dimensional metric has $N(N-1)/2$ degrees of
freedom (see \cite{Coll:2001wy} for details).

With this
definitions in mind, let us consider the main properties of
deforming matrices.
Let us take into account a four dimensional space-time with
Lorentzian signature. A family of matrices
$\Phi^{A}_{\phantom{A}C}(x)$ such that
\begin{equation}\label{fi}
    \Phi^{A}_{\phantom{A}C}(x)\in GL(4)\, \forall x,
\end{equation}
are defined on such a space-time.
These functions are not necessarily continuous and can connect
space-times with different topologies. A singular scalar field
introduces a deformed manifold $\widetilde{\mathcal{M}}$ with a
space-time singularity.

As it is well known, the Lorentz matrices
$\Lambda^{A}_{\phantom{A}C}$ leave the Minkowski metric invariant
and then

\begin{equation}
\mathbf{g}=\eta_{EF}\Lambda^{E}_{\phantom{E}A}\Lambda^{F}_{\phantom{F}B}\Phi^{A}_{\phantom{A}C}\Phi^{B}_{\phantom{B}D}\,
\omega^{C}
\omega^{D} =\eta_{AB}\Phi^{A}_{\phantom{A}C}\Phi^{B}_{\phantom{B}D}\,\omega^{C}
\omega^{D}.\label{deformlambda2}
\end{equation}
It follows that $\Phi^{A}_{\phantom{A}C}$ give rise to right
cosets of the Lorentz group, {\it i.e.} they are the elements of  the
quotient group $GL(4,\mathbf{R})/SO(3,1)$. On the other hand,  a
right-multiplication of $\Phi^{A}_{\phantom{A}C}$ by a Lorentz
matrix induces a different deformation matrix.

The inverse deformed metric is
\begin{equation}\label{inversemetric}
     \widetilde{g}^{\alpha\beta}=\eta^{CD}{\Phi^{-1}}^{A}_{\phantom{A}C}{\Phi^{-1}}^{B}_{\phantom{B}D}e_{A}^{\alpha}e_{B}^{\beta}
\end{equation}
where ${\Phi^{-1}}^{A}_{\phantom{A}C}{\Phi}^{C}_{\phantom{C}B}=\Phi^{A}_{\phantom{A}C}{\Phi^{-1}}^{C}_{\phantom{C}B}=
\delta^{A}_{B}$.

Let us decompose now the matrix $\Phi_{AB}=\eta_{AC}\,
\Phi^{C}_{\phantom{C}B}$ in its symmetric and antisymmetric parts
\begin{equation}\label{decomposition1}
     \Phi_{AB}= \Phi_{(AB)}+\Phi_{[AB]}= \Omega\,\eta_{AB} +  \Theta_{AB} + \phi_{AB}
\end{equation}
where $ \Omega= \Phi^{A}_{\phantom{A}A}$,  $ \Theta_{AB} $ is the
traceless symmetric part   and $ \phi_{AB}$ is the skew
symmetric part of  the first deformation matrix respectively. Then
standard conformal transformations are nothing else but
deformations with $\Theta_{AB}=\phi_{AB}=0$ \cite{wald}.

Finding the inverse matrix ${\Phi^{-1}}^{A}_{\phantom{A}C}$ in
terms of $\Omega$, $\Theta_{AB}$  and $\phi_{AB}$ is not
immediate, but as above, it can be split in the three terms
\begin{equation}\label{inversesplitting}
      {\Phi^{-1}}^{A}_{\phantom{A}C}=\alpha\delta^{A}_{\phantom{A}C}+\Psi^{A}_{\phantom{A}C}+\Sigma^{A}_{\phantom{A}C}
\end{equation}
where $\alpha$, $\Psi^{A}_{\phantom{A}C}$ and
$\Sigma^{A}_{\phantom{A}C}$ are  respectively the trace, the
traceless symmetric part and the antisymmetric part of the inverse
deformation matrix. The second deformation matrix, from the above
decomposition, takes the form
\begin{equation}\label{secondmatrix}
  \gamma_{AB}= \eta_{CD}(\Omega\, \delta_{A}^{C}+
  \Theta_{\phantom{C}A}^{C}+ \phi_{\phantom{C}A}^{C})(\Omega\, \delta_{B}^{D}+
  \Theta_{\phantom{D}B}^{D}+ \phi_{\phantom{D}B}^{D})
\end{equation}
and then

\begin{eqnarray}\label{secondmatrix1}
   \gamma_{AB}&=&  \Omega^{2}\,\eta_{AB} + 2\Omega\,\Theta_{AB}+  \eta_{CD}\, \Theta_{\phantom{C}A}^{C}\,\Theta_{\phantom{D}B}^{D} +\nonumber\\&&+ \eta_{CD}\, (\Theta_{\phantom{C}A}^{C}\,\phi_{\phantom{D}B}^{D}
 +   \phi_{\phantom{C}A}^{C}\,\Theta_{\phantom{D}B}^{D}) + \eta_{CD}\,\phi_{\phantom{C}A}^{C}\,\phi_{\phantom{D}B}^{D}.\nonumber\\
\end{eqnarray}
In general, the deformed metric can be split as
\begin{equation}\label{splits}
     {\tilde{g}}{_{\alpha\beta}}=\Omega^{2}{g}{_{\alpha\beta}}+{\gamma}{_{\alpha\beta}}
\end{equation}
where
\begin{eqnarray}\label{acca}
{\gamma}{_{\alpha\beta}}&=&\left( 2\Omega\,\Theta_{AB}+  \eta_{CD}\, \Theta_{\phantom{C}A}^{C}\,\Theta_{\phantom{D}B}^{D} + \eta_{CD}\, (\Theta_{\phantom{C}A}^{C}\,\phi_{\phantom{D}B}^{D}+\right. \nonumber\\ &&  \left. +   \phi_{\phantom{C}A}^{C}\,\Theta_{\phantom{D}B}^{D})+ \eta_{CD}\,\phi_{\phantom{C}A}^{C}\,\phi_{\phantom{D}B}^{D}\right){\omega}{^{A}_{\alpha}}{\omega}{^{B}_{\beta}}
  \end{eqnarray}
In particular,  if $ \Theta_{AB}=0$, the deformed metric
simplifies to
\begin{equation}\label{split}
    \widetilde{g}_{\alpha\beta}=\Omega^{2}g_{\alpha\beta}+\eta_{CD}\,\phi_{\phantom{C}A}^{\,C}\,\phi_{\phantom{D}B}^{\,D}
    \omega^{A}_{\phantom{A}\alpha}\omega^{B}_{\phantom{B}\beta}
\end{equation}
and, if $\Omega=1$, the deformation of a metric consists in adding
to the background metric a tensor $\gamma_{\alpha\beta}$. We have to
remember that all these quantities are not independent as, by the
theorem mentioned in \cite{Coll:2001wy}, they have to form at most
six independent functions in a four dimensional space-time.
Similarly the controvariant deformed metric can be always
decomposed in the following way
\begin{equation}\label{controvariantdecomposition}
    \widetilde{g}^{\alpha\beta}= \alpha^{2}g^{\alpha\beta}+ \lambda^{\alpha\beta}
\end{equation}
Let us find the relation between ${\gamma}{_{\alpha\beta}}$ and $
\lambda^{\alpha\beta}$. By using
$\widetilde{g_{\alpha\beta}}\widetilde{g^{\beta\gamma}}=\delta_{\alpha}^{\gamma}$, we obtain
\begin{equation}\label{relationgammalambda}
     \alpha^{2}\Omega^{2}\delta_{\alpha}^{\gamma}+ \alpha^{2}\gamma_{\alpha}^{\gamma}+\Omega^{2}\lambda_{\alpha}^{\gamma}+
     {\gamma}{_{\alpha\beta}}\lambda^{\beta\gamma}=\delta_{\alpha}^{\gamma}
\end{equation}
if the deformations are conformal transformations, we  have
$\alpha=\Omega^{-1}$, so  assuming such a condition, one obtain
the following matrix equation
\begin{equation}\label{relationgammalambda1}
    \alpha^{2}\gamma_{\alpha}^{\gamma}+\Omega^{2}\lambda_{\alpha}^{\gamma}+
     {\gamma}{_{\alpha\beta}}\lambda^{\beta\gamma}=0\,,
\end{equation}
and
 \begin{equation}\label{lambdaaa}
 (\delta_{\alpha}^{\beta}+
     \Omega^{-2}{\gamma}_{\alpha}^{\beta})\lambda_{\beta}^{\gamma}=-\Omega^{-4}\gamma_{\alpha}^{\gamma}
 \end{equation}
 and finally
 \begin{equation}\label{lambdaaaa}
 \lambda_{\beta}^{\gamma}=-\Omega^{-4}{{(\delta+
     \Omega^{-2}{\gamma})^{-1}}}{^{\alpha}_{\beta}}\gamma_{\alpha}^{\gamma}
 \end{equation}
where   ${(\delta+ \Omega^{-2}{\gamma})^{-1}}$ is the inverse
tensor of $(\delta_{\alpha}^{\beta}+
     \Omega^{-2}{\gamma}_{\alpha}^{\beta})$.

To each matrix $\Phi^{A}_{\phantom{A}B}$,  we can associate a
(1,1)-tensor $\phi^{\alpha}_{\phantom{\alpha}\beta}$  defined by
\begin{equation}\label{3.1}
  \phi^{\alpha}_{\phantom{\alpha}\beta}= \Phi^{A}_{\phantom{A}B}\omega^{B}_{\beta}e_{A}^{\alpha}
\end{equation}
such that
\begin{equation}\label{3.2}
     \widetilde{g}_{\alpha\beta}=g_{\gamma\delta}\phi^{\gamma}_{\phantom{\gamma}\alpha}\phi^{\delta}_{\phantom{\delta}\beta}\,.
\end{equation}
Vice-versa from a
(1,1)-tensor $\phi^{\alpha}_{\phantom{\alpha}\beta}$, we can define a matrix of
scalar fields as
\begin{equation}\label{3.3}
     \phi^{A}_{\phantom{A}B} =  \phi^{\alpha}_{\phantom{\alpha}\beta} \omega_{\alpha}^{A}e_{B}^{\beta}.
\end{equation}
In summary, space-time deformations are endowed with the conformal-affine structure
of $GL(N)$ groups (in particular $GL(4)$) and, passing from a given metric to another,
 gives rise to induced scalar fields that, as we shall show below, can be generated by
  a dimensional reduction mechanism. As we shall show below, this feature can
  be considered as a sort of  gravitational Higgs-like  mechanism capable of inducing symmetry breakings.

\section{ The 5D-space and the reduction to 4D-dynamics}
\label{cinque}

\subsection{The 5D-formalism}
\label{cinque.uno}

In this section, we are going to define the curvature invariants,
the field equations and the conservation laws in the 5D-space. In
general, we ask for a space which is a smooth manifold,
singularity free and, first of all, defined in such a way that
 conservation laws on it are
valid. Technically given,  a 2-rank tensor of the form $T^{AB}$, the relation
$\nabla_A T^{A}_{B}=0$ must be always valid and never  singular in
the sense that it is preserved by  diffeomorphisms in any
coordinate frame. The 5D-Riemann tensor is

\be\label{a2}
 R^{D}_{ABC}=\partial_{B}\Gamma^{D}_{AC}-
\partial_{C}\Gamma^{D}_{AB}+\Gamma^{D}_{EB}\Gamma^{E}_{AC}-\Gamma^{D}_{EC}\Gamma^{E}_{AB}\,.
\ee
The number of independent components of such a tensor,
after the full derivation and thanks to the Petrov classification
\cite{weinberg}, is ${\displaystyle \frac{1}{12}N^2(N^2-1)=50}$.
The Ricci tensor and scalar are derived from the contractions

\be\label{a3}R_{AB}=R^{C}_{ACB}\,,\qquad ^{(5)}R=R^{A}_{A}\,.
\ee
The field equations can be derived from the 5D--Hilbert--Einstein action

  \be\label{a4} ^{(5)}{\cal
A}=-\frac{1}{16\pi \,^{(5)}G}\int
d^{5}x\sqrt{-g^{(5)}}\left[^{(5)}R\right]\,,\ee
where, using the
standard notation,  $^{(5)}G$ is the 5D-gravitational coupling
constant and $g^{(5)}$ is the determinant of the 5D-metric. The
variational principle
 \be\label{a5} \delta\int
d^{5}x\sqrt{-g^{(5)}}\left[^{(5)}R\right]=0\,,\ee
 gives the
5D-field equations which are \be \label{a9}
G_{AB}=R_{AB}-\frac{1}{2}g_{AB}\,^{(5)}R=0\,,\ee so that at least
the Ricci-flat space is always a solution. Let us define now the
5D-stress-energy tensor:

\be \label{b11} T_{AB}=\nabla_{A}\Phi\nabla_{B}\Phi-\frac{1}{2}
g_{AB}\nabla_{C}\Phi\nabla^{C}\Phi\,, \ee
 where only the kinetic
terms are present. As standard, such a tensor can be derived from
a variational principle
\be\label{b111}
T^{AB}=\frac{2}{\sqrt{-g^{(5)}}}\frac{\delta\left(\sqrt{-g^{(5)}}{\cal
L}_{\Phi}\right)}{\delta g_{AB}}\,, \ee
where ${\cal L}_{\Phi}$
is a Lagrangian density connected with the scalar field $\Phi$ whose physical meaning will be clear below.
Because of the definition of 5D space itself \cite{conservation}, it
is important to stress now that no self-interaction potential
$V(\Phi)$ has been taken into account so that $T_{AB}$ is a
completely symmetric object and $\Phi$ is, by definition, a
cyclic variable. This fact guarantees that Noether theorem
holds for $T_{AB}$ and a conservation law intrinsically exists.
With these considerations in mind, the field equations can now
assume the form
 \be\label{a11}
R_{AB}=\chi\left(T_{AB}-\frac{1}{2}g_{AB}T\right)\,, \ee
 where
$T$ is the trace of $T_{AB}$ and $\chi=8\pi \,^{(5)}G$, being
$\hbar=c=1$. The form (\ref{a11}) of field equations is useful in
order to put in evidence the role of the scalar field $\Phi$, if
we are not simply assuming Ricci-flat 5D-spaces. As we said,
 $T_{AB}$ is a symmetric tensor for which the relation

\be \label{b12} T_{[A,B]}=T_{AB}-T_{BA}=0\,,\ee
 holds. Due to
the choice of the metric and to the symmetric nature of the
stress-energy tensor $T_{AB}$ and of the Einstein field equations
$G_{AB}$, the contracted Bianchi identities
 \be \label{b13} \nabla_{A}T^{A}_{B}=0\,,\qquad \nabla_{A}G^{A}_{B}=0\,, \ee
 hold. Developing  the stress-energy tensor, we have

\ba
\nabla_{A}T^{A}_{B}&=&\nabla_{A}\left(\partial_{B}\Phi\partial^{A}\Phi-\frac{1}{2}
\delta_{B}^{A}\partial_{C}\Phi\partial^{C}\Phi
\right)=\nonumber\\
&=&\left(\nabla_{A}\Phi_{B}\right)\Phi^{A}+\Phi_{B}\left(\nabla_{A}\Phi^{A}\right)-
\nonumber\\ &&+\frac{1}{2}\left(\nabla_{B}\Phi_{C}\right)\Phi^{C}-
\frac{1}{2}\Phi_{C}\left(\nabla_{B}\Phi^{C}\right)=\nonumber\\
&=&\left(\nabla_{A}\Phi_{B}\right)\Phi^{A}+\Phi_{B}\left(\nabla_{A}\Phi^{A}\right)-
\Phi_{C}\left(\nabla_{B}\Phi^{C}\right)\label{ciccio1}\nonumber\\ \,.\ea
Since our 5D-space is a Riemannian manifold, it is \be
\nabla_{A}\Phi_{B}=\nabla_{B}\Phi_{A}\,, \ee and then

\ba
&&\Phi^{A}\left(\nabla_{A}\Phi_{B}\right)-\Phi_{C}\left(\nabla_{B}\Phi^{C}\right)=\nonumber\\&&
=\Phi^{A}\left(\nabla_{B}\Phi_{A}\right)-\Phi_{C}\left(\nabla_{B}\Phi^{C}\right)=0\,.
\ea
 In this case, partial and covariant derivatives coincide for
the scalar field $\Phi$. Finally
 \be
\nabla_{A}T_{B}^{A}=\Phi_{B}\,^{(5)}\Box\Phi\,,\ee where
$^{(5)}\Box$ is the 5D d'Alembert operator defined as
$\nabla_{A}\Phi^{A}\equiv g^{AB}\Phi_{,A;B}\equiv \,^{(5)}\Box
\Phi$. The general result is that the conservation of the
stress-energy tensor $T_{AB}$ (\ie the contracted Bianchi
identities) implies the Klein-Gordon equation which assigns the
dynamics of $\Phi$, that is

\be \label{klein1} \nabla_{A} T_{A}^{B}=0 \qquad
\Longleftrightarrow \qquad \,^{(5)}\Box \Phi=0\,, \ee assuming
$\Phi_{B}\neq 0$ since we are dealing with a non-trivial physical field.
Let us note again the absence of self-interaction (\ie potential)
terms. As we shall see below, the relation (\ref{klein1}), being
a field equation,   gives a physical meaning to the fifth
dimension.

The reduction to the 4D-dynamics can accomplished by  taking into
account the Campbell theorem \cite{campbell}. This theorem states
that it is always possible to consider a 4D Riemannian manifold,
defined by the line element $ds^2=g\dab dx\ua dx\ub$, in a 5D one
with $dS^2=g_{AB}dx^A dx^B$.

We have $g_{AB}=g_{AB}(x\ua,x^4)$
with $x^4$ the yet unspecified extra coordinate. As we discussed
above,  $g_{AB}$ is covariant under the group of 5D coordinate
transformations $x^A\rightarrow \overline{x}^A(x^B)$, {\it but
not} under the (restricted) group of 4D transformations
$x\ua\rightarrow\overline{x}\ua(x\ub)$. This   fact has,
as a relevant consequence, that the choice of 5D coordinates results as
the {\it gauge} necessary  to specify the 4D physics also in its
non-standard aspects.
Vice-versa, in specifying the 4D physics, the bijective embedding
process in 5D gives physical meaning to the fifth coordinate
$x^4$.   In other words,  the fifth
coordinate, in 4D can assume the physical meaning of the mass.

\subsection{The reduction to 4D}
\label{cinque.due}

Let us replace the variational
principle (\ref{a5})
 with \be\label{a6} \delta\int
d^{(5)}x\sqrt{-g^{(5)}}\left[^{(5)}R+\lambda(g_{44}-\epsilon\Phi^2)\right]=0\,,\ee
where $\lambda$ is a Lagrange multiplier, $\Phi$ a  scalar
field and $\epsilon=\pm 1$. This approach is completely general
and  used in theoretical physics when we want to put in evidence
some specific feature \cite{lagrange}. In this case, we need it in order to derive
the physical gauge for the 5D--metric.
We can write down the metric as
\ba\label{a1"}
dS^2&=&g_{AB}dx^{A}dx^{B}= g\dab dx\ua dx\ub+g_{44}(dx^4)^2\nonumber\\ &&=g\dab
dx\ua dx\ub+\epsilon\Phi^2(dx^4)^2\,,\ea
from which we  obtain
directly particle-like solutions $(\epsilon=-1)$ or wave-like
solutions $(\epsilon=+1)$ in the 4D-reduction procedure. The
standard signature of 4D-component of the metric is $(+\,-\,-\,-)$
and $\alpha,\beta=0,1,2,3$. Furthermore, the 5D-metric can be
written in a Kaluza--Klein fashion as the matrix

\begin{equation}
\label{gmatrix}
 g_{AB}=\left(
\begin{array}{cc}
g\dab & 0\\ 0 & \epsilon\Phi^2
\end{array}
\right)\,,
\end{equation}
and the 5D-curvature Ricci tensor is
 \ba \label{a8}
^{(5)}R\dab&=&R\dab-\frac{\Phi_{,\alpha;\beta}}{\Phi}+\frac{\epsilon}{2\Phi^2}
\left(\frac{\Phi_{,4}g_{\alpha\beta,4}}{\Phi}-
g_{\alpha\beta,44}+\right.\nonumber\\ &&\left.+g^{\lambda\mu}g_{\alpha\lambda,4}
g_{\beta\mu,4}-\frac{g\umunu g_{\mu\nu,4}g_{\alpha\beta,4}}{2}
\right)\,,\ea
 where $R\dab$ is the 4D-Ricci tensor.   The
expressions for $^{(5)}R_{44}$ and $^{(5)}R_{4\alpha}$ can be
analogously derived. After the projection
from 5D to 4D, $g\dab$, derived from $g_{AB}$, no longer
explicitly depends on $x^4$, so, from Eq.(\ref{a8}),  a
useful  expression for the Ricci scalar can be derived:
\be
\label{ricci} ^{(5)}R=R-\frac{1}{\Phi}\Box \Phi\,,
\ee
 where the
dependence on $\epsilon$ is explicitly disappeared and $\Box$ is
the 4D-d'Alembert operator which gives $\Box\Phi\equiv
g\umunu\Phi_{,\mu ;\nu}$. The action in Eq.(\ref{a6}) can be
recast in a 4D-reduced Brans-Dicke  action of the form
\be\label{e1} {\cal A}=-\frac{1}{16\pi G_N}\int
d^{4}x\sqrt{-g}\left[\Phi R+{\cal L}_{\Phi}\right]\,, \ee where
the Newton constant is given by \be \label{450}
G_N=\frac{^{(5)}G}{2\pi l}\ee with $l$ a characteristic length in
5D which can be related to a suitable Compton length. Defining a generic function of a 4D-scalar
field $\phi$ as
\be\label{pippo} -\frac{\Phi}{16\pi G_N}=\f\,, \ee we get, in 4D, a
general action in which gravity is non-minimally coupled to a
scalar field, that is
\ba \label{2.1.1} {\cal A}&=&\int_{\cal
M}d^{4}x\sqrt{-g}\left[\f R+\frac{1}{2}\gu\p\ddemu\p\ddenu
-\v\right]+\nonumber\\ &&+ \int_{\pa {\cal M}}d^{3}x\sqrt{-b}K\,, \ea
where  the
form and the role of  $\v$ are still general. The second integral
is a boundary term where
 $K\equiv h^{ij}K_{ij}$ is the trace of the extrinsic curvature tensor
 $K_{ij}$ of the hypersurface
$\pa {\cal M}$ which is embedded in the 4D-manifold ${\cal M}$;
$b$ is the metric determinant of the 3D-manifold.

The Einstein field equations can be derived by varying with respect to the
4D-metric $\gd$
 \be \label{2.1.2} G\dmunu=R\dmunu-\frac{1}{2}\gd R=\tilde{T}\dmunu\,, \ee
where
\ba \label{2.1.4}
\tilde{T}\dmunu&=&\frac{1}{\f}\left\{-\frac{1}{2}\p\ddemu\p\ddenu
+\frac{1}{4}\gd\p\ddea\p\udea+\right.\nonumber\\&&\left.-\frac{1}{2}\gd\v
-\gd\Box\f+\f\ddemunu\right\} \ea
 is the  effective
stress--energy tensor containing the non-minimal coupling
contributions, the kinetic terms and the potential of the scalar
field $\p$. In the case in which $\f$ is a constant $F_{0}$ (in
our units, $F_{0}=-1/(16\pi G_N)$), we get the
stress--energy tensor of  a scalar field  minimally coupled to
gravity, that is \be \label{2.1.5}
T\dmunu=\p\ddemu\p\ddenu-\frac{1}{2}\gd\p\ddea\p\udea+\gd\v\,.
\ee By varying with respect to $\p$, we get the 4D-Klein--Gordon
equation \be \label{2.1.6} \Box\p-R\fp+\vp=0, \ee where
$\fp=d\f/d\p$ and  $\vp=d\v/d\p$.  It
is possible to show that Eq.(\ref{2.1.6})  is nothing else but the
contracted Bianchi identity.
This feature shows that the effective stress--energy tensor at
right hand side of (\ref{2.1.2}) is a zero--divergence tensor and
this fact is fully compatible with Einstein theory of gravity also
if we started from a 5D-space. Specifically, the reduction
procedure, which we have used, preserves the standard features of
GR since we are in the realm of the conformal-affine structure discussed above.

In order to physically
identify  the fifth dimension,  let us recast
the above Klein-Gordon equation (\ref{2.1.6}) as \be
\label{kg1} \left(\Box + m_{eff}^2\right)\p=0\,,\ee where
\be\label{kg2}m_{eff}^2=\left[\vp-R\fp\right]\phi^{-1}\,,\ee is the
effective mass, \ie a function of $\phi$, where self-gravity
contributions, $R\fp$, and  scalar field self-interactions, $\vp$,
 are taken into account.   In any quantum field theory
formulated on curved space-times, these contributions, at
one-loop level, have the same "weight" \cite{birrell}. We want to show that a "natural" way to generate
particle  masses  can be achieved starting from a 5D picture. In other words,  the
concept of {\it mass} can be derived from a  geometric viewpoint.

\section{Extended Theories of Gravity}
\label{sei}

\subsection{Effective gravity in 4D}
\label{sei.uno}

The above scalar-tensor  action  for gravitational field is a particular case of a general
class of  effective theories in four dimensions which can be achieved by  the reduction procedure. In general,
starting from any higher-dimensional theory, we can derive the class of effective actions \cite{book,odino}
  \begin{eqnarray} \label{3.1.1.} {\cal A}&=&\int
d^{4}x\sqrt{-g}\left[F(R,\Box R,\Box^{2}R,..\Box^kR,\p)+\right. \nonumber\\ &&\left.
 -\frac{\epsilon}{2}
g\umunu \phi\ddemu \phi\ddenu+ {\l}_{m}\right], \end{eqnarray}
 where $F$ is
a generic function   of curvature invariants\footnote{Other curvature invariants like $R_{\mu\nu}R^{\mu\nu}$,
$R_{\mu\nu\alpha\beta}R^{\mu\nu\alpha\beta}$,
$C_{\mu\nu\alpha\beta}C^{\mu\nu\alpha\beta}$  are also
possible, as we will show below. }
and  scalar
field $\p$.   It is interesting to point out that this theory looks similar to non-local gravities of
general sort introduced in \cite{iodin} and \cite{iodur}.
The term ${\l}_{m}$, as above, is the minimally
coupled ordinary matter contribution. We shall use physical units
$8\pi G=c=\hbar=1$;
 $\epsilon$ is a constant which specifies the theory. Actually its
 values can be $\epsilon =\pm 1,0$ fixing the nature and the
 dynamics of the scalar field which can be a standard scalar
 field, a phantom field or a field without dynamics
 \cite{valerio,CP} .

In the metric approach, the field equations are obtained by
varying (\ref{3.1.1.}) with respect to  $\gd$.  We get \begin{eqnarray}
\label{3.2.2.} G\umunu&=&\frac{1}{{\cal
G}}\left[T\umunu+\frac{1}{2}\gu (F-{\cal G}R)+
(g^{\mu\lambda}g^{\nu\sigma} -\right.\nonumber\\&&\left.+ \gu g^{\lambda\sigma})
{\cal G}_{;\lambda\sigma} +\frac{1}{2}\sum_{i=1}^{k}\sum_{j=1}^{i}(\gu
g^{\lambda\sigma}+\right.\nonumber\\ &&\left.+
  g^{\mu\lambda} g^{\nu\sigma})(\Box^{j-i})_{;\sigma}
\left(\Box^{i-j}\frac{\pa F}{\pa \Box^{i}R}\right)_{;\lambda}-\right.\nonumber\\ && \left. \gu g^{\lambda\sigma}\left((\Box^{j-1}R)_{;\sigma}
\Box^{i-j}\frac{\pa F}{\pa \Box^{i}R}\right)_{;\lambda}\right]\,,
\end{eqnarray}
 where $G\umunu$ is the above Einstein tensor and \beq
\label{3.4}
  {\cal G}\equiv\sum_{j=0}^{n}\Box^{j}\left(\frac{\pa F}{\pa \Box^{j} R}
\right) \eeq
is a scalar function which determines the coupling.

The differential Eqs.(\ref{3.2.2.}) are of order
$(2k+4)$. The stress-energy tensor is due to the kinetic part of
the scalar field and to the ordinary matter: \beq \label{3.5}
T\dmunu=T^{m}\dmunu+\frac{\epsilon}{2}[\p\ddemu\p\ddenu-
\frac{1}{2}\p\udea\p\ddea]\;. \eeq The (possible) contribution of
a potential $\v$ is contained in the definition of $F$. From now
on, we shall indicate by a capital $F$ a Lagrangian density
containing also the contribution of a potential $\v$ and by
$F(\p)$, $f(R)$, or $f(R,\Box R)$ a function of such fields
 without potential.

By varying with respect to the scalar field $\p$, we obtain the
Klein-Gordon equation \beq \label{3.6} \epsilon\Box\p=-\frac{\pa
F}{\pa\p}\,. \eeq

The simplest class of (\ref{3.1.1.}) theories  extending  GR is achieved assuming \beq\label{fr}
F=f(R)\,,\qquad \epsilon=0\,,\eeq in the action (\ref{3.1.1.});
 $f(R)$ is an arbitrary (analytic) function of the Ricci
curvature scalar $R$.   The standard
Hilbert-Einstein action is  recovered for $f(R)=R$.
Varying with respect to $g\dab$, we get the field equations
\begin{equation}\label{h2}
f'(R)R\dab-\frac{1}{2}f(R)g\dab=f'(R)^{;\umunu}\left(
g_{\alpha\mu}g_{\beta\nu}-g\dab\gd\right)\,, \end{equation} which
are fourth-order equations due to the term $f'(R)^{;\mu\nu}$; the
prime  indicates the derivative with respect to  $R$.
Eq.(\ref{h2}) is also the equation for $T\dmunu=0$ when the matter
term is absent.

By a suitable manipulation, the above equation can be rewritten
as:
 \begin{eqnarray}\label{h4}
G\dab&=&\frac{1}{f'(R)}\left\{\frac{1}{2}g\dab\left[f(R)-Rf'(R)\right]
+f'(R)\ddeab -\right.\nonumber\\ &&\left.+g\dab\Box f'(R)\right\}+ \frac{T^{m}_{\alpha
\beta}}{f'(R)}=T^{curv}_{\alpha\beta}+\frac{T^{m}_{\alpha
\beta}}{f'(R)}\,,\nonumber\\
\end{eqnarray}
where $T^{curv}_{\alpha\beta}$ is an effective stress-energy
tensor constructed by the extra curvature terms and standard matter contribution has been also considered.  In the case of
GR,   $T^{curv}_{\alpha\beta}$ identically vanishes while the
standard, minimal coupling is recovered for the matter
contribution. The peculiar behavior of GR, that is $f(R)=R$, is  due to the
particular form of the Lagrangian itself which, even though it is
a second order Lagrangian, can be non-covariantly rewritten as the
sum of a first order  Lagrangian plus a pure divergence term. The
Hilbert-Einstein Lagrangian can be recast as follows:
 \begin{eqnarray}
{\cal L}_{HE}   &=&\sqrt{-g}\Big[ p^{\alpha \beta}
(\Gamma^{\rho}_{\alpha \sigma} \Gamma^{\sigma}_{\rho
\beta}-\Gamma^{\rho}_{\rho \sigma} \Gamma^{\sigma}_{\alpha
\beta})+\Big.\nonumber\\ && \Big.+ \nabla_\sigma (p^{\alpha \beta} {u^{\sigma}}_{\alpha
\beta}) \Big]\,,
\end{eqnarray}
\noindent where:
\begin{equation}
 p^{\alpha \beta} =\sqrt{-g}  g^{\alpha \beta} = \frac{\pa {\cal{L}}_{HE}}{\pa R_{\alpha \beta}}\,,
\end{equation}
$\Gamma$ is the Levi-Civita connection of $g$ and
$u^{\sigma}_{\alpha \beta}$ is a quantity constructed from the
variation of $\Gamma$ \cite{weinberg}. Since $u^{\sigma}_{\alpha
\beta}$ is not a tensor, the above expression is not covariant;
however a standard procedure has been studied to recast covariance
in the first order theories \cite{FF1}. This clearly shows that
the field equations should consequently be second order  and the
Hilbert-Einstein Lagrangian is thus degenerate. In other words, we can say that the only
degenerate theory of gravity (with respect to the class of $f(R)$-gravity) is GR being its Hessian determinant null.

From the action (\ref{3.1.1.}), it is possible to obtain the
 case discussed in the previous section by choosing \beq F=\f R-V(\phi)\,,\qquad \epsilon
=-1\,,\eeq and then
\begin{equation} \label{s1} {\cal A}=  \vol \left[F(\phi) R+ \half \gu
\phi\ddemu \phi\ddenu- V(\phi) \right]\,. \end{equation}
Several other interesting cases can be obtained by suitable choices of the function $F$. However, all these models can be related by conformal-affine transformations.

\subsection{Conformal Transformations}
\label{sei.due}

It is possible to show that any
higher-order or scalar-tensor theory, in absence of ordinary
matter, e.g. a perfect fluid,  is conformally equivalent to an
Einstein theory plus minimally coupled scalar fields (see \cite{book,odino} for details). If standard
matter is present, conformal transformations allow to transfer
non-minimal coupling to the matter component \cite{magnano-soko}.
The conformal transformation on the metric $\gd$ is
\begin{equation} \label{s6}
\tilde{g}\dmunu= e^{2 \ome} \gd\,,
\end{equation}
where $e^{2 \ome}$ is the conformal factor. Under this
transformation,  the Lagrangian  in (\ref{s1}) becomes
\begin{eqnarray} \label{s7}
&& \sqrt{-g} \left(F R+ \half \gu \phi\ddemu \phi\ddenu- V\right
) = \sqrt{-\tilde{g}} e^{-2 \ome} \left(F \tilde{R}-\right.\nonumber\\ &&\left. + 6 F \Box_{\tilde{g}}
\ome  -6 F \ome\ddea \ome\udea+ \half \tilde{g}\umunu
\phi\ddemu \phi\ddenu- e^{-2 \ome} V\right)\,,\nonumber\\
\end{eqnarray}
in which $\tilde{R}$ and $\Box_{\tilde{g}}$ are the Ricci scalar
and the d'Alembert operator  relative to the metric $\tilde{g}$.
Requiring the theory in the metric $\tilde{g}\dmunu$ to appear as
a standard Einstein theory  \cite{conf1},  the conformal factor
has to be related to $F$, that is
\begin{equation} \label{s8} e^{2 \ome}= -2
F. \end{equation} where $F$ must be negative in order to restore
physical coupling in our adopted signature. Using this relation and
 introducing a new scalar field $\tilde{\phi}$ and a new potential $\tilde{V}$,
defined respectively by
\begin{equation} \label{s10} \tilde{\phi}\ddea=
\sqrt{\frac{3F_{\phi}^2- F}{2 F^2}}\, \phi\ddea,~~~
\tilde{V}(\tilde{\phi}(\phi))= \frac{V(\phi)}{4 F^2(\phi)},
\end{equation}  the Lagrangian  (\ref{s7})
becomes
\begin{eqnarray} \label{s11} \sqrt{-g}
&&\left( F R+ \half \gu \phi\ddemu \phi\ddenu- V\right) =\nonumber\\ &&=
\sqrt{-\tilde{g}} \left( -\half \tilde{R}+ \half \tilde{\phi}\ddea
\tilde{\phi}\udea- \tilde{V}\right)\,, \nonumber \end{eqnarray}
 which
is the standard Hilbert-Einstein Lagrangian plus the
Lagrangian  of a minimally coupled  scalar field $\tilde{\phi}$. Therefore,
every non-minimally coupled scalar-tensor theory, in absence of
ordinary matter, e.g. perfect fluid,  is conformally equivalent to
an Einstein theory, being the conformal transformation and the
potential suitably defined by (\ref{s8}) and (\ref{s10}). The
converse is also true: for a given $F(\phi)$, such that $3 F_{\phi}^2-
F> 0$, we can transform a standard Einstein theory into a
non-minimally coupled scalar-tensor theory. This means that, in
principle, if we are able to solve the field equations in the
framework of the Einstein theory in presence of a scalar field
with a given potential, we should be able to get the solutions for
the scalar-tensor theories, assigned by the coupling $F(\phi)$, via
the conformal transformation  (\ref{s8}) with the constraints
given by (\ref{s10}). Following the standard terminology, the
``Einstein frame'' is the framework of the Einstein theory with
the minimal coupling and the
 ``Jordan frame'' is the framework of the non-minimally coupled theory
\cite{magnano-soko}.

Performing the conformal
transformation (\ref{s6}) in the case of $f(R)$-gravity, the field equations (\ref{h4}) become:
\begin{eqnarray}\label{hl4}
\tilde{G}\dab&=&\frac{1}{f'(R)}\left\{\frac{1}{2}g\dab\left[f(R)-
Rf'(R)\right] +f'(R)\ddeab- \right.\nonumber\\ && \left.+g\dab\Box f'(R)\right\}+
2\left(\omega_{;\alpha ;\beta}
+g\dab\Box \omega -\omega_{;\alpha}\omega_{;\beta}+\right.\nonumber\\ && \left.+
\frac{1}{2}g\dab\omega_{;\gamma}\omega^{;\gamma}\right)\,.\nonumber\\
\end{eqnarray}
The conformal factor is
 \begin{equation}\label{h5}
\omega=\frac{1}{2}\ln |f'(R)|\,, \end{equation} which has  to
be substituted into (\ref{h4}). Rescaling $\omega$ in such a way
that
\begin{equation}\label{h6} k\phi =\omega\,, \end{equation} and $k=\sqrt{1/6}$, we obtain
the Lagrangian equivalence \begin{equation} \label{h7} \sqrt{-g}
f(R)= \sqrt{-\tilde{g}} \left( -\half \tilde{R}+ \half
{\phi}\ddea {\phi}\udea- \tilde{V}\right)\,, \end{equation}
and the Einstein equations in standard form
\begin{equation}\label{h8} \tilde{G}\dab=
\phi\ddea\phi\ddeb-\frac{1}{2}\tilde{g}\dab\phi_{;\gamma}\phi^{;\gamma}
+\tilde{g}\dab V(\phi)\,, \end{equation}  with the potential
\begin{eqnarray}\label{h9}
V(\phi)&=&\frac{e^{-4k\phi}}{2}\left[{\cal P}(\phi)-{\cal
N}\left(e^{2k\phi}\right)e^{2k\phi}\right]
=\nonumber\\ &&=\frac{1}{2}\frac{f(R)-Rf'(R)}{f'(R)^{2}}\,.
\end{eqnarray}
Here ${\cal N}$ is the inverse function of
 ${\cal P}'(\phi)$ and ${\cal P}(\phi)=\int \exp (2k\phi) d{\cal N}$. However, the
problem is
 completely solved if
${\cal P}'(\phi)$ can be analytically inverted. In summary, a
fourth-order theory is conformally equivalent to the standard
second-order Einstein theory plus a scalar field.

This procedure can be always extended to more general theories. If the
theory is assumed to be higher than fourth order, we may have
Lagrangian densities of the form,
\begin{equation}\label{h10} {\cal L}={\cal L}(R,\Box R,...\Box^{k} R)\,. \end{equation}
As we have seen, any $\Box$ operator introduces two further terms of derivation
into the field equations. For example a theory like
\begin{equation}\label{h11} {\cal L}=R\Box R\,, \end{equation} is a sixth-order theory
and the above approach can be pursued by considering a conformal
factor of the form
 \begin{equation}
\label{h12} \omega=\frac{1}{2}\ln \left|\frac{\pa {\cal L}}{\pa R}
+\Box\frac{\pa {\cal L}}{\pa \Box R}\right|\,.
\end{equation}
In general,  increasing two orders of derivation in the field
equations (\ie for every term $\Box R$), corresponds to adding a
scalar field in the conformally transformed frame. A sixth-order theory can be reduced to an
Einstein theory plus two minimally coupled scalar fields; a
$2n$-order theory can be, in principle, reduced to an Einstein
theory plus $(n-1)$-scalar fields. On the other hand, these
considerations can be directly generalized to
higher-order-scalar-tensor theories in any number of dimensions.   Due to the conformal transformations (and the conformal--affine invariance), the physical information remains the same. This means that any two further  degrees of freedom   can be recast as a scalar field (of gravitational origin)  whose dynamics is given by an effective Klein-Gordon equation. We want to show that such further gravitational degrees of freedom can play a fundamental role inducing spontaneous symmetry breaking and then generating the mass of observed particles.

\section{ The generation of masses}
\label{sette}


 \subsection{Particle masses  as eigenstates from 5D}
 \label{sette.uno}

The above considerations allow a straightforward mechanism for the generation of masses that can be related to the projection from 5D to 4D-manifolds.
Let us start with a toy model that can immediately illustrate the mechanism. In a flat 5D-space-time,
 the 5D d'Alembert operator can be split, following the metric definition
 (\ref{a1"}) for particle-like solutions, as:
\be ^{(5)}\Box=\Box-{\partial_4}^2 \,,\ee   selecting the
value $\epsilon=-1$  in the metric.
 Introducing the scalar field $\Phi$,  generated by the projection through the Lagrange multiplier in the action (\ref{a5}), we have
\be\label{embedding}
^{(5)}\Box\Phi=\left[\Box-{\partial_4}^2\right]\Phi=0 \,,\ee and
then \be\label{splitt} \Box\Phi={\partial_4}^2\Phi\,. \ee
The
problem is solvable by separation of variables since the metric matrix (\ref{gmatrix} ) is diagonal in the fifth component. We split
the scalar field $\Phi$ into two functions
\be\label{split1}
\Phi=\p(t,\vec{x})\psi(x_4)\,,
\ee
where the field $\p$ depends
on the ordinary space-time coordinates, while $\chi$ is a function
of the fifth coordinate $x_4$. Inserting (\ref{split1}) into
Eq.(\ref{splitt}), we get \be\label{split2}
\frac{\Box\p}{\p}=\frac{1}{\psi}\left[\frac{d^2\psi}{dx_4^2}\right]=-k_n^2\,,
\ee where $k_n$ must be a constant for consistency. From
Eq.(\ref{split2}), we obtain the two equations of motion \be
\label{kg5} \left(\Box + k_n^2\right)\p=0\,,\ee and

\be \label{oscillation}\frac{d^2 \psi}{dx_4^2}+k_n^2\psi=0\,.\ee
Eq.(\ref{kg5}) is the evolution equation for  the further gravitational degrees of freedom that we will discuss below.
Eq.(\ref{oscillation}) describes a harmonic oscillator whose
general solution is
\be \label{massol}\psi(x_4)=c_1
e^{-i{k_n}{x_4}}+c_2 e^{i{k_n}{x_4}}\,.
\ee
The constant $k_n$
has the physical dimension of the inverse of a length and,
assigning boundary conditions, we can derive the eigenvalue
relation  \be k_n=\frac{2\pi}{l}n\,,\ee where $n$ is an integer
and $l$ a length which we have previously defined in
Eq.(\ref{450}). As a result, in standard units, we  recover
 the Compton length
\be\lambda_n=\frac{\hbar}{2\pi m_n c}=\frac{1}{k_n}\,, \ee which
 assigns the mass of a particle. It has to be stressed
that, the eigenvalues of Eq.(\ref{oscillation}) are the masses of
particles which are generated by the reduction process
from 5D to 4D.
On the other hand, different values of $n$ fix the families of particles,
while, for any given value of $n$, the parameters
$c_{1,2}$ distinguish a particle and an antiparticle  within a family.
This toy model can be refined by considering other quantum numbers of particles. For the moment, we are interested only in the problem of mass generation.

 \subsection{ 4D-dynamics  of the scalar field $\Phi$}
 \label{sette.due}

Dynamics of  4D-component of the induced scalar field $\Phi$, see Eq.(\ref{split1}),  can be related to the space-time deformations  discussed above. In this way,
the role of $GL(4)$-group of diffeomorphism will result of fundamental importance.  Let us start our consideration by showing how particles can acquire mass by deformations and let us relate the procedure with the above reduction mechanism.
 As standard,  a particle with  zero mass is characterized by the invariant relation
\begin{equation}\label{rm1}
     \eta^{\alpha\beta}p_{\alpha}p_{\beta}=0\,,
\end{equation}
in the Minkowski space.
Deforming the space-time  and considering the above operators, one has
\begin{equation}\label{rm2}
     \eta^{AB}\Phi_{A}^{C}\Phi_{B}^{D}e_{C}^{\alpha}e_{D}^{\beta}p_{\alpha}p_{\beta}=g^{\alpha\beta}p_{\alpha}p_{\beta}=0\,,
\end{equation}
so we have defined two frames, one, the Minkowski space-time, defined by the metric $\eta^{\alpha\beta}$ and the other defined by the metric $g^{\alpha\beta}$, generated by the projection from 5D. The two frames are related by the matrices of deformation functions $\Phi_{A}^{C}(x)$. In both frames the massless particle follow a null path, but we observe that using the decomposition (\ref{controvariantdecomposition}) the particle does not appear massless with respect to the first frame. As  matter of fact, Eq. (\ref{rm2}) becomes
\begin{equation}\label{rm3}
    \Omega^{-2}\eta^{\alpha\beta}p_{\alpha}p_{\beta}+\chi^{\alpha\beta}p_{\alpha}p_{\beta}=0\,,
\end{equation}
so that
\begin{equation}\label{rm33}
     \eta^{\alpha\beta}p_{\alpha}p_{\beta}=-\Omega^{2}\chi^{\alpha\beta}p_{\alpha}p_{\beta} \neq 0.
\end{equation}
and now in the first frame we are able to define a rest reference system for the particle.

For a massless particle it is not possible to define a rest reference system since considering the invariant relation,
\begin{equation}\label{rm4}
    g^{00}p^{2}_{0}+ 2g^{0i}p_{0}p_{i}+g^{ij}p_{i}p_{j}=0\,,
\end{equation}
and defining as  rest frame the system in which $p_{j}=0$,  the  solution exists  only for  $p_{0}=0$ {\it i.e.} only the trivial solution  $p_{\alpha}=0$ satisfies the `` rest reference frame'' condition.

On the other hand, if we consider massive particle, then
\begin{equation}\label{rm5}
     g^{00}p^{2}_{0}+ 2g^{0i}p_{0}p_{i}+g^{ij}p_{i}p_{j}=m^{2}\,,
\end{equation}
the rest frame is characterized by the conditions $p_{j}=0$ and consequently $g^{00}p^{2}_{0}=m^{2}$. This means that  the (squared) mass is proportional to the (squared) energy
and the solution is no more trivial.

To overcome this problem, let us fix the deformation such that
\begin{equation}\label{rm6}
     -\Omega^{2}\chi^{\alpha\beta}p_{\alpha}p_{\beta} = m^{2}\,,
\end{equation}
$m$ being the mass "attribute" to the particle.

In the second frame we have
\begin{equation}\label{rm7}
  p^{2}_{0}-{\vec{p}}^{\;2}+\Omega^{2}\chi^{00} p^{2}_{0} + 2\chi^{0i}p_{0}p_{i}+\chi^{ij}p_{i}p_{j}=0\,,
\end{equation}
if $\vec{p}=0$ then
\begin{equation}\label{rm8}
     p^{2}_{0}(1+\Omega^{2}\chi^{00})=1\,,
\end{equation}
which implies, besides the trivial solution, also the condition $\Omega^{2}\chi^{00}=-1$ which, together with Eq. (\ref{rm6}),  gives
\begin{equation}\label{rm9}
    p^{2}_{0}=m^{2},
\end{equation}
the rest system condition in the first frame.

It is also possible to  show that there is an equivalence between deforming a metric and giving mass to a  massless particle.
As we have shown,  we have described the space-time deformation by using  matrices of scalar fields.  The same problem can be
addressed in terms of space-time  tensors
\begin{equation}
\label{(36)}
\eta_{AB}\Phi^A_C\Phi^B_D e^C_\mu e^D_\nu =
 g_{\alpha\beta}\Phi^\alpha_\mu \Phi^\beta_\nu = \tilde g_{\mu\nu}\,,
\end{equation}
where we have used the tetrad
\begin{equation}
\label{rm10}
e^\mu_A\, e^B_\mu = \delta^B_A\,.
\end{equation}
It should be observed that the $\Phi^\alpha_\mu$ do not represent coordinates transformations
as far as they cannot be reduced to Jacobian matrices. This means that $\Phi^\alpha_\mu$ have a fundamental physical meaning.
Starting from  $\tilde g$ we observe that if $g_{\mu\nu}\,p^\mu\, p^\nu = 0$
then $\tilde g_{\mu\nu}\,p^\mu\, p^\nu \neq 0$ when $\tilde g$ is a general deformation that can be related to the  conformal transformations  of $g$.
Eq. (\ref{(36)}) tells us that, equivalently, we can read the deformation as a transformation
of the 4-momentum of the particle
\begin{equation}
\label{rm11}
\tilde g_{\mu\nu}\,p^\mu\, p^\nu =g_{\alpha\beta}\Phi^\alpha_\mu \Phi^\beta_\nu  p^\mu\, p^\nu  \equiv  g_{\alpha\beta}\,\tilde p^\alpha\, \tilde p^\beta\,,
\end{equation}
in such a way there exists deformations of space-time which give mass to massless particles. In other words, deformations, that  are elements of the conformally-invariant $GL(4)$-group can be related, in principle, to the generation of   the masses of particles.

\subsection{A Lagrangian Approach for mass generation from deformations }
\label{sette.tre}

So far we have considered the geometrical definition of mass from the point of view of relativistic mechanics.
Now we would like to extend this result to classical field theory with the aim to extend it to quantum field theory.

Let us consider a scalar free massless particle. It can be described by the Lagrangian
\begin{eqnarray}
\label{la1}
{\cal L} &=& \frac{1}{2}\sqrt{-g}g^{\mu\nu}(\partial_\mu \phi )(\partial_\nu \phi )\,=\nonumber\\ && =\frac{1}{2}\sqrt{-g}\left(\eta^{\mu\nu}+\chi^{\mu\nu}\right)(\partial_\mu \phi )(\partial_\nu \phi ).
\end{eqnarray}
It is well-known that the free propagator has a pole in $p^2=0$.
In order   to introduce a mass term in the Lagrangian  and in the field equations, we have to eliminate a divergence from it, that is
considering
\begin{eqnarray}\label{ft1}
&&  \partial_{\mu}\left[ \partial_{\nu} \sqrt{-g}\,\chi^{\mu\nu}\phi^{2}\right]=\partial_{\mu}\partial_{\nu}\left(\sqrt{-g}\,\chi^{\mu\nu}\right)\phi^{2}
  +\nonumber\\ &&+4\,\partial_{\nu}\left(\sqrt{-g}\,\chi^{\mu\nu}\right)\,\phi\,\partial_{\mu}\,\phi
  +2\sqrt{-g}\,\chi^{\mu\nu}\partial_{\mu}\phi\;\partial_{\nu}\phi+\nonumber\\ &&+2\;\sqrt{-g}\,\chi^{\mu\nu}\phi\;\partial_{\mu}\partial_{\nu}\phi\,,
\end{eqnarray}
 the Lagrangian  takes the form
\begin{eqnarray}\label{ft2}
\widetilde{{\cal L}} &=& \frac{1}{2}\sqrt{-g}\left(\eta^{\mu\nu} \right)(\partial_\mu \phi )(\partial_\nu \phi )- \frac{1}{4}\partial_{\mu}\partial_{\nu}\left(\sqrt{-g}\,\chi^{\mu\nu}\right)\phi^{2}
  -\nonumber\\ &&+ \,\partial_{\nu}\left(\sqrt{-g}\,\chi^{\mu\nu}\right)\,\phi\,(\partial_{\mu}\,\phi)
  - \frac{1}{2}\;\sqrt{-g}\,\chi^{\mu\nu}\phi\;\partial_{\mu}\partial_{\nu}\phi.\nonumber\\
\end{eqnarray}
In this way a ``mass'' term $m^{2}=\frac{1}{2}\partial_{\mu}\partial_{\nu}\left(\sqrt{-g}\,\chi^{\mu\nu}\right)$ can be defined in a new Lagrangian.
On the other hand,  considering  an action
\begin{equation}\label{ft3}
    {\cal A}=\int\sqrt{- g}\widetilde{\mathcal{L}}\,.
\end{equation}
%
 %
 and a variational principle
 \begin{equation}\label{ft4}
    \frac{\delta {\cal A}}{\delta \phi}= 0\,,
 \end{equation}
 implies  the equation
 \begin{equation}\label{ft5}
    \frac{\partial \widetilde{\mathcal{L}}}{\partial \phi}-\partial_{\alpha}\frac{\partial \widetilde{\mathcal{L}}}{\partial (\partial_{\alpha}\phi)}+\partial_{\alpha}\partial_{\beta}\frac{\partial \widetilde{\mathcal{L}}}{\partial (\partial_{\alpha}\partial_{\beta}\phi)}=0\,,
 \end{equation}
 which gives
 \begin{equation}\label{ft6}
    \partial_{\mu}\left(\sqrt{-g}g^{\mu\nu}\partial_\nu \phi \right)=0\,,
 \end{equation}
or
\begin{equation}\label{ft7}
    \Box\phi=\eta^{\mu\nu}\partial_{\mu}\partial_{\nu}\phi+\Omega^{2}\chi^{\mu\nu}\partial_{\mu}\partial_{\nu}\phi=0\,.
 \end{equation}
It is well-known that this equation cannot give, in general, a potential or a mass term, except when we take as a solution a plane wave $\phi=\exp i k_{\mu} x^{\mu}$. In this case the $\chi$ part of the equation can be interpreted as a mass term according to
\begin{equation}\label{ft8}
    \chi^{\mu\nu}\partial_{\mu}\partial_{\nu}\phi = -\Omega^{2}\chi^{\mu\nu}k_{\mu}k_{\nu} \exp i k_{\mu} x^{\mu}=m^{2}\phi\,.
\end{equation}
This equation can be compared with  Eq. (\ref{rm6}), previously derived  for a relativistic particle. We can extend this result to each function which can be expressed by a Fourier transform,
\begin{equation}\label{ft9}
\phi(x)=\int\exp (i k_{\mu} x^{\mu}) \tilde{f}(k)d^{4}k\,,
\end{equation}
where the space-time dependent mass term is defined by the equation
\begin{equation}\label{ft10}
    m^{2}(x)=-\int \chi^{\mu\nu}k_{\mu}k_{\nu}\exp (i k_{\mu} x^{\mu}) \tilde{f}(k)d^{4}k.
\end{equation}
With these restrictions, the Klein-Gordon equation for a massless particle in the $g$-frame is seen in the $\eta$-frame as a {\it massive particle} as soon as the plane wave solutions are considered.  This result makes more sense when the mass is interpreted as a  quantum effect.

In the case  a curved space-time, the above arguments can be implemented by the substitutions of operators
 \begin{eqnarray}
 \nonumber   \eta &\to&  g \\
  \nonumber   \partial&\to& \nabla\,.
 \end{eqnarray}
It is important to note that the scalar field equation is not conformally invariant. In order to have a conformally invariant scalar field equation, it is necessary to introduce, in the Lagrangian density a non-minimal coupling between geometry and field. A possible choice is
\begin{equation}
\label{nmc1}
{\cal L} = \frac{1}{2}\sqrt{-g}g^{\mu\nu}(\partial_\mu \phi )(\partial_\nu \phi )-\xi R(x) \phi^{2}\,,
\end{equation}
that perfectly fits requirements of action (\ref{3.1.1.}) with suitable choices .
By varying with respect to   $\phi$, the Klein-Gordon equation
\begin{equation}\label{nmc2}
    \Box \phi + \xi R \phi=0\,,
\end{equation}
is recovered.
Minimal coupling is obtained for $\xi=0$. Conformal coupling is recovered for
\begin{equation}\label{nmc3}
    \xi=\frac{1}{4}\left[ \left(N-2\right)\left(N-1\right)\right]\,,
\end{equation}
in an $N$ dimensional space-time \cite{birrell}.
By analogy with the previous considerations, a deformation with constant curvature $R=m^{2}$
gives a mass term  in the original frame. However, we need also to  interpret   the other   terms appearing in the new frame.
Expanding  Eq. (\ref{nmc2}),
\begin{equation}\label{nmc4}
     \eta^{\mu\nu} \partial_{\mu}\partial_{\nu} \phi +  \chi^{\mu\nu} \partial_{\mu}\partial_{\nu} \phi +
     \left(\eta^{\mu\nu}+\chi^{\mu\nu}\right)\Gamma_{\mu\nu}^{\lambda}\partial_{\lambda} \phi +\xi R \phi=0\,,
\end{equation}
we need that the connection is compatible with the metric tensor $(\eta^{\mu\nu}+\chi^{\mu\nu}) $, that is
\begin{equation}\label{nmc5}
     \nabla \left(\eta^{\mu\nu}+\chi^{\mu\nu}\right)=0.
\end{equation}
In order to have a deformation defining a mass $m^{2}=R$ in the original frame, two conditions must be satisfied,
\begin{equation}\label{nmc6}
   R=R_0 >0
\end{equation}
where $R_0$ is a constant and
\begin{equation}\label{nmc66}
   \chi^{\mu\nu} \partial_{\mu}\partial_{\nu} \phi +
     \left(\eta^{\mu\nu}+\chi^{\mu\nu}\right)\Gamma_{\mu\nu}^{\lambda}\partial_{\lambda} \phi=0\,,
\end{equation}
which is a restriction on the deformation tensor $\chi^{\mu\nu}$. This condition allows to determine the mass by a conformal transformation.
In fact, in the new frame, the equation is
\begin{equation}\label{nmc7}
    g^{\alpha\beta}\nabla_{\alpha}\partial_{\beta}\phi+\xi R \phi=0\,,
\end{equation}
which, in the case of a conformal transformation
\begin{equation}\label{nmc8}
    g_{\alpha\beta}=\Omega^{2}\eta_{\alpha\beta}\,,
\end{equation}
becomes
\begin{equation}\label{nmc9}
    \Omega^{-2}\eta^{\alpha\beta}\left(\partial_{\alpha}\partial_{\beta}\phi + 2 \Omega^{-1}\partial^{\gamma}\Omega\partial_{\gamma}\phi\right)-6\xi\eta^{\alpha\beta}\Omega^{-3}\partial_{\alpha}\partial_{\beta}\Omega=0\,.
\end{equation}
It is equivalent to a massive scalar field equation (in the first frame),  if the condition
\begin{equation}\label{nmc10}
    2 \Omega^{-1}\partial^{\gamma}\Omega\partial_{\gamma}\phi -6\xi\eta^{\alpha\beta}\Omega^{-1}\partial_{\alpha}\partial_{\beta}\Omega=m^{2}\,,
\end{equation}
holds.
Since Eq. (\ref{nmc9}) is linear in $\phi$,  we can define $\phi=e^{i\lambda_{\alpha}x^{{\alpha}}}$. Eq. (\ref{nmc10}) becomes
\begin{equation}\label{nmc11}
  2 \Omega^{-1}\partial^{\gamma}\Omega \lambda_{\gamma} -6\xi\eta^{\alpha\beta}\Omega^{-1}\partial_{\alpha}\partial_{\beta}\Omega=m^{2}\,,
\end{equation}
then a solution is $ \Omega=e^{ik_{\alpha}x^{{\alpha}}} $ and the mass depends on the combination
\begin{equation}\label{nmc12}
    m^{2}=6\xi \eta^{\alpha\beta}k_{\alpha}k_{\beta}-2\eta^{\alpha\beta}k_{\alpha}\lambda_{\beta}.
\end{equation}
In this  example, the deformation is a specific complex function, but we can extend this result to any function  by taking
\begin{equation}\label{nmc13}
\Omega(x)=\int \tilde{\Omega}(k)e^{ikx}d^{4}k\,,
\end{equation}
and also  non-constant  masses can be  is obtained being
\begin{equation}\label{nmc14}
    m^{2}=\frac{\int \tilde{\Omega}(k)(6\xi \eta^{\alpha\beta}k_{\alpha}k_{\beta}-2\eta^{\alpha\beta}k_{\alpha}\lambda_{\beta})e^{ikx}d^{4}k}{\int \tilde{\Omega}(k)e^{ikx}d^{4}k}\,.
\end{equation}
These results mean that a geometrical  definition of  mass is always possible. It can be induced by  the conformal transformation $\Omega$ which is a restriction of the
deformations group $GL(4)$.  Summarizing we have shown  that:
\begin{itemize}
\item the fifth dimension of a reduction mechanism from 5D to 4D-manifolds can be interpreted as a mass generator:
\item the further degrees of freedom coming out from Extended Theories of Gravity ({\it i.e.} extensions of GR) have a physical meaning and cannot be simply gauged away;
\item as soon as particles acquire masses,  $GL(4)$ can induce symmetry breakings giving rise to the observed interactions.
\end{itemize}
In the next section, we will discuss and classify  the further degrees of freedom related to the  Extended Gravity. The aim is to show how they could be related to a sort of Higgs-like mechanism.

\section{The classification of gravitational modes in Extended Gravity}
\label{otto}

\subsection{Massive and massless modes of gravitational field}
\label{otto.uno}

The above considerations demonstrate that any reduction scheme from 5D to 4D induces effective theories of gravity where further gravitational modes have to be taken into account.
In this view, GR is an exception including only massless tensor modes. Furthermore,  GR is a degenerate theory where Hessian determinant is null preventing any quantization approach. With this situation is mind, it is natural to take into account  further gravitational modes, emerging from Extended Gravity,  that are  usually discarded. Such modes  could have interesting  effects at ultra-violet and infra-red scales and could play an important role both in  the Standard Model of particles  and in gravitational radiation \cite{felix}. In particular, they could  be connected to the symmetry breaking  in the  high energy limit (and  investigated at LHC) and have a signature in  the cosmological  stochastic background of gravitational waves \cite{francaviglia}.  We want to show that such statements are very general and only GR shows two polarization modes (being a degenerate, constrained theory).  As we are going to demonstrate, a generic Extended Theory of Gravity, constructed with a function of curvature invariants, shows six gravitational polarizations  consistently with the Riemann theorem \cite{riemann}.

Assuming  any  curvature invariant other than the Ricci scalar \footnote{ We restrict  to fourth-order theories which  give the main contributions in a renormalization process, but we can extended the following considerations to any higher-order theory involving generic powers of the $\Box$-operator.}
 a generic 4D-effective action for the gravitational interaction is
 \be {\cal A}=\int
d^4x\sqrt{-g} f(R,P,Q) \ee where \ba && P\equiv R_{\alpha\beta}R^{\alpha\beta}\nn \\
&&Q\equiv R_{\alpha\beta\gamma\delta}R^{\alpha\beta\gamma\delta} \ea
Varying with respect to the metric one gets the field equations:
 \ba FG_{\mu\nu}&=&\frac{1}{2}g_{\mu\nu}\left(f-
R~F\right)-(g_{\mu\nu}\Box-\na_\mu\na_\nu)F\nn\\&&
-2\left(f_P R^\alpha_\mu
R_{\alpha\nu}+f_Q~R_{\alpha\beta\gamma\mu}R^{\alpha\beta\gamma}_{~~~\nu} \right)\nn\\&&-
g_{\mu\nu}\na_\alpha\na_\beta(f_P R^{\alpha\beta})-\Box (f_P
R_{\mu\nu})\nn\\ &&+2\na_\alpha\na_\beta\left(f_P~R^\alpha_{~(\mu}\delta^beta_{~\nu)}+2
f_Q~R^{\alpha~~~~\beta}_{~(\mu\nu)}\right)\nn\\\label{fieldeqs}\ea
where we have set
  \be F\equiv\frac{\D f}{\D R}, ~~~f_P\equiv\frac{\D f}{\D P}, ~~~f_Q\equiv\frac{\D
f}{\D Q} \ee and, as above,  $\Box=g^{\alpha\beta}\na_\alpha\na_\beta$ is the d'Alembert
operator. The notation $T_{(\mu\nu)}=\frac{1}{2}(T_{\mu\nu}+T_{\nu\mu})$
denotes symmetrization with respect to the indices $(\mu,\nu)$.

Taking the trace of Eq. (\ref{fieldeqs}),  we find:
 \ba && \Box\left(F+\frac{f_P}{3}
R\right)=\nonumber\\&&
\frac{1}{3}\left[ 2 f-RF-2 \na_a\na_b((f_P+2f_Q)R^{\alpha\beta})+\right.\nonumber\\
&&\left.-2
(f_P P+f_Q Q)\right] \label{trace}\ea
Expanding the third term on the r.h.s. of (\ref{trace}) and using the
Bianchi identity $G^{\alpha\beta}_{~~;\beta}=0$, we get: \ba
&&\Box\left(F+\frac{2}{3}(f_P+f_Q) R\right)=
\frac{1}{3}\times\nn\\ && [2
f-RF-2R^{\alpha\beta}\na_\alpha\na_\beta(f_P+2f_Q)-R\Box(f_P+2f_Q)\nn\\ &&-2 (f_P
P+f_Q Q)] \label{trace1}\ea
If we define \ba \phi & \equiv&
F+\frac{2}{3}(f_P+f_Q) R \label{phidef} \\ && \textrm{and} ~~~ \nn
\\\frac{dV}{d\phi} & \equiv& \textrm{r.h.s. ~of~ (\ref{trace1})}\nn \ea
then we get a Klein-Gordon equation for the scalar field $\phi$:
\be \Box \phi = \frac{dV}{d\phi}\,. \ee
Considering the discussion in previous section, it is  clear that the scalar field $\phi$, defined in 4D,
assumes the role of a field induced by the further degrees of freedom of Extended Gravity.
Obviously, $\phi$ is identically zero in GR.

In order to classify  the gravitational
modes that can be obtained from this approach,  we need to linearize around the Minkowski background.
As discussed above, this means to take into account deformations. According to the above results, we assume
\ba g_{\mu\nu}&=&\eta_{\mu\nu}+h_{\mu\nu} \nn \\
\phi&=&\phi_0+\delta \phi \ea
Then, from Eq. (\ref{phidef}), we get
\be \delta \phi=\delta F+\frac{2}{3}(\delta f_P+\delta f_Q) R_0+
\frac{2}{3}(f_{P0}+f_{Q0}) \delta R \label{pertphi1}\ee
where $R_0
\equiv R(\eta_{\mu\nu})=0$ and similarly $f_{P0}=\frac{\D f}{\D
P}|_{\eta_{\mu\nu}}$  which is either constant or zero.  Note that the index  0 indicates evaluation with respect to
the Minkowski metric, that means no deformation.  $\delta
R$  denotes the first order perturbation on the Ricci scalar
which, along with the perturbed parts of the Riemann and Ricci
tensors, are given by:

\ba \delta R_{\mu\nu\rho\sigma}&=&\frac{1}{2}\left(\D_\rho \D_\nu
h_{\mu \sigma}+\D_\sigma \D_\mu h_{\nu \rho}-\D_\sigma \D_\nu
h_{\mu
\rho}-\D_\rho \D_\mu h_{\nu \sigma} \right) \nn\\
\delta R_{\mu\nu} &=& \frac{1}{2}\left(\D_\sigma \D_\nu
h^\sigma_{~\mu}+\D_\sigma \D_\mu h^\sigma_{~\nu}-\D_\mu \D_\nu
h-\Box h_{\mu \nu} \right)\nn\\
\delta R &=& \D_\mu \D_\nu h^{\mu \nu}-\Box h\nn\ea where
$h=\eta^{\mu \nu} h_{\mu \nu}$. The first term of Eq.
(\ref{pertphi1}) is \be \delta F=\frac{\D F}{\D R}|_0~\delta
R+\frac{\D F}{\D P}|_0~\delta P+\frac{\D F}{\D Q}|_0~\delta Q \ee
However, since $\delta P$ and $\delta Q$ are second order, we get
$\delta F\simeq F_{,R0}~ \delta R $ and
 \be
 \delta \Phi
=\left(F_{,R0} +\frac{2}{3} (f_{P0}+f_{Q0})\right) \delta R
\label{pertphi2}
\ee
Finally, from Eq. (\ref{trace1}) we get the
Klein-Gordon equation for the scalar perturbation $\delta \phi$
\ba \Box \delta \phi&=&\frac{1}{3}\frac{F_0}{F_{,R0} +\frac{2}{3}
(f_{P0}+f_{Q0})}\delta \phi-\nn\\ &&\frac{2}{3}\delta
{R}^{\alpha\beta}\D_\alpha\D_\beta(f_{P0}+2f_{Q0})-\frac{1}{3}\delta
{R}\Box(f_{P0}+2f_{Q0})\nn \\ &=& m_s^2 \delta \phi
\label{kgordon1}\nn\\ \ea The last two terms in the first line are
actually  zero since the terms $f_{P0}$, $f_{Q0}$ are constants
and we have defined the scalar mass as $m_s^2\equiv
\frac{1}{3}\frac{F_0}{F_{,R0} +\frac{2}{3} (f_{P0}+f_{Q0})}$.

Perturbing the field equations (\ref{fieldeqs}) we get: \ba &&
F_0(\delta{R}_{\mu\nu}-\frac{1}{2}\eta_{\mu\nu}
\delta{R})=\nn\\ &&-(\eta_{\mu\nu}\Box -\D_\mu\D_\nu)(\delta
\phi-\frac{2}{3}(f_{P0}+f_{Q0})\delta{R})\nn
\\&&-\eta_{\mu\nu} \D_\alpha\D_\beta (f_{P0} \delta{R}^{\alpha\beta})-\Box(f_{P0}
\delta{R}_{\mu\nu})\nn\\ &&+2
\D_\alpha\D_\beta(f_{P0}~\delta{R}^\alpha_{~(\mu}\delta^\beta_{~\nu)}+2
f_{Q0}~\delta{R}^{\alpha~~~~\beta}_{~(\mu\nu)})\nn\\ \ea
It is convenient to
work in Fourier space where the following substitutions have to be operated:  $\D_\gamma
h_{\mu\nu}\rightarrow i k_\gamma h_{\mu\nu}$ and $\Box h_{\mu\nu}
\rightarrow -k^2 h_{\mu\nu}$.
The above equation becomes
\ba
&& F_0(\delta{R}_{\mu\nu}-\frac{1}{2}\eta_{\mu\nu}
\delta{R})=\nn\\ &&(\eta_{\mu\nu}k^2 -k_\mu k_\nu)(\delta
\phi-\frac{2}{3}(f_{P0}+f_{Q0})\delta{R})\nn
\\&&+\eta_{\mu\nu} k_\alpha k_\beta (f_{P0} \delta{R}^{\alpha\beta})+k^2(f_{P0}
\delta{R}_{\mu\nu})\nn\\ &&-2 k_a
k_b(f_{P0}~\delta{R}^a_{~(\mu}\delta^b_{~\nu)})-4 k_\alpha k_\beta(
f_{Q0}~\delta{R}^{\alpha~~~~\beta}_{~(\mu\nu)})\nn\\ \label{fields2}\ea
We
can rewrite the metric perturbation as \be
h_{\mu\nu}=\bar{h}_{\mu\nu}-\frac{\bar{h}}{2}~
\eta_{\mu\nu}+\eta_{\mu\nu} h_f \label{gauge}\ee and impose  the standard gauge conditions
$\D_\mu
\bar{h}^{\mu\nu} =0$ and $\bar{h}=0$. The first of these
conditions implies that $k_\mu \bar{h}^{\mu\nu} =0$ while the
second gives \ba h_{\mu\nu}&=&\bar{h}_{\mu\nu}+\eta_{\mu\nu} h_f \nn \\
h&=&4 h_f\ea
Inserting into the perturbed curvature quantities, we get
 \ba \delta
R_{\mu\nu}&=&\frac{1}{2}\left(2k_\mu k_\nu h_f+k^2 \eta _{\mu\nu}
h_f+k^2 \bar{h}_{\mu\nu}\right) \nn\\
\delta R &=& 3k^2 h_f\nn\\
k_\alpha k_\beta ~\delta
R^{\alpha~~~~~\beta}_{~~(\mu\nu)~}&=&-\frac{1}{2}\left((k^4
\eta_{\mu\nu}-k^2 k_\mu k_\nu)h_f+k^4 \bar{h}_{\mu\nu}\right)\nn\\
k_\alpha k_\beta~\delta{R}^\alpha_{~(\mu}\delta^\beta_{~\nu)}&=&\frac{3}{2}k^2k_\mu
k_\nu h_f \nn\\ \label{results1}\ea
Substituting  Eqs.
(\ref{gauge})-(\ref{results1}) into (\ref{fields2}) and after some
algebra we get: \ba
&&\frac{1}{2}\left(k^2-k^4\frac{f_{P0}+4f_{Q0}}{F_0}\right)\bar{h}_{\mu\nu}=\nn\\
&&(\eta_{\mu\nu}k^2 -k_\mu k_\nu)\frac{\delta \phi}{F_0}
+(\eta_{\mu\nu}k^2 -k_\mu k_\nu)h_f \nn\\\ea
Defining $h_f\equiv
-\frac{\delta \phi}{F_0}$ we find the equation for the
perturbations:
\be
\left(k^2+\frac{k^4}{m^2_{spin2}}\right)\bar{h}_{\mu\nu}=0
\label{solution} \ee
 where  $m^2_{spin2}\equiv
-\frac{F_0}{f_{P0}+4f_{Q0}}$.
 From Eq. (\ref{kgordon1}) we
get: \be \Box h_f=m_s^2 h_f \label{kgordon3}\ee
From Eq.
(\ref{solution}) it is easy to see that we have a modified
dispersion relation which corresponds to a massless spin-2 field
($k^2=0$) and a massive spin-2  field
$k^2=\frac{F_0}{\frac{1}{2}f_{P0}+2f_{Q0}}\equiv -m^2_{spin2}$.
 To see this better this point, let us  note that the propagator for
$\bar{h}_{\mu\nu}$ can be rewritten as \be G(k) \propto
\frac{1}{k^2}-\frac{1}{k^2+m^2_{spin2}} \ee Clearly the second
term has the opposite sign, which indicates the presence of a
ghost energy mode (see also
\cite{Nunez:2004ts,Chiba:2005nz,Stelle:1977ry}).

As a "sanity check", we can see that for the Gauss-Bonnet term
$\mathcal{L}_{GB}=Q-4P+R^2$ we have $f_{P0}=-4$ and $f_{Q0}=1$.
Then, Eq. (\ref{solution}) simplifies to $k^2
\bar{h}_{\mu\nu}=0$ and, in this case, we have no negative
energy modes as expected.

The solutions of Eqs. (\ref{solution}) and (\ref{kgordon3}) can be
written in terms of plane waves \ba \bar{h}_{\mu\nu}&=&A_{\mu\nu}
(\overrightarrow{p}) \cdot  exp(ik^\alpha x_\alpha)+cc \label{pw1}
\ea \ba h_f &=& a(\overrightarrow{p}) \cdot exp(iq^\alpha
x_\alpha)+cc\label{pw2} \ea where

\begin{equation}
\begin{array}{ccc}
k^{\alpha}\equiv(\omega_{m_{spin2}},\overrightarrow{p}) &  & \omega_{m_{spin2}}=\sqrt{m_{spin2}^{2}+p^{2}}\\
\\q^{\alpha}\equiv(\omega_{m_s},\overrightarrow{p}) &  & \omega_{m_s}=\sqrt{m_s^{2}+p^{2}}.\end{array}\label{eq: k e q}\end{equation} and
where $m_{spin2}$ is zero (non-zero) in the case of massless
(massive) spin-2 mode and the polarization tensors $A_{\mu\nu}
(\overrightarrow{p})$ can be found in Ref. \cite{vanDam:1970vg}
(see equations (21)-(23)).  Eqs. (\ref{solution}) and
(\ref{pw1}), contain the equation and the solution for the standard gravitational waves
of GR \cite{gravitation} plus massive spin 2 terms.
Eqs. (\ref{kgordon3}) and (\ref{pw2}) are respectively the
equation and the solution for the massive scalar mode (see also
\cite{felix}).

The fact that the dispersion law for the modes of the massive
field $h_{f}$ is not linear has to be emphasized. The velocity of
every {}``ordinary'' (arising from GR)
mode $\bar{h}_{\mu\nu}$ is the light speed $c$, but the dispersion
law (the second of Eq. (\ref{eq: k e q})) for the modes of $h_{f}$
is that of a massive field which can be seen as a
wave-packet. Also, the group-velocity of a
wave-packet of $h_{f}$, centered in $\overrightarrow{p}$, is

\begin{equation}
\overrightarrow{v_{G}}=\frac{\overrightarrow{p}}{\omega},\label{eq: velocita' di gruppo}\end{equation}
which is exactly the velocity of a massive particle with mass $m$
and momentum $\overrightarrow{p}$.
From the second of Eqs. (\ref{eq: k e q}) and Eq. (\ref{eq: velocita' di gruppo})
it is straightforward to obtain:

\begin{equation}
v_{G}=\frac{\sqrt{\omega^{2}-m^{2}}}{\omega}.\label{eq: velocita' di gruppo 2}\end{equation}
This means that the  speed of the wave-packet is
\begin{equation}
m=\sqrt{(1-v_{G}^{2})}\omega.\label{eq: relazione massa-frequenza}
\end{equation}
Summarizing these results, we can say that considering Extended Theories of Gravity (which we have generically  assumed as  analytic functions of curvature invariants)  more gravitational modes than the standard massless ones of GR have to be taken into account.  These further modes can be derived from metric deformations and characterized as propagating particles.

%
%
%

%
%
%
%
%
%

\subsection{Polarization states}
\label{otto.due}

As we have seen,   there are
solutions of Eq. (\ref{solution})  depending on the value of
$k^2$. We have  $k^2=0$ modes that corresponds to a
massless spin-2 field with two independent polarizations (the standard polarization states of GR). If  $k^2\neq0$, we have  massive spin-2
ghost modes and there are five independent polarization tensors. The number of  polarizations can be easily achieved by the formula of spin degeneration $d=(2s+1)$
\cite{vanDam:1970vg}. A further scalar mode comes out from Eq.(\ref{kgordon3}). Due to the above formula for spin degeneration, it gives one polarization state.

Let us first consider the case where the
spin-2 field is massless.
Assuming $\overrightarrow{p}$ in the $z$ direction, a gauge in which
only $A_{11}$, $A_{22}$, and $A_{12}=A_{21}$ are different from zero
can be chosen. The condition $\bar{h}=0$ gives $A_{11}=-A_{22}$.
 In this frame,  we may take the bases of  polarizations defined as\footnote{The polarizations are
defined in the 3-space, not in a space-time. Each polarization mode is orthogonal to one
another and is normalized $e_{\mu\nu}e^{\mu\nu} =2\delta$. Note that other modes are not traceless, in contrast to the ordinary
plus and cross polarization modes in GR.}
\begin {equation}
e_{\mu\nu}^{(+)}=\frac{1}{\sqrt{2}}\left(
\begin{array}{ccc}
1 & 0 & 0 \\
0 & -1 & 0 \\
0 & 0 & 0
\end{array}
\right),\nonumber\qquad e_{\mu\nu}^{(\times)}=\frac{1}{\sqrt{2}}\left(
\begin{array}{ccc}
0 & 1 & 0 \\
1 & 0 & 0 \\
0 & 0 & 0
\end{array}
\right)\nonumber
\end{equation}
\begin {equation}
e_{\mu\nu}^{(s)}=\frac{1}{\sqrt{2}}\left(
\begin{array}{ccc}
0 & 0 & 0 \\
0 & 0 & 0 \\
0 & 0 & 1
\end{array}\right)
\end{equation}


Inserting these equations in Eq. (\ref{gauge}), it results

\ba h_{\mu\nu}(t,z)&=&A^{+}(t-z)e_{\mu\nu}^{(+)}
+A^{\times}(t-z)e_{\mu\nu}^{(\times)}\nn
\\&+&h_{s}(t-v_{G}z)e_{\mu\nu}^{s}\label{eq: perturbazione
totale}\ea
The terms
$A^{+}(t-z)e_{\mu\nu}^{(+)}+A^{\times}(t-z)e_{\mu\nu}^{(\times)}$
describe the two standard polarizations of gravitational waves
which arise from GR, while the term
$h_{s}(t-v_{G}z)\eta_{\mu\nu}$ is the massive field arising from
a generic Extended Theory of Gravity where further degrees of freedom can be represented by a scalar field.

When the spin-2 field is massive, we have six polarization (five due to the spin 2 and one due to the spin 0).
 Possible bases of    polarizations are
\begin {equation}
e_{\mu\nu}^{(+)}=\frac{1}{\sqrt{2}}\left(
\begin{array}{ccc}
1 & 0 & 0 \\
0 & -1 & 0 \\
0 & 0 & 0
\end{array}
\right),\nonumber\qquad e_{\mu\nu}^{(\times)}=\frac{1}{\sqrt{2}}\left(
\begin{array}{ccc}
0 & 1 & 0 \\
1 & 0 & 0 \\
0 & 0 & 0
\end{array}
\right)\nonumber
\end{equation}

\begin {equation}
e_{\mu\nu}^{(B)}=\frac{1}{\sqrt{2}}\left(
\begin{array}{ccc}
0 & 0 & 1 \\
0 & 0 & 0 \\
1 & 0 & 0
\end{array}
\right),\nonumber\qquad e_{\mu\nu}^{(C)}=\frac{1}{\sqrt{2}}\left(
\begin{array}{ccc}
0 & 0 & 0 \\
 0 & 0 & 1 \\
0 & 1 & 0
\end{array}
\right)\nonumber
\end{equation}
\begin {equation}
e_{\mu\nu}^{(D)}=\frac{\sqrt{2}}{3}\left(
\begin{array}{ccc}
\frac{1}{2} & 0 & 0 \\
0 & \frac{1}{2} & 0 \\
0 & 0 & -1
\end{array}
\right),\nonumber\qquad e_{\mu\nu}^{(s)}=\frac{1}{\sqrt{2}}\left(
\begin{array}{ccc}
0 & 0 & 0 \\
0 & 0 & 0 \\
0 & 0 & 1
\end{array}
\right)\label{tensorpol}\nonumber\,.
\end{equation}
The total  amplitude can be written in
terms of the 6 polarization states as

\ba
&&h_{\mu\nu}(t,z)=A^{+}(t-v_{G_{s2}} z)e_{\mu\nu}^{(+)}+A^{\times}(t-v_{G_{s2}} z)e_{\mu\nu}^{(\times)}\nn\\
&&+B^{B}(t-v_{G_{s2}} z)e_{\mu\nu}^{(B)}+C^{C}(t-v_{G_{s2}} z)e_{\mu\nu}^{(C)}\nn\\
&&+D^{D}(t-v_{G_{s2}}
z)e_{\mu\nu}^{(D)}+h_{s}(t-v_{G}z)e_{\mu\nu}^{s}.\nn\\\ea
where
$v_{G_{s2}}$ is the group velocity of the massive spin-2 field and
is given, as above,  by \be
v_{G_{s2}}=\frac{\sqrt{\omega^{2}-m_{s2}^{2}}}{\omega}.\label{spin2group}\ee

The first two polarizations are the same as in the massless case,
inducing tidal deformations on the x-y plane. In Fig.1, we
illustrate how each  polarization could affect test masses arranged
along a circle.
\begin{figure}
\begin{center}
\leavevmode
\centerline{\epsfig{figure=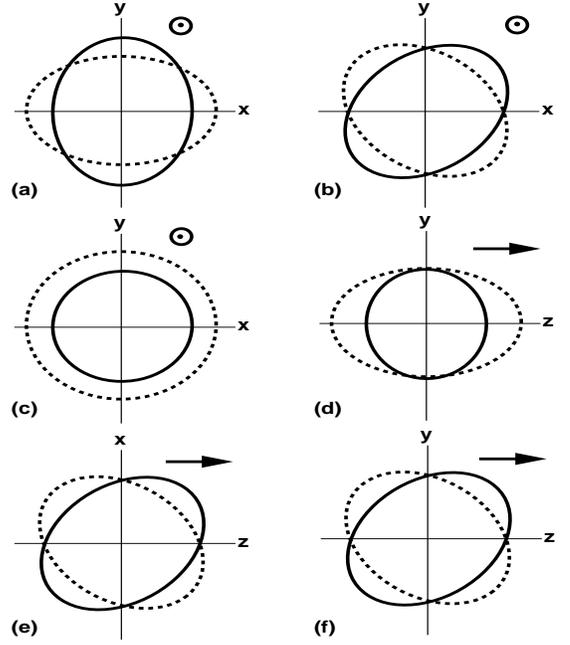,angle=360,height=8.5cm,width=7.25cm}}
\caption{The possible polarization modes of  gravitons. The picture shows
the displacement that each mode induces on a
sphere of test particles.   The mode propagates out of the plane in (a), (b), (c), and
it propagates in the plane in (d), (e) and (f).  In (a) and
(b) we have respectively the plus mode and cross mode, in (c) the
scalar mode, in (d), (e) and (f) the extra D, B and C modes induced by massive spin-2 gravitons.
This behavior could be, in principle, observed also for a beam of particles undergoing this interaction at suitable scales.}
\end{center}
\label{fig1}
\end{figure}

\subsection{The interpretation of gravitational  ghost modes }
\label{otto.tre}

The presence of  negative energy modes  could seem
 a pathology  from a purely quantum-mechanical
point of view. However, there are several interpretations that have to be taken into account for such phenomena.
A ghost mode can be
viewed as either a particle state of positive energy and negative
probability density, or a positive probability density state with
a negative energy. In the first case, allowing the presence of
such a particle will quickly induce violation of unitarity. The
negative energy scenario leads to a theory where there is no energy
minimum  and the system thus becomes unstable with
growing amplitudes. The vacuum can decay into pairs of
ordinary and ghost gravitons leading to an
instability. In other words, such a dynamics induces a symmetry breaking and this fact can be directly related to the fact that $GL(4)$ is  a non-unitary group. In order to regularize the theory,  dynamical ghost modes can be cancelled out taking into account  yet higher derivative terms. This  leads to a higher
order effective action  as the one in  Eq.(\ref{3.1.1.})  which, in principle, is an infinite order effective action  \footnote{In this case, we should consider higher order powers in the $\Box$-operator.}.

One way out of such problems is to ask for a very weak coupling of
the ghost with  the other particles in the theory. In this case,
the decay rate of the vacuum becomes comparable to the inverse
of a suitable length scale\footnote{The largest  length is the Hubble scale. This gives rise to the weakest physical coupling. }. If such a scale is extremely large,
the present vacuum state will  appear to
be sufficiently stable. This option is   viable   when
 ghost states, due to a different interaction length,
couple differently than the ordinary massless gravitons  with the other forms of matter.

Another interpretation is to assume that this picture does not hold up to
arbitrarily high energies but, at some cutoff scale
$M_{cutoff}$,  the theory modifies appropriately  to ensure a
ghost-free behavior and a stable ground state. This could  happen, for
example, if the Lorentz invariance is violated at a certain
$M_{cutoff}$, thereby restricting any potentially harmful decay
rates. This possibility could be extremely interesting for the investigations at LHC,
 being a sort of gravitationally induced Higgs-like mechanism \cite{Emparan:2005gg, achucarro}. .

However, we have to point out that  Extended Theories of Gravity could not
 hold up to arbitrary high energies. Such models are plagued, at  fundamental quantum level,
by the same problems as ordinary GR, but Extended Theories of Gravity are
renormalizable  at one loop-level \cite{birrell} . They are not proper
  candidates for a full quantum gravity theory (in canonical sense)  but
the corresponding ghost  particle interpretation (virtual
massive modes)  is a useful approach to address, at lower energies than Planck scales,  shortcomings of Standard Model.

At semi-classical
level, the perturbation $h_{\mu \nu}$ (deformation) is a tensor representing the ``stretching'' of space-time
away from flatness. A ghost mode then makes sense as just another
way of propagating this perturbation of the space-time geometry,
one which carries the opposite sign in the propagator than an
ordinary massive graviton would.

Viewed in this way, the presence of the massive ghost graviton
will induce the same effects as an ordinary
massive graviton transmitting the perturbation, but with the
opposite sign in the displacement. Tidal stretching from a
polarized wave on the polarization plane will be turned into
shrinking and vice-versa. This signal will, at the end, be a
superposition of the displacements coming from the ordinary
massless spin-2 graviton and the massive ghost. Since these modes  induce
two competing effects, this fact would lead to a less pronounced signal
than the one we would expect if the ghost mode was absent, setting
in this way less severe constraints on the theory.  On
the other hand,  treating ghost modes  as small perturbations
could be not sensible. As stated above,  the dynamical ghost modes
could be cancelled by other higher derivative terms. In that case,
nonetheless, it might still make sense to analyze the impact on
propagation owing to virtual massive mode effects (Yukawa terms)
on the massless modes. However, the presence of
the new modes will also affect the total energy density carried by
the graviton and could induce new effects at ultraviolet scales, as we will discuss below.
To conclude, effects and symmetry breaking discussed here could give rise to relevant signatures in colliding beams of particles (e.g. hadrons)  at TeV scales.

\section{Massive Gravitational States  and the induced symmetry breaking}
\label{nove}
\subsection{A gravitational cut-off at TeV scales}
\label{nove.uno}

The above results could be interesting  to investigate quantum gravity effects  and symmetry breaking in the range between  GeV and TeV scales.
Such scales are actually investigated by  the experiments at LHC.
It is important to stress that any ultra-violet  model of  gravity (e.g. at TeV scales)  have to explain also the observed weakness of gravitational effects at  largest (infra-red)  scales.
This means that  massless (or quasi-massless)  modes have to be considered in any case.

The above  5D-action (\ref{a4})  is  an example of  higher dimensional action
where the effective  gravitational energy scale (Planck scale)  can be "rescaled" according to Eqs. (\ref{450}) and (\ref{pippo}). In terms of mass, being $M_p^2=\frac{c\hbar}{G_N}$  the constraint coming from the ultra-violet limit of the theory ($10^{19}$ GeV) ,   we can set   $M_p^2 = M_{\sharp}^{D-2} V_{D-4}$, where $V_{D-4}$ is the "volume" coming from  the extra dimension.  It is easy to see that  $V_{D-4}$, in the 5D case,  is  related to the fifth component of $\Phi$, that is $\psi(x_4)$ in Eq.(\ref{split1}).  $M_{\sharp}$  is the above discussed cut-off mass that becomes relevant as soon as the Lorentz invariance is violated. Such a scale, in the context discussed here could be  of the order TeV.

As we have shown in Sec. \ref{quattro},  it is  quite  natural  to obtain effective theories  containing  scalar  fields of gravitational origin. In this sense,  $M_{\sharp}$  is  the result of  dimensional reduction. To be more explicit, the  4D dynamics is led by the effective potential $V(\phi)$  and the non-minimal coupling $F(\phi)$. Such functions   could be experimentally tested since related to massive states.
In particular,  the effective Extended Gravity, produced by the reduction mechanism from 5D to 4D, can be chosen  as
\begin{eqnarray}
{\cal A}=\int d^4x ~\sqrt{-g} ~\Big [ -\frac{\phi^2}{2}  R
~+\frac{1}{2} g^{\mu\nu}\partial_\mu\phi \partial_\nu\phi
-V\Big]
\label{model}
\end{eqnarray}
plus contributions of  ordinary matter terms.  The potential for $\phi$  can be assumed as
\begin{eqnarray}
V(\phi) = \frac{M_{\sharp}^2}{2} \phi^2 + \frac{\lambda}{4} \phi^4\,,
\end{eqnarray}
where  a massive  term and the self-interaction  term are present. This is the standard choice of quantum field theory which perfectly fits with the arguments of dimensional reduction.
Let us recall again that  the scalar field  $\phi$ is not put {\it by hand} into dynamics but it is given by the extra degrees of freedom of  gravitational field generated by the reduction process in 4D.
It is easy to derive the  vacuum expectation value of $\phi$, being
\begin{eqnarray}
M_{\sharp}^2 = 2\lambda M_{p}^2\,,
\end{eqnarray}
which is a fundamental scale of the theory.

Some considerations are in order at this point. Such a scale has to be confronted with Higgs vacuum expectation value which is 246 GeV and then with the {\it hierarchy problem}.
 If $M_{\sharp}$ is larger than Higgs mass, the problem is obviously circumvented.
It is important to recall that  hierarchy problem occurs when couplings and  masses of effective  theories  are very different  than the parameters measured by experiments. This  happen since measured parameters are related to the fundamental parameters by renormalization and  fine cancellations between the fundamental quantities and the quantum corrections are necessary. The hierarchy problem is  essentially a fine-tuning problem.

In particle physics, the question is  why the weak force is stronger and stronger  than gravity. Both of these forces involve constants of nature, Fermi's constant for the weak force and Newton's constant for gravity. From  the Standard Model, it appears that Fermi's constant is unnaturally large and should be closer to Newton's constant, unless there is a fine cancellation between the bare value of Fermi's constant and the quantum corrections to it.

More technically, the question is why the Higgs boson is so much lighter than the Planck mass (or the grand unification energy). In fact, researchers are searching for Higgs masses ranging from 115 up to 350 GeV with different selected decay channels from $b\bar{b}$ to $t\bar{t}$ (see for example \cite{camy} and refereces therein).   One would expect that the large quantum contributions to the square of the Higgs boson mass would inevitably make the mass huge, comparable to the scale at which new physics appears, unless there is an incredible fine-tuning cancellation between the quadratic radiative corrections and the bare mass.
With this state of art, the problem cannot  be formulated in the  context of the Standard Model where the Higgs mass cannot be calculated. In a sense, the problem is solvable if, in a given effective theory of particles, where the Higgs boson mass is calculable, there are no  fine-tunings.
If one accepts the {\it big-desert}  assumption and  the existence of a hierarchy problem, some new mechanism (at Higgs scale) becomes necessary to avoid  fine-tunings.

 The  model which we are discussing  contains a "running"  scale that could avoid to set precisely the Higgs scale. If the mass of the field $\phi$ is in  TeV region, there is no hierarchy problem
 being $\phi$ a gravitational scale. In this case, the Standard Model holds up plus an extended  gravitational sector derived from the fifth dimension.

 In other words,  the Planck scale can be   dynamically derived from  the vacuum expectation value of  $\phi$.   In some sense,  our model, in its low energy realization, works like the   model proposed by Antoniadis et al \cite{Antoniadis:2001sw}. The Planck scale can be recovered, as soon as the coupling $\lambda$ is of the order $10^{-31}$. Action (\ref{model}) (and the other Extended Gravity Theories discussed above)  as  effective models  valid up to a cutoff scale of a few $M_\sharp \sim$ TeV (see also \cite{Minkowski:1977aj,Zee:1978wi,calmet}). The tiny value of
 $\lambda$, coming from the extra dimension,  is in agreement with the above considerations allowing the presence of physical (quasi-) massless gravitons with very large interaction lengths.

 As in \cite{Antoniadis:2001sw}, gravity is weak  although the gravitational scale  is  low.
 Also the string theory limit corresponds to a large scalar field  vacuum expectation value at TeV.
 It is important to stress that, by a
conformal  transformation from the Jordan frame to the Einstein frame, the Planck scale is decoupled from the vacuum expectation of the scalar field $\phi$. However the scalar field redefinition has to preserve the vacuum of the theory. Besides,   the gauge couplings and masses  depend on the vacuum expectation value of  $\phi$ and are dynamically determined. This means that Standard Model and Einstein Gravity (in the above conformal-affine sense) could  be recovered {\it without the hierarchy problem}. As discussed in \cite{calmet1},  it is possible to show that
the operators generated by the self-interaction of the scalar field are of the form
 \begin{equation}
 \frac{1}{M_\sharp^{N-4}}\lambda^\frac{N}{2} \phi^N
 \end{equation}
 and  they are
always suppressed by the small parameter $\lambda$ and  do not destabilize the potential of the theory. This result holds also for perturbative corrections coming from quantum gravity.

Considering again  the problem of mass  generation,   one can assume  that particles of Standard Model  have sizes related to the cut off, that is
$M_\sharp^{-1}$, and  their collisions could lead to the
formation of bound states as in   \cite{Antoniadis:2001sw, Antoniadis:1998ig}.  Potentially, such a phenomenon could   mimic the
decay of semi-classical  quantum  black holes and, at lower energies,
it could be useful to investigate
substructures  of    Standard Model.
This means that  we should expect some strong scattering effects in the TeV region  involving the coupling of  $\phi$  to the Standard Model fields. The "signature" of this phenomenon could lead to polarization effects of the particle beam as discussed in the previous section.  Furthermore the  strong dynamics derived from the phenomenon  could resemble compositeness as discussed in \cite{Meade:2007sz}.  Furthermore,  bounds on the production of mini-black holes can be derived from astroparticle physics
\cite{Feng:2001ib,Anchordoqui:2001cg,Anchordoqui:2003jr,Ringwald:2001vk,Kowalski:2002gb}.  In \cite{Anchordoqui:2001cg}  a bound on the cross-section is
\begin{eqnarray}
\sigma_{\nu N \to BH + X} <  \frac{0.5}{\mbox{TeV}^2}.
\end{eqnarray}
Assuming, in our case,  the cross-section  $\sigma = M_\sharp^{-2}$, we get  a bound of TeV-order. If the fundamental scale of our theory  is of this order, strong scattering processes  at  LHC would have the cross-section
\begin{eqnarray}
 \sigma_{(pp  \to  grav. ghosts + X)}\sim 1 \times 10^7 \mbox{fb}
\end{eqnarray}
and would  dominate the cross-sections expected from the Standard Model. In this case,   the Higgs boson could not be detected and no hierarchy problem would be present.

\subsection{Gravitational and electroweak interactions}
\label{nove.due}

The Higgs mechanism is an approach that allows: $i)$ to generate the masses of  electroweak gauge bosons; $ii)$ to preserve  the perturbative unitarity of the S-matrix;   $iii)$ to preserve
the renormalizability of the theory.
 The masses of  the electroweak bosons can be written in a gauge invariant form using either the non-linear sigma model  \cite{CCWZ2} or a gauge invariant formulation of the electroweak bosons.  However  if there is no propagating Higgs boson, quantum field  amplitudes describing  modes of the electroweak bosons grow too fast violating the unitarity  around  TeV scales
  \cite{LlewellynSmith:1973ey,Lee:1977yc,Lee:1977eg,Vayonakis:1976vz}. There are several ways in which unitarity could be restored  but the Standard Model without a Higgs boson is
  non-renormalizable  at  perturbative level.

 A possibility is that the weak interactions  become strongly coupled at TeV scales and then  the related gauge theory becomes unitary   at non-perturbative level. Another possibility for models without a Higgs boson consists in introducing weakly coupled new particles to delay the unitarity problem into the multi TeV regime where the UV limit of the Standard Model is expected to become relevant. Dvali et al. \cite{Dvali:2010jz}  proposed that, as black holes in gravitational scattering, classical objects could form in the scattering of longitudinal W-bosons leading to unitary scattering amplitude.

 These  ideas are very intriguing and show  several features  of  electroweak interactions. First of all, the Higgs mechanism is strictly necessary  to generate masses for the electroweak bosons.
  Beside, some mechanisms can be  unitary but not renormalizable or vice-versa. In summary, the paradigm is that  three different  criteria should be fulfilled: $i)$ a gauge invariant generation of masses of electroweak bosons, $ii)$ perturbative unitarity;  $iii)$ renormalizability of the theory.

 Here we have proposed an alternative approach, based on Extended Theories of Gravity deduced from a 5D-manifold where the Standard Model is fully recovered enlarging the gravitational sector but avoiding the Higgs boson and the hierarchy problem.

It is important to point out that, in both the non-linear sigma model   and in gauge invariant formulation of Standard Model, it is possible to define an action in terms of an expansion in the scale of the electroweak interactions $v$. The action  can be written  as
  \begin{eqnarray} \label{effaction}
{\cal A}={\cal A}_{SM w/o Higgs}+\int d^4 x  \sum_i \frac{C_i}{v^N} O^{4+N}_i\,,
\end{eqnarray}
 where $O^{4+N}_i$ are operators compatible with the symmetries of the model. The electroweak bosons are gauge invariant fields defined by
 \begin{eqnarray}
\underline W^i_\mu&=& \frac{i}{2g} \mbox{Tr} \ \Omega^\dagger \stackrel{\leftrightarrow}
{D_\mu} \Omega \tau^i\,,
\end{eqnarray}
with $D_\mu=\partial_\mu - i  g  B_\mu(x)$ and
  \begin{eqnarray}
\Omega=\frac{1}{\sqrt{\phi^\dagger
    \phi}}\left(\begin{array}{cc}  \phi_2^* & \phi_1
    \\ -\phi_1^* & \phi_2
  \end{array}
\right )\,,
\end{eqnarray}
 where
 \begin{eqnarray}
\phi=\left(\begin{array}{c}  \phi_1 \\
   \phi_2
  \end{array}
\right ).
\end{eqnarray}
is a $SU(2)_L$ doublet scalar field which is considered to be a dressing field and does not need to propagate.  The same approach can be applied to fermions \cite{tHooft:1998pk,'tHooft:1980xb,Visnjic:1987pj}.

The analogy between the effective action for the electroweak interactions (\ref{effaction}) and that of Extended Gravity is striking. Considering only the leading terms, the above theory can be written as a Taylor series of the form
\begin{equation}
\label{effaction2}
f(R)\simeq \Lambda+f_0' R+\frac{1}{2!} f_0''R^2+\frac{1}{3!}f_0'''R^3+.....
\end{equation}
where the coefficients are the derivatives of $f(R)$ calculated at a certain value of $R$. Clearly,  as shown in previous sections,, the extra gravitational degrees of freedom
 can be suitably transformed in a scalar field $\phi$ which allows to avoid the hierarchy problem.
Both electroweak theory and  Extended Gravity have a dimensional energy scale which defines the strength of the interactions. The Planck mass sets the strength of  gravitational interactions
 while the weak scale $\lambda$ determines the range and the  strength of the electroweak interactions.  As shown in the previous subsection, these scales can be compared at TeV energies.

 In other words, the electroweak bosons are not gauge bosons in standard sense but they can be "derived" from the above further gravitational degrees of freedom. The local $SU(2)_L$ gauge symmetry is imposed at the level of the quantum fields. However there is a residual global $SU(2)$ symmetry, {\it i.e.} the custodial symmetry. In the case of gravitational theories formulated as the $GL(4)$-group of diffeomorphisms,  tetrads are an unavoidable feature necessary to construct the theory. They are gauge fields which transform under the local Lorentz
transformations $SO(3,1)$ and under general coordinate transformations, the metric $g_{\mu\nu}= e^a_\mu e^b_\nu \eta_{ab}$ which is the field that is being quantized, transforms under general coordinate transformations which is the equivalent of the global $SU(2)$ symmetry for the weak interactions (in our case the residual $GL(2)\supset SU(2)$) . Such an analogy between the tetrad fields and the Higgs field is extremely relevant. As shown above for deformations,   we can say that the Higgs field has the same role of the tetrads for the electroweak interactions while the electroweak bosons have the same role  of the metric. Dynamics is given by deformations.

A gravitational action like (\ref{effaction2})  is, in principle,   non-perturbatively renormalizable  if,  as shown by Weinberg,    there is a non-trivial fixed point which makes the gravity
asymptotically free \cite{fixedpoint}. This scenario implies that only a finite number of the Wilson coefficients in the effective action would need to be measured and the theory would thus be predictive and probed at LHC.

Measuring the strength of the electroweak interactions in the electroweak W-boson scattering could easily reveal a non-trivial running of the electroweak scale $v$. If an electroweak fixed point exists,  an increase in the strength of the electroweak interactions could be found, as in the strongly interacting W-bosons scenario, before the electroweak interactions become very weak and eventually irrelevant in the fixed point regime.  In analogy to the non-perturbative running of the non-perturbative Planck mass,  it is possible to introduce an effective weak scale
 \begin{eqnarray}
v_{eff}^2=v^2\left(1+\frac{\omega}{8\pi} \frac{\mu^2}{v^2} \right)\,,
\end{eqnarray}
where $\mu$ is an arbitrary mass scale, $\omega$ a non-perturbative parameter which determines the running of the effective weak scale  and $v$ is the weak scale measured at low energies. If $\omega$ is positive, the electroweak interactions would become weaker with increasing center of mass energy. This  asymptotically free  weak interaction  would be renormalizable at the non-perturbative level without having a propagating Higgs boson again in analogy to Extended Gravity.

The asymptotically free weak interaction scenario could also solve the unitarity problem of the standard model without a Higgs boson.
In the standard model without a Higgs boson, there are five amplitudes contributing  at tree-level to the scattering of two longitudinally polarized electroweak W-bosons. Summing these five amplitudes, one finds at order $s/M_W^2$
\begin{eqnarray}
&&{\cal A}(W^+_L+W_L^-  \to W^+_L+W_L^-)= \frac{s}{v_{eff}^2} \left ( \frac{1}{2} + \frac{1}{2} \cos \theta \right )\,,\nonumber\\
\end{eqnarray}
where $s$ is the center of mass energy squared and $\theta$ is the scattering angle. Clearly if $v_{eff}$ grows fast enough with energy, the  ultra-violet  behaviour of these amplitudes can be compensated and the summed amplitude can remain below the unitary bound. A similar proposal has been made to solve problems with unitarity in extra-dimensional models \cite{Hewett:2007st}.

It is important to stress that our approach does not require  new physics but to take only into account the whole budget of gravitational degrees of freedom. The  monitoring of the strength of the electroweak interactions in the W-bosons scattering at LHC could establish the existence of a fixed-point in the weak interactions. Using the one-loop renormalization group of the weak scale could help in formalizing this picture \cite{Arason:1991ic}. To be more precise, let us consider the scale of electroweak interactions
\begin{eqnarray}
v(\mu)=v_0 \left (\frac{\mu}{\mu_0} \right)^\frac{\gamma}{16 \pi^2}\,,
\end{eqnarray}
where
\begin{eqnarray}
\gamma=\frac{9}{4}\left ( \frac{1}{5} g_1^2 + g_2^2\right) -Y_2(S)
\end{eqnarray}
and
\begin{eqnarray}
Y_2(S)= \mbox{Tr} (3 \mbox{Y}_u^\dagger  \mbox{Y}_u + 3 \mbox{Y}_d^\dagger  \mbox{Y}_d + \mbox{Y}_e^\dagger  \mbox{Y}_e)\,,
\end{eqnarray}
where $\mbox{Y}_i$ are the respective Yukawa matrices. If the theory is in the perturbative regime e.g. at $m_W$, the Yukawa coupling of the top dominates since at this scale $g_1=0.31$ and $g_2=0.65$ and $\gamma$ is negative. In this case, the scale of the weak interactions become smaller. If the weak interactions become strongly coupled at TeV region, $g_2$ becomes large and $\gamma$ is expected to become positive.  We obtain the expected running and  the weak scale becomes larger. This is not possible in the framework of a perturbative approach. This result could represent a "signature" for   the approach presented here. However, we stress once again that there are indications of a non-trivial fixed point for the non-linear sigma model using exact renormalization group techniques \cite{Fabbrichesi:2010xy,Percacci:2009fh}.
In conclusion,  the unitarity problem of the weak interactions could be fixed by a non-trivial fixed point in the renormalization group of the weak scale. A similar mechanism could also fix the unitarity problem for fermions masses \cite{Appelquist:1987cf,Maltoni:2000iq,Maltoni:2001dc,Dicus:2004rg,Dicus:2005ku}  if their masses are not generated by the standard Higgs mechanism but in the same way considered here (let us remind that also $SU(3)$ could be generated by the splitting of $GL(4)$-group). In the case of  electroweak interactions this approach could  be soon checked   at LHC but good indications are also available for QCD \cite{atlas}.

\section{Discussion and conclusions}
\label{dieci}

The goal of this work is to give a unification
scheme of fundamental interactions based on a well defined
non-perturbative dynamics, the non-introduction of ad hoc hypotheses and the
consideration of the minimal necessary number of free parameters
and dimensions.

As general preliminaries, we have discussed the conformal-affine structure of fiber bundles showing that, in principle,  different  Extended Theories of Gravity can be conformally related each other.
After, we have discussed the group structure in 5D and in 4D-spaces showing how  the group of diffeomorphisms $GL(4)$ works from a reduction procedure from 5D to 4D. Such a group can be suitably split generating the fundamental groups of physical interactions.  In this respect, the group splitting is
\begin{equation}
 \underbrace{GL(4)}_{\underbrace{4\times 4}_{\mbox{diffeom.}}}\supset \underbrace{SU(3)}_{\underbrace{3^2-1}_{\mbox{gluons}}} \otimes \underbrace{SU(2)}_{\underbrace{2^2-1}_{\mbox{vec. bosons}}} \otimes \underbrace{U(1)}_{\underbrace{1}_{\mbox{photon}}}\otimes\underbrace{ GL(2)}_{\underbrace{2\times 2}_{\mbox{gravitons}}}
 \end{equation}
with further gravitational degrees of freedom.

After we have shown that the space-time deformations in 4D have a conformal  structure and the  $GL(4)$-algebra.

Starting with these mathematical tools, we  proposed a unification
scheme based on the assumption that a 5D-space can be defined
where conservation laws are always and absolutely conserved. Such
a  General Conservation Principle
\cite{conservation} holds since we ask for the
validity of the 5D-Bianchi identities which must be always
non-singular and invariant for every diffeomorphism.
The 5D-space is a smooth, connected and compact manifold where we
can derive field equations, geodesic equations and a globally defined
 Lorentz  structure.
The standard physics emerges as soon as
 we reduce from 5D to 4D-space recovering the $GL(4)$-group of diffeomorphisms. By a reduction procedure one is
capable  of generating  the masses of particles and their
organization in families.  The byproduct is an effective theory of gravity in 4D where further gravitational degrees of freedom naturally emerge, induced by  the  fifth dimension.
By this dynamics, we do not recover the Standard GR but Extended Theories of Gravity where non-minimal coupling, scalar field self-interaction potentials and higher-order curvature terms have to be considered.
These theories can be confronted and related, as we have seen, by conformal transformations.

The main feature is that higher-order terms or induced scalar fields enlarge the gravitational sector giving rise to massless, massive spin-2 gravitons and massive spin-0 gravitons. Such gravitational modes results in 6 polarizations, according to the prescription of the Riemann theorem stating that in a given $N$-dimensional space, $N(N-1)/2$ degrees of freedom are possible. The massive spin-2 gravitational states are ghost particles.  Their role result relevant as soon as we can define a cut-off mass at TeV scale (the vacuum state of the scalar field) that allows both to circumvent the hierarchy problem and the detection of the Higgs boson. In such a case, the Standard Model of particles should be confirmed without recurring to perturbative, renormalizable schemes involving new particles. The weakness of self-interaction  coupling would guarantee the fact that gravity  could  be compared, at TeV scale, with  electroweak interaction.

However, some crucial points have to be considered  in order to improve of the proposed approach.
 The main goal of our scenario is that the Standard Model of particles could be generated by the effective
gravitational interactions coming  from higher dimensions.  In particular the gauge symmetries and mass
generations could be achieved starting from conservation laws in 5D.  It is important to stress   that the Standard Model does
not mean only the gauge interaction but also quarks and leptons with their mass matrices that have to be exactly addressed.
In particular, the fermion sector has to be recovered.

It is well-known that the standard gauge interactions
contains the chiral gauge interactions, which, in our picture,
have to be generated from the gravitational interactions otherwise
there is no possibility to distinguish between the left-handed and
the right-handed particles. In particular, the $SU(2)$ part of the
standard gauge interactions, generated from $GL(4)$, has to be
chiral and, consequently,  fermions acquire a chiral
representation. To this end, torsion fields have to be
incorporated for the following reasons. As discussed in
\cite{Hehl1,Hehl2}, the Cartan torsion tensor plays the role of
spin source in the gravitational field equations where the affine
connection is not simply Levi-Civita. Furthermore, as demonstrated
in  \cite{bundle, CCSV1,CCSV3,CV4}, torsion plays an important
role in Extended Theories of Gravity since brings  further
gravitational  degrees of freedom responsible of chiral
interactions. In other words, torsion is not only the source of
spin but, thanks to the non-trivial structure of connections
$\Gamma_{\beta\gamma}^{\alpha}$, can  give rise to chiral
interactions of geometric origin \cite{cosimo}.  This means that
the $SU(2)_L$ and $SU(3)$ could be related to  a gravitationally
induced symmetry breaking process  where torsion plays a
fundamental role. Due to these facts,  the present approach has to
be generalized including torsion fields. Phenomenological studies
considering torsion and fermion interactions are already reported
in \cite{shapiro}.

Furthermore, as shown in \cite{cosimo}, space-like, time-like and null torsion tensors,
generated by non-trivial combinations of vector and bi-vector fields,
can be classified and represented by matrices which could explain
mass matrices of fermions and the hierarchy in the generation of quarks.
This approach agrees with other approaches where the effects of gluonic condensates
in holographic QCD can be encoded in  suitable deformations of 5D metrics (see e.g. \cite{cappiello}).

A detailed study in this sense will be  the argument of
forthcoming studies.

The validity of the presented scheme could be reasonably
checked at LHC in short time, due to the  increasing luminosities
of the set up. In fact,  the LHC experiments (in particular ATLAS
and CMS) are indicating, very preliminary,  the presence of  resonances and condensate
states that confirm the Standard Model but, up to now, cannot be
considered as evidences for the Higgs boson \cite{atlas}. Similar very preliminar
results, but with larger integrated luminosity, are reported also
by the CDF collaboration at Fermi Lab \cite{cdf}. The
interpretation of such data could be that the gravitational ghosts
(discussed here) would induce the formation of  resonances and
condensates without the presence of any Higgs field as discussed
in the previous sections.

 A final remark deserves the treatment of ghost states
that we have introduced here. According to the  Becchi-Rouet-Stora
(BRS) formulation of  gauge theories, negative norm states or
ghosts are allowed to propagate and are eliminated by applying a
particular   condition (BRS-condition) onto the state vectors
\cite{kaku}.  The approach can be expressed in terms of path
integrals.  As consequence, the vacuum is the ground state and
ghosts do not generate the negative energy states  without
conflicting with the Copenhagen Interpretation of  Quantum
Mechanics. In our case,  we are in a more general context. Our
gravitational ghosts are related to non-local effects connected to
the fact that  $GL(4)$ is a non-unitary group where some
sub-groups can be unitary. Such ghosts are related to the
gravitationally-induced symmetry breaking process and to the
further degrees of freedom emerging in the Extended Theories of
Gravity. In other words, they are not the  ghosts of the standard
BRS mechanism but further gravitational modes related to
non-locality. In a very general sense, we are compatible with the
Many Worlds Interpretation of Quantum Mechanics \cite{halliwell}
since any projection from the 5D manifold realizes a different
universe and then a different effective theory of gravity. This
point will be discussed in details in further works.

\begin{acknowledgments}
The Authors wish to thank Sergio Bertolucci for useful comments, common discussions and suggestions which allowed to improve the paper.
\end{acknowledgments}


\end{document}